\documentclass[fleqn,usenatbib]{mnras}

\usepackage{newtxtext,newtxmath}

\usepackage[T1]{fontenc}
\usepackage{ae,aecompl}

\usepackage{graphicx}	
\usepackage{amsmath}	
\usepackage{amssymb}	
\usepackage{bm}

\expandafter\def\expandafter\UrlBreaks\expandafter{\UrlBreaks
  \do\a\do\b\do\c\do\d\do\e\do\f\do\g\do\h\do\i\do\j%
  \do\k\do\l\do\m\do\n\do\o\do\p\do\q\do\r\do\s\do\t%
  \do\u\do\v\do\w\do\x\do\y\do\z\do\A\do\B\do\C\do\D%
  \do\E\do\F\do\G\do\H\do\I\do\J\do\K\do\L\do\M\do\N%
  \do\O\do\P\do\Q\do\R\do\S\do\T\do\U\do\V\do\W\do\X%
  \do\Y\do\Z}
  
\title[Relativistic effects: correlation-function dipole]{Imprints of relativistic effects on the asymmetry of the halo cross-correlation function: from linear to non-linear scales}

\author[Breton et al.]{
Michel-Andr\`es Breton,$^{1}$\thanks{E-mail: michel-andres.breton@obspm.fr}
Yann Rasera,$^{1,2}$
Atsushi Taruya,$^{2,3}$
Osmin Lacombe,$^{2,4}$ 
\newauthor
~Shohei Saga$^{2}$
\\
$^{1}$LUTH, Observatoire de Paris, PSL Research University, CNRS, Universit\'e Paris Diderot, Sorbonne Paris Cit\'e \\
5 place Jules Janssen, 92195 Meudon, France\\
$^{2}$Center for Gravitational Physics, Yukawa Institute for Theoretical Physics, Kyoto University, Kyoto 606-8502, Japan\\
$^{3}$Kavli Institute for the Physics and Mathematics of the Universe (WPI),
The University of Tokyo Institutes for Advanced Study\\
The University of Tokyo, 5-1-5 Kashiwanoha, Kashiwa, Chiba 277-8583, Japan\\
$^{4}$\'Ecole polytechnique, 91128 Palaiseau Cedex, France\\
}

\date{Accepted 2018 November 21. Received 2018 November 21; in original form 2018 March 12.}

\pubyear{2018}

\begin{document}
\label{firstpage}
\pagerange{\pageref{firstpage}--\pageref{lastpage}}
\maketitle

\begin{abstract}
The apparent distribution of large-scale structures in the universe is sensitive to the velocity/potential of the sources as well as the potential along the line-of-sight through the mapping from real space to redshift space (redshift-space distortions, RSD). Since odd multipoles of the halo cross-correlation function vanish when considering standard Doppler RSD, the dipole is a sensitive probe of relativistic and wide-angle effects. We build a catalogue of ten million haloes (Milky-Way size to galaxy-cluster size) from the full-sky light-cone of a new  ``RayGalGroupSims" N-body simulation which covers a volume of ($2.625~h^{-1}$Gpc)$^3$  with $4096^3$ particles. Using ray-tracing techniques, we find the null geodesics connecting all the sources to the observer. We then self-consistently derive all the relativistic contributions (in the weak-field approximation) to RSD: Doppler, transverse Doppler, gravitational, lensing and integrated Sachs-Wolfe. It allows us, for the first time, to disentangle all contributions to the dipole from linear to non-linear scales. At large scale, we recover the linear predictions dominated by a contribution from the divergence of neighbouring line-of-sights. While the linear theory remains a reasonable approximation of the velocity contribution to the dipole at non-linear scales it fails to reproduce the potential contribution below $30-60~h^{-1}$Mpc (depending on the halo mass). At scales smaller than $\sim 10~h^{-1}$Mpc, the dipole is dominated by the asymmetry caused by the gravitational redshift. The transition between the two regimes is mass dependent as well. We also identify a new non-trivial contribution from the non-linear coupling between potential and velocity terms. 
\end{abstract}




\section{Introduction}

Late time structure formation is a non-linear process which is very sensitive to the underlying cosmology. However we do not observe large-scale structures in themselves but rather an image of these objects via messengers (photons, neutrinos, gravitational waves, etc.). Most of our observations come from light, but the information transported by photons is altered during their path from the source to the observer. This leads to several distortions of the image and spectrum of the objects we are interested in. Because of lensing \citep{schneider1992gravitational,bartelmann2001weak}, the angular position as well as the shape/luminosity of objects can be modified: this is related to the bending of light near local energy overdensities and tidal deformations of light beams respectively.  The observed redshift of an object is also perturbed by its proper motion, its gravitational potential and light propagation in time-varying potentials. As a consequence, the comoving radial distance inferred from redshift (assuming a given homogeneous cosmology) is also perturbed. The apparent distribution of structures is therefore modified by redshift perturbations and lensing: this effect is called \emph{redshift-space distortions} (RSD\footnote{We stick to the terminology, RSD, as used widely in the community although we admit that this term is ambiguous. It should be taken as a synonymous of  ``Observed-Space Distortions" including all distortions induced by the presence of an observer.}) \citep{kaiser1987clustering,hamilton1992measuring}. RSD modify the position and properties of objects but also carry relevant cosmological information: studying these effects is a major contemporary challenge. Early RSD studies only took into account peculiar velocities at linear order in the distant-observer approximation. However with the improvement of data precision more subtle effects need to be accounted for: in \citet{szapudi2004wide,reimberg2016redshift} the distant-observer approximation has been relaxed leading to a wide-angle calculation. In \citet{papai2008nonperturbative, raccanelli2016doppler} a more sophisticated treatment of Doppler terms has been implemented and, in \citet{mcdonald2009} the effect of gravitational redshift has been included leading to an imaginary part of the power spectrum. 

However linear theory does not provide a fully satisfactory prediction since small scales are dominated by non-linear clustering and large-scale modes can also be affected by smaller scale modes through mode coupling. In fact, non-linear effects are already visible in real space at $100~h^{-1}$Mpc scales in Baryon Acoustic Oscillations (BAO) \citep{crocce2008nonlinear,taruya2012direct,rasera2014cosmic}. Moreover the mapping from real space to redshift space also becomes non-linear. The question of non-linearities has been addressed through semi-analytical approaches: either using different flavours of perturbation theory \citep{crocce2006renormalized,matsubara2008resumming,carlson2009critical,taruya2010baryon,taruya2012direct,taruya2013precision,carlson2013convolution}, using streaming models \citep{scoccimarro2004redshift,reid2011towards}, Effective Field Theory \citep{carrasco2012effective} or halo model \citep{tinker2008toward}. Alternatively, N-body simulations have been performed \citep{tinker2006redshift,tinker2007redshift} to investigate RSD beyond the quasi-linear regime. One important limitation of most of these works is that only standard RSD have been considered (distant-observer approximation and no relativistic effects).

Recently, several authors computed the observed galaxy number count including all relativistic effects at first order in the weak field approximation \citep{yoo2009new, yoo2010general, bonvin2011what,challinor2011linear}. 
Given the complexity of the calculation, all the terms were computed using the linear regime approximation and, assuming a linear mapping between real space and redshift space. These works allow to better understand the relative amplitude of all the contributions to the multipoles of the observed galaxy power spectrum (or two-point correlation function) at large linear scales. Measuring the so-called relativistic effects (i.e. beyond standard RSD effects) in a galaxy survey would be exciting since it would provide alternative ways of testing the nature of gravity and of the dark sector. Unfortunately the (usually studied) even multipoles of the observed halo correlation function are dominated by standard RSD and the detection of such effects might be challenging.

While standard RSD generate only even multipoles, the relativistic and wide-angle effects generate an asymmetry in the observed galaxy distribution (i.e. odd multipoles) when cross-correlating two halo populations living in different environments (or more, \citealt{bonvin2016optimising}). Using a multipole expansion of the linear cross-correlation function including all relativistic terms, \citet{bonvin2014asymmetric} has shown that the dipole is dominated by relativistic terms (which scale as $\mathcal{H}/k$, where $\mathcal{H}$ is the conformal Hubble constant and $k$ is the wavenumber of interest). Without confirmation by simulation, the validity of these results was however limited to large linear scales ($>100$~Mpc). At these scales the dipole generated by the gravitational potential is cancelled out because of velocity terms present in the Euler equation.

On the other hand at halo scale ($<$ few Mpc), the asymmetry of the distribution of galaxies caused by the gravitational potential of galaxy clusters has been investigated through analytical models \citep{cappi1995gravitational, kaiser2013measuring,zhao2013testing}
and simulations \citep{cai2017gravitational}. In these studies the relative shift between the mean redshift of two galaxy populations was considered instead of the dipole of the galaxy cross-correlation. \citet{wojtak2011gravitational} claimed a detection of this effect by stacking galaxy clusters: this has provided an alternative way to test gravity in cluster although the exact interpretation of the measurement is debated \citep{kaiser2013measuring,zhao2013testing}.

\citet{croft2013gravitational} proposed to use the same concept at larger scales in order to probe the gravitational redshift outside galaxy clusters. They introduce a new estimator since a clear boundary cannot be defined in the universe as is the case in galaxy clusters. The shell-estimator measures the relative shift of galaxy's redshift within spherical shells centred on galaxy clusters. In a recent paper, they have measured this estimator from \emph{snapshots} of N-body simulations \citep{zhu2017nbody}. Because of the noise related to the limited size of the simulation, they have used an artificial boost factor to increase the signal-to-noise ratio. Interestingly, \citet{alam2017relativistic} have claimed a detection of this effect within the SDSS survey. However, exact predictions for this estimator remains difficult at all scales \citep{giusarma2017relativistic}.  While predictions of the dipole at large linear scales are already well established, a clean dipole measurement from linear to non-linear scales within simulations (or observations) is still missing \citep{zhu2017nbody, gaztanaga2017measurement, alam2017relativistic}. 

In this paper, we directly measure the cross-correlation between two halo populations within the \emph{full-sky light-cone} of a larger and more resolved simulation. We use sophisticated ray-tracing techniques to self-consistently include all relativistic effects at first order in the weak-field approximation. For the first-time, we fill in the gap between large-scale linear predictions of the dipole (dominated by the contribution from the divergence of the line-of-sights) and small-scale non-linear expectations (dominated by the contribution from gravitational redshift). We also decompose all the contributions to the dipole, compare them to linear theory and shed light on to new ones.

The paper is organized as follows. In Section~\ref{sec:Theory} we review the theoretical predictions for dipole in the linear and non-linear regimes. We then present in Section~\ref{sec:Methods} the methodology used to compute halo cross-correlation function from our simulated light-cone. In Section~\ref{sec:Datavalid} we describe our halo catalogues and test our measurements. Finally in Section~\ref{sec:Results} we show the results of the dipole from linear to non-linear scales.

\section{Theory}
\label{sec:Theory}

In this paper, we will consider a well defined mass-limited collection of haloes within a given cosmological volume. We will not consider observational effects such as selection effects, magnification-bias, absorption/diffusion of light, redshift errors or the fact that galaxies can be hidden if they are aligned along the line-of-sight.

\subsection{Apparent halo overdensity: from real space to redshift space}
We consider scalar perturbations of the Friedmann-Lema\^itre-Robertson-Walker (FLRW) metric in conformal Newtonian gauge. The metric reads \citep{ma1995cosmological}

\begin{equation}
 \label{eq:metric}
   g_{\mu\nu}\textrm{d}x^{\mu}\textrm{d}x^{\nu} = a(\eta)^2 \left[-(1 + 2\psi/c^2)c^2\textrm{d}\eta^2 + (1 - 2\phi/c^2)\delta_{ij}\textrm{d}x^i\textrm{d}x^j \right],
\end{equation}
where $a$ is the expansion factor, $c$ is the speed of light, $\delta_{ij}$ is the Kronecker delta, $\psi$ and $\phi$ the Bardeen potentials \citep{bardeen1980gauge}, $x$ is the comoving position and $\eta$ the conformal time. Using $k_{\nu}k^{\nu} = 0$ (where $k_\nu$ are the components of the wavevector) and the lensing deviation equation we know that the apparent comoving position of a source is \citep{challinor2011linear}
 \begin{equation}
 \bm{s} = \chi\bm{n} + \frac{c}{H}\delta z\bm{n} - \bm{n}\int^{\chi}_{0}(\phi+\psi)/c^2 \textrm{d}\chi' - \int^{\chi}_{0}(\chi - \chi')\mathbf{\nabla}_{\perp}(\phi+\psi)/c^2 \textrm{d}\chi',
 \label{eq:distance}
\end{equation}
where $\bm{n}$ is a unit vector pointing towards the direction from which the unperturbed photon is coming and $\chi$ is the unperturbed comoving distance of the source. On the right hand side the first term $\bm{x}=\chi\bm{n}$ is the unperturbed comoving position of the source, the second term is the distance perturbation along the line-of-sight due to redshift perturbation $\delta z$. The third term is the (small) Shapiro effect and the last term is the transverse displacement due to lensing.

For the redshift perturbation we will consider the usual first order terms plus the special relativistic transverse Doppler effect that can be a non negligible fraction of the gravitational redshift at small scales \citep{zhao2013testing}
 \begin{eqnarray}
  \label{eq:perturbedredshift}
  \delta z = \frac{a_0}{a}\left \{\frac{\bm{v}\cdot\bm{n}}{c} - \frac{(\psi - \psi_0)}{c^2} +
  				\frac{1}{2}\left(\frac{v}{c}\right)^{2} - \frac{1}{c^2}\int^{\eta_0}_{\eta} \frac{\partial(\phi+\psi)}{\partial\eta}\textrm{d}\eta' \right \},
 \end{eqnarray}
 where $\bm{v}$ is the velocity and $\phi$ the potential. Quantities with the subscript $``0"$ are evaluated at the observer location today. In the above expression, we have assumed a comoving observer.
Assuming mass conservation gives
 \begin{eqnarray}
  \left(1 + \delta^{(\rm s)}(\bm{s})\right)\textrm{d}^3\bm{s} = \left(1 + \delta^{(\rm x)}(\bm{x})\right)\textrm{d}^3\bm{x},
  \label{eq:mass_conservation}
 \end{eqnarray}
 where $\delta^{(\rm s)}$ and $\delta^{(\rm x)}$ are the matter density contrast respectively in redshift space and real space. We have 
 \begin{eqnarray}
  \delta^{(\rm s)} = \left(1 + \delta^{(\rm x)}\right)|J|^{-1} - 1,
  \label{eq:nonlinearjacobian}
 \end{eqnarray}
 where $J$ is the Jacobian of the transformation from real space to redshift space.

\subsection{Two-point halo-halo cross-correlation function: linear theory}
\label{sec:lineartheory}
The halo-halo cross-correlation function between two halo populations $h_1$ and $h_2$ is given by 
\begin{eqnarray}
\xi_{h_1 h_2}(\vec{r_2}-\vec{r_1})=\langle \delta_{h_1}(\vec{r_1})  \delta_{h_2}(\vec{r_2}) \rangle,
\label{crosscorreq}
\end{eqnarray}
with $\delta_{h_i}(\vec{r_i})$ the overdensity of population $i$ and $\langle \rangle$ the ensemble average.
The cross-correlation function is related to the cross power spectrum through
\begin{equation}
 \label{eq:fouriertransform}
\begin{split}
 \langle \delta_{h_1}^{(\rm s)}(\bm{k}_1) \delta_{h_2}^{(\rm s)\ast}(\bm{k}_2) \rangle = \int \textrm{d}^3\bm{s}_1 \textrm{d}^3\bm{s}_2 
 \langle e^{-i\bm{k}_1\bm{s}_1} e^{i\bm{k}_2\bm{s}_2} \delta_{h_1}^{(\rm s)}(\bm{s}_1) \delta_{h_2}^{(\rm s)}(\bm{s}_2) \rangle ,
\end{split}    
\end{equation}
where $\delta^{(\rm s)}(\bm{k}_i)$ is the Fourier transform of $\delta^{(\rm s)}(\bm{s}_i)$. 
To rewrite this expression in terms of real space quantities we can use Eqs.~\eqref{eq:mass_conservation} and \eqref{eq:nonlinearjacobian}.
Eq.~\eqref{eq:fouriertransform} is the general formula for the power spectrum in redshift space but this leads to complicated mode couplings \citep{zaroubi1996clustering}.
In the linear regime, we can linearise the mapping between real and redshift space,
\begin{eqnarray}
    \Delta^{(\rm s)} = \Delta^{(\rm r)} + 1 - |J|,
    \label{eq:linearjacobian}
\end{eqnarray}
where we use $\Delta$ to denote the galaxy number count as an observable thus gauge invariant quantity.
Assuming no velocity bias, the observed galaxy number count is given by the sum of the following terms \citep{challinor2011linear,bonvin2011what,bonvin2014asymmetric,tansella2017}
\begin{eqnarray}
\label{eq:delta_first}
 \Delta^{\rm std} &=& b\delta - \frac{1}{\mathcal{H}} \nabla_{r} (\bm{v}\cdot\bm{n}), \\
 \Delta^{\rm acc} &=& \frac{1}{\mathcal{H}c} \dot{\bm{v}}\cdot\bm{n},   \\ 
 \Delta^{\rm q} &=& -\frac{\dot{\mathcal{H}}}{c\mathcal{H}^2} \bm{v}\cdot\bm{n},   \\  
 \Delta^{\rm div} &=& -\frac{2 }{\mathcal{H}\chi} \bm{v}\cdot\bm{n},   \\ 
 \Delta^{\rm pot, (1)} &=& \frac{1}{\mathcal{H}c} \bm{\nabla}_{r} \psi \cdot \bm{n},  \\ 
 \Delta^{\rm pot, (2)} &=& \left(\frac{\dot{\mathcal{H}}}{\mathcal{H}^2} + \frac{2 c}{\mathcal{H}\chi} \right) \psi/c^2 - \frac{1}{\mathcal{H}c^2}\dot{\psi},  \\ 
 \Delta^{\rm shapiro} &=& (\phi + \psi)/c^2 ,  \\ 
 \Delta^{\rm lens} &=& - \frac{1}{c^2} \int^{\chi}_{0}\frac{(\chi - \chi')\chi'}{\chi}\nabla^2_{\perp}(\phi+\psi) \textrm{d}\chi', \\
  \Delta^{\rm isw} &=& \frac{1}{\mathcal{H}c^2}(\dot{\phi}+\dot{\psi}), \\
 \Delta^{\rm LC} &=& \bm{v}\cdot\bm{n}/c,
 \label{eq:delta_last}
\end{eqnarray}
with $\delta$ the matter density contrast and $b$ a scale-independent bias.
 $\Delta^{\rm std}$ is the standard contribution to RSD \citep{kaiser1987clustering}, $\Delta^{\rm acc}$ the contribution from the acceleration of sources,  $\Delta^{\rm q}$ the contribution related to the acceleration of the expansion of the universe,  $\Delta^{\rm div}$ the contribution from the divergence of line-of-sights due to a finite observer, $\Delta^{\rm pot, (1)}$ the contribution from the gravitational redshift at first order in $\mathcal{H}/k$, $\Delta^{\rm pot, (2)}$ the contribution of the dominant terms in $\left(\mathcal{H}/k\right)^2$ to the gravitational redshift, $\Delta^{\rm shapiro}$ the contribution from the Shapiro time delay, $\Delta^{\rm isw}$ the contribution from the Integrated Sachs-Wolfe effect, $\Delta^{\rm lens}$ the lensing contribution equal to the lensing convergence as light-beam deformations modify the apparent source distribution, and $\Delta^{\rm LC}$ the light-cone contribution as the observed position of sources on the light-cone is different from their position on constant-time hypersurfaces due to peculiar velocities \citep{bonvin2014asymmetric}. A more refined calculation of this effect is given in \citet{kaiser2013measuring}.
We neglect the following terms, which are the subdominant $\left(\mathcal{H}/k\right)^2$ terms:

\begin{equation}
\label{eq:delta_neglect}
\begin{split}
\Delta_{\rm neglect} =& \left(\frac{\dot{\mathcal{H}}}{\mathcal{H}^2} + \frac{2 c}{\mathcal{H}\chi} \right)\frac{1}{c^2}\int^{\eta_0}_{\eta} \frac{\partial(\phi+\psi)}{\partial\eta}\textrm{d}\eta' \\
                       &+ \frac{2}{\chi c^2}\int^{\chi}_{0}(\phi+\psi) \textrm{d}\chi'.               
\end{split}
\end{equation}
It is straightforward to see that these terms are subdominant: for the dipole the only contribution of integrated terms comes from the integration between $\chi-r/2$ and $\chi+r/2$ where $\chi$ is the pair centre and $r$ the pair separation for the correlation function. We do not consider the term from transverse Doppler effect because it does not contribute to the correlation function in the linear regime. We also neglect higher order terms in redshift (see \citealt{bendayan2012second, umeh2014nonlinear}).
The linear correlation function of two different populations of galaxies is given by $\xi = \langle \Delta_1^{\rm s} \Delta_2^{\rm s}\rangle$. In the following we will focus on the terms that generate an asymmetry 
\begin{eqnarray}
  \xi^{\rm A} &=& \sum_i \langle\Delta_1^{\rm std} \Delta_2^{\rm A} \rangle + (1) \leftrightarrow (2),
\end{eqnarray}
where $\Delta^{\rm A}$ is given by Eqs.~\eqref{eq:delta_first} to~\eqref{eq:delta_last}. In the following, we will omit the subscript which indicates the halo population and we will implicitly assume that we perform the summation over the two populations (first and second term in the right-hand side).
Odd multipoles come from an asymmetry in the correlation function via exchange of pairs. If the position of each object of a pair is given by $\bm{x}_1$ and $\bm{x}_2$ we choose the convention $\bm{x} = (\bm{x}_1 + \bm{x}_2)/2$, $\bm{r} = \bm{x}_2 - \bm{x}_1$ and $\mu = \hat{\bm{x}}\cdot\hat{\bm{r}}$, where a hat denotes a unit vector. The angle defined this way is symmetric under exchange of pairs and we therefore do not need any additional geometrical term for the dipole due to the choice of angle \citep{reimberg2016redshift,gaztanaga2017measurement}.

The contributions to the dipole are the following \citep{bonvin2014asymmetric,tansella2017},
\begin{eqnarray}
\label{eq:potdipolelinear}
\langle \xi^{\rm acc}\rangle  &=& -\left(f^2 + \frac{\dot{f}}{\mathcal{H}} + \frac{\dot{\mathcal{H}}}{\mathcal{H}^2} f\right) \mathcal{G} \gamma^1_1(r),\\
\langle \xi^{\rm q}\rangle  &=& \frac{\dot{\mathcal{H}}}{\mathcal{H}^2} f \mathcal{G} \gamma^1_1(r),\\
\langle \xi^{\rm div}\rangle &=& \frac{2c}{\mathcal{H}\chi} f \mathcal{G} \gamma^1_1(r),\\
\langle \xi^{\rm pot, (1)}\rangle  &=& \frac{3}{2} \Omega_m(z) \mathcal{G} \gamma^1_1(r), \\
\langle \xi^{\rm pot, (2)}\rangle  &=& -\frac{3}{2} \Omega_m(z)\frac{\mathcal{H}}{\mathcal{H}_0} \mathcal{G}\gamma^2_1(r)
 \left(\frac{2c}{\mathcal{H}\chi} +\frac{\dot{\mathcal{H}}}{\mathcal{H}^2} - f + 1 \right) , \\
\langle \xi^{\rm shapiro}\rangle  &=& -3 \Omega_m(z) \frac{\mathcal{H}}{\mathcal{H}_0} \mathcal{G} \gamma^2_1(r), \\
\langle \xi^{\rm isw}\rangle  &=& 3 \Omega_m(z)\frac{\mathcal{H}}{\mathcal{H}_0} \mathcal{G}\gamma^2_1(r)
 \left(1-f \right), 
\end{eqnarray}
where we have $f = \partial\ln D_+/\partial\ln a$ the linear growth rate, $D_+$ the linear growth factor, $\mathcal{G} = (b_1 - b_2)\frac{\mathcal{H}}{\mathcal{H}_0}$ and
\begin{eqnarray}
 \gamma^m_{\ell}(r) = \frac{1}{2\pi^2} \left( \frac{\mathcal{H}_0}{c} \right)^m\int \textrm{d}k k^{2-m}j_{\ell}(kr)P(k,z).
\end{eqnarray}
with $j_{\ell}$ the $\ell^{th}$ spherical Bessel function and $P(k,z)$ the linear power spectrum at redshift $z$. A dot denotes a derivative w.r.t conformal time
The wide-angle term, coming from the fact that two haloes have different line-of-sights, reads
\begin{eqnarray}
 \langle \xi^{\rm wa}\rangle =  -\frac{2}{5}(b_1-b_2)f\frac{r}{\chi}\gamma^0_2(r).
\end{eqnarray}
The lensing term is given by \citep{matsubara2000gravitational,hui2007anisotropic,hui2008anisotropic}
\begin{eqnarray}
\langle \xi^{\rm lens}\rangle =    \langle \Delta^{\rm std}\Delta^{\rm lens}\rangle = -\frac{9}{4}\Omega_m(z)  \frac{r \mathcal{H}}{c} \mathcal{G} \varpi(r,z),
\end{eqnarray}
with 
\begin{eqnarray}
  \varpi(r,z) = \frac{1}{2\pi}\left(\frac{\mathcal{H}_0}{c}\right)\int_{-1}^1 \mu^2 \textrm{d}\mu \int \textrm{d}k k J_0\left(kr\sqrt{1-\mu^2}\right)P(k,z).
\end{eqnarray}
$J_0$ is the Bessel function of the first kind of order 0. The lensing effect is due to the fact that galaxies on the far side of a massive halo look more spread out than the ones in front of the halo due to light bending. This generates an observed under-dense region on the far side of the halo which leads to a negative dipole.
Following \citet{bonvin2014asymmetric}, the light-cone term due to peculiar velocities is given by 

\begin{eqnarray}
\langle \xi^{\rm LC}\rangle = -f \mathcal{G} \gamma^1_1(r).
\end{eqnarray}
The bias and growth factor are not constant. This leads to evolution terms 
\begin{equation}
\label{eq:evo1}
\langle \xi^{\rm evo1}\rangle =\frac{r}{6}\left\{\left[(b_1-b_2)f'-f(b_1'-b_2')\right]\left[\gamma_0^0(r) - \frac{4}{5}\gamma^0_2(r)\right] \right\}, 
\end{equation}
\begin{equation}
\langle \xi^{\rm evo2}\rangle =\frac{r}{2} \left(b_1 b_2' - b_1' b_2\right) \gamma_0^0(r),
\label{eq:evo2}
\end{equation}
where an apostrophe denotes a derivative w.r.t comoving distance. We find it convenient to split the evolution term in two parts because they appear in different configurations (depending on the velocity field) as we will see in Section~\ref{sec:Results}. 
In computing Eqs.~\eqref{eq:biassheth} and \eqref{eq:derivbiassheth}, while we use the bias parameter directly measured from our simulation, the comoving distance derivative of the bias is estimated from theoretical model, since the measured comoving distance derivative is basically noisy (see Fig.~\ref{fig:snapshotbias_derivbias}). Here, we specifically use the model given by \citet{sheth1999large}:

\begin{eqnarray}
 b_{\rm{ST}} &=& 1 + \frac{d \nu^2 - 1}{\delta_c} + 2 \frac{p/\delta_c}{1+(d\nu^2)^p},
 \label{eq:biassheth}
\end{eqnarray}
where $p = 0.3$, $d = 0.707$, $\delta_c = 1.673$ for the $\Lambda CDM$ cosmology that we used, $\nu = \delta_c/\sigma(M,z)$ and $\sigma(M,z)$ = $\sigma(M,0)D_+(z)$. Finally we get 
\begin{eqnarray}
 b'_{\rm{ST}} = \frac{2\mathcal{H}f}{c\delta_c}\left[ d\nu^2 - 2 \frac{p^2(d\nu^2)^p}{(1+(d\nu^2)^p)^2}\right].
  \label{eq:derivbiassheth}
\end{eqnarray}
We verified that this value is in agreement with the derivative computed from our simulation (see Appendix \ref{sec:appendix_bias}).

Note that in order to compute the full correlation function it is also possible to use the pressureless Euler equation $\dot{\bm{v}}\cdot\bm{n} + \mathcal{H}\bm{v}\cdot\bm{n} + \nabla_{\bm{r}}\psi = 0$. \\

We sum up in Table~\ref{tab:predictiontable} the linear regime contributions to the dipole predictions used in this paper.
\begin{table*}
	\centering
	\caption{Dipole prediction table for the linear regime. This table indicates which terms to consider when predicting the dipole for a specific choice of angle ($\theta$ or $\beta$) and redshift ($z_i$) which are given by Eqs.~\eqref{eq:firstnewredshift} to~\eqref{eq:lastredshift}. A cross shows if a term should be added to the prediction, while a zero indicates that the term should not be taken into account. }
	\label{tab:predictiontable}
	\begin{tabular}{|l|c|c|c|c|c|c|c|c|c|c|c|c|c|} 
		\hline
		 & $\xi^{\rm pot,(1)}$ & $\xi^{\rm pot,(2)}$&  $\xi^{\rm acc}$ & $\xi^{\rm q}$ & $\xi^{\rm div}$ & $\xi^{\rm wa}$ & $\xi^{\rm LC}$ & $\xi^{\rm isw}$& $\xi^{\rm lens}$&$\xi^{\rm evo1}$ & $\xi^{\rm evo2}$\\
		\hline
		flrw$=(\beta, z_0)$  & 0 & 0 & 0& 0 & 0 & 0 & $\times$ & 0& 0 & 0 & $\times$ \\       
		\hline
		$(\beta, z_1)$-flrw : \textbf{Potential only (1)}   & $\times$ & 0 & 0& 0 & 0 & 0 & 0 &0 & 0 & 0 & 0 \\       
		\hline         
		$(\beta, z_1)$-flrw : \textbf{Potential only (2)}   & $\times$ & $\times$ &  0 &0& 0 & 0 & 0 &0 & 0 & 0 & 0 \\       
		\hline  
		$(\beta, z_2)$-flrw : \textbf{Doppler only}  & 0  & 0 &$\times$ &$\times$ & $\times$ & $\times$ & 0 & 0 & 0  & $\times$ & 0 \\       
		\hline 
		$(\beta, z_3)$-flrw : \textbf{Transverse Doppler only}  & 0 & 0 & 0 &0& 0 & 0 & 0 & 0& 0 & 0 & 0 \\       
		\hline 
		$(\beta, z_4)$-flrw : \textbf{ISW only}  & 0 & 0 & 0 & 0 & 0&0 & 0 & $\times$ & 0 & 0 & 0 \\       
		\hline 
		$(\theta, z_0)$-flrw : \textbf{Lensing only}  & 0 & 0 & 0 & 0 &0& 0 & 0 & 0 &$\times$ & 0 & 0 \\       
		\hline         
		$(\theta, z_5)$-flrw : \textbf{All}\footnotemark & $\times$ & $\times$ & $\times$ & $\times$ &$\times$ & $\times$ & 0 & $\times$ & $\times$ & $\times$ & 0 \\       
		\hline         
	\end{tabular}
\end{table*}
The dominant term for large separation is $\xi ^{\rm div}$. It is related to a geometrical effect due to the divergence of line-of-sights for an observer at finite distance. However \emph{this is not} a  ``wide-angle" term. Indeed, even if we consider the pairs to be aligned the effect does not vanish. It comes from the fact that an element of volume seen under a given solid angle is perceived as less dense when receding from us and similarly is perceived as denser when coming towards us. This generates an overall positive dipole while the usual wide-angle term produces a negative dipole.

To infer the final prediction, we computed the linear theory prediction at 200 redshifts between the limits of our survey and took the volume average. Indeed, for some of the dipole terms, it is not equivalent to computing the prediction at the volume averaged redshift. Throughout the paper we always consider the case $b_1 > b_2$.

\footnotetext[2]{Using the Euler equation, this term becomes: \\All = $\xi^{\rm pot,(2)} + \xi^{\rm q} +\xi^{\rm div}+\xi^{\rm wa}-\xi^{\rm LC}+\xi^{\rm isw}+\xi^{\rm lens}+\xi^{\rm evo1}$}
\subsection{Two-point halo-halo cross-correlation function: non-linear regime}

Starting from Eq.~\eqref{crosscorreq}, two difficulties arise in the non-linear regime of structure formation. The evolution of the matter fields (density, velocity and potential) becomes non-linear. Moreover, the mapping from real space to redshift space becomes non-linear too. A vast literature has addressed these questions in the context of the standard RSD with distant observer approximation (i.e. Doppler effect along one fixed direction).  A naive perturbation theory expansion provides poor results because of Finger-of-God effect. \citet{taruya2010baryon} have developed a perturbation-based theory where the damping is characterised by a univariate function with one single free parameter. This model performs well in the quasi-linear regime. It however needs to be extended to include wide-angle effect and other relativistic effects. We plan to work on these aspects in the future. Another approach to the non-linear regime is to rewrite redshift-space distortions in the context of the streaming model. Following \citet{scoccimarro2004redshift} the observed (Doppler-only) correlation function $\xi_s^{std}$ is given (in the distant observer approximation) by
\begin{eqnarray}
1+\xi_s^{\rm std} (s_{\perp},s_{\parallel})=\int \left[1+\xi (r_{\perp},r_{\parallel})\right] \mathcal{P}(v^{12}_{\parallel} | r_{\perp},r_{\parallel}) \textrm{d} v^{12}_{\parallel},
\end{eqnarray}
where the apparent position in redshift space is decomposed into a component perpendicular to the line-of-sight ($s_{\perp}$) and along the line-of-sight ($s_{\parallel}$), the real-space position is also decomposed as $r_{\perp}=s_{\perp}$ and $r_{\parallel}=s_{\parallel}-v_{\parallel}^{12}/\mathcal{H}$, $v_{\parallel}^{12}$ is the pairwise velocity along the line-of-sight and $\mathcal{P}(v_{\parallel}^{12} | r_{\perp},r_{\parallel})$ is the pairwise-velocity Probability Distribution Function (PDF) at the position $(r_{\perp},r_{\parallel})$ in real space. 
This decomposition is exact even at non-linear scales. However one still needs to predict the PDF of the pairwise velocity using halo model or perturbation theory. Again an extension to wide angles is still missing as it is quite challenging. The detailed analytical modelling of the non-linear velocity PDF  which is highly non-Gaussian \citep{scoccimarro2004redshift} (plus possible wide-angle effects at intermediate scales) is beyond the scope of this paper and we leave it for future work.
However, as we will see the contribution from gravitational redshift dominates the dipole at small non-linear scales: we have therefore focused on this specific contribution. Taking into account the PDF of the gravitational potential at a given pair separation $\mathcal{P}(\phi_{12} | r_{\perp},r_{\parallel})$, a general expression to the halo-halo correlation function with potential-only RSD is given by
\begin{eqnarray}
1+\xi_s^{\rm pot} (s_{\perp},s_{\parallel})=\int \left[1+\xi (r_{\perp},r_{\parallel})\right] \mathcal{P}(\phi_{12} | r_{\perp},r_{\parallel}) \textrm{d} \phi_{12},
\label{eq:ksipotstream}
\end{eqnarray}
where $r_{\perp}=s_{\perp}$, $r_{\parallel}=s_{\parallel}+\phi_{12}/(c\mathcal{H})$ and $\phi^{12}$ is the difference of potential between the two halo populations.
We assume a simple spherical model to derive the potential difference as a function of radius. Following \citet{croft2013gravitational}, the PDF of the potential difference is a single-valued function which depends only on pair separation. The contribution from the potential difference $\phi_{12}$ to the halo-matter correlation function (no velocity), is given by
\begin{eqnarray}
M_{12}(<r)&=&4 \pi \bar{\rho} \int_0^r (\xi_{h_1 m}(x)- \xi_{h_2 m}(x)) x^2 \textrm{d}x, \\
\phi_{12}(R) &=&-G \int^R_0 \frac{M_{12}(<r)}{r^2} \textrm{d}r,\\
\xi^{\rm pot,sing}_{s} (s_{\perp},s_{\parallel})&=& \xi (r_{\perp},r_{\parallel} ),
\label{eq:ksipotstreamsingle}
\end{eqnarray}
where $\xi_{h m}$ is the monopole of the halo-matter correlation function, $M_{12}(<r)$ is the enclosed mass, G is the gravitational constant, $R=\sqrt{r_{\perp}^2+r_{\parallel}^2}$, $r_{\perp}=s_{\perp}$ and $r_{\parallel}=s_{\parallel}+\phi(R)/(c\mathcal{H})$. $\xi_{h m}$ is taken as the maximum between the linear prediction $b \xi_{mm}$ and a spherical NFW profile \citep{navarro1997universal} with a concentration parameter given by \citet{zhao2009accurate}. This model is an approximation as the distribution of matter is not spherical \citep{cai2017gravitational}, the PDF of the pairwise potential is not single-valued and the halo profile can deviate from NFW \citep{balmes2014imprints}.

\citet{croft2013gravitational} have also taken into account both the standard and (single-valued) potential contributions to the dipole with a simple streaming model. However they have neglected wide-angle effects, other relativistic effects and they have assumed a simple exponential model for the pairwise velocities. To conclude, a full model of the cross-correlation function with all relativistic effects in the non-linear regime is still missing. We will use cosmological simulations to address this regime including all contributions.

\section{Methods}
\label{sec:Methods}
The numerical setup is described in the three following subsections. In the first subsection, we introduce the new large N-body simulation that we have performed. It is part of the \textsc{Raygalgroupsims}\footnote{Ray-tracing Galaxy Group Simulations, soon available at \url{http://cosmo.obspm.fr/}} suite of simulations dedicated to ray-tracing studies (Breton et al., in prep). 
This simulation gives us Lagrangian and Eulerian quantities within snapshots (volume at constant times) and light-cones (volume as seen from an observer located at the centre of the box, a point further away from the centre being seen further back in time). In the second subsection, we use \textsc{Magrathea}\footnote{\url{https://github.com/vreverdy/magrathea-pathfinder/} \\ The version used for this work is currently very different from the master branch, however they will be synchronised soon} \citep{reverdy2014propagation}, a ray-tracing library integrating photon paths using geodesic equations on the light-cones. Searching for the geodesics connecting a given observer to all detected haloes, we compute the \emph{seen} angle and the \emph{observed} redshift (which are different from the \emph{true} angle and \emph{comoving} redshift provided directly by the simulation). The ray-tracing gives us catalogues of apparent redshift and angular positions $(z, \theta, \phi)$. In the last subsection, we introduce the correlation function estimator.

We would like to point out that in the following we use the Newtonian gauge to interpret the data from N-body simulation. In principle this can lead to errors in redshift  due to the fact that in this gauge there are relativistic corrections at the horizon scale. To be perfectly rigorous we should interpret the position of particles in the N-body gauge \citep{fidler2015general,fidler2016relativistic}. The difference of interpretation comes from the fact that contrary to the Newtonian gauge, the N-body gauge leaves the spatial volume unperturbed in a similar way to Newtonian simulations. To account for relativistic corrections in Newtonian gauge we could also apply a time-independent displacement on the particle position \citep{chisari2011connection} which depends on the initial particle distribution. However as shown by \citet{adamek2016gevolution,adamek2016general,adamek2017perturbed} these gauge effects are small compared to the effects we are interested in, meaning for this work we can safely use the Newtonian gauge to interpret the data from our Newtonian simulation.

\subsection{N-body simulation}

The simulation used in this work consists in a dark-matter only simulation with $4096^3$ particles within a volume of (2625~$h^{-1}$Mpc)$^3$. The corresponding particle mass-resolution is $1.88 \times 10^{10}h^{-1}$M$_\odot$. The final number of AMR (Adaptive Mesh Refinement) cells is 0.4 trillion and the spatial resolution reaches $5~h^{-1}$kpc. We use \textsc{CAMB} \citep{lewis2000efficient} to compute the initial linear matter power spectrum for a $\Lambda$CDM cosmology (seven-year \textsc{WMAP} data, \citealt{komatsu2011seven}) with Hubble parameter $h = 0.72$, matter density $\Omega_m = 0.25733$, baryon density $\Omega_b = 0.04356$, radiation density $\Omega_r = 8.076\times 10^{-5}$, slope of the primordial power spectrum $n_s = 0.963$ and normalisation $\sigma_8 = 0.801$. Initial conditions were generated with a second-order Lagrangian perturbation theory (2LPT) version of \textsc{MPGRAFIC} \citep{prunet2008initial} at a redshift $z_{\textrm{start}} = 46$. 

Dark matter particles were evolved with an improved version of the particle-mesh adaptive-mesh-refinement (PM-AMR)  N-body code \textsc{Ramses} \citep{teyssier2002cosmological}. We borrow the Triangular Shape Cloud assignment routine from \citet{li2012ecosmog} in order to make the density, potential and force more isotropic than with the standard Cloud In Cell assignment.

During the simulation, $\sim$50 snapshots (particles) and 3 light-cones (particles and gravity cells) with different depths and apertures were written. The light-cones were built using the onion-shells technique \citep{fosalba2008onion,fosalba2015miceA,fosalba2015miceB, teyssier2009fullsky}. In this article, we focus on snapshots between $z=0.5$ and $z=0$ and the full-sky light-cone up to $z_{\rm max}=0.5$. Choosing this maximum redshift ensures that we avoid any replica in the light-cone. The light-cone consists in $\sim 300$ shells (i.e. every coarse time-step): this ensures a good time-resolution. 

 We use pFoF\footnote{\url{https://gitlab.obspm.fr/roy/pFoF}} \citep{roy2014pfof}, a parallel friend-of-friend algorithm to detect haloes both in snapshots and light-cones. We adopt a standard linking-length of $b=0.2$ times the mean inter-particle separation and we only pick haloes with more than $100$ particles (to guarantee that haloes are sufficiently sampled).
 
Last, we note that we use a Newtonian simulation and therefore the two Bardeen potentials are equal.

\subsection{Ray-tracing}

The goal of ray-tracing is to make a connection between sources and observers. There are many  approaches to ray-tracing. The most basic approximation (widely used in analytical works) is the Born approximation where deflections and lensing are computed from integral along the undeflected path. Within simulations a commonly used method consists in splitting the universe in several thin lenses orthogonal to the direction of observation (multiple lens formalism, \citealt{hilbert2009raytracing}). The light is assumed to move in straight lines between lenses while being deflected when crossing the lenses. Lenses are usually flat and plane-parallel, thus inducing error for large angles. Moreover the potential is often computed by solving a 2D Poisson equation on the lens (instead of a 3D one). In order to deal with wide-angle effects an onion-shells technique has been implemented \citep{fosalba2008onion,fosalba2015miceA,fosalba2015miceB, teyssier2009fullsky} with spherical lenses. More recently the ray-tracing algorithm has been ported within \textsc{Ramses}  in order to use the accurate 3D potential from the simulation \citep{barreira2016ray}. In most of the works, lensing is computed using integral along the light-ray while redshift perturbations are computed independently, only considering the Doppler effect. 

Our method consists in unifying deflection, lensing and redshift calculation by directly solving geodesics equations. While this approach is usually done in the field of general relativity, it is not common within cosmological simulations. To our knowledge, this approach has been introduced by \citet{killedar2012gravitational} using a fixed grid resolution and limited-size simulation. Here we use \textsc{Magrathea}, a hybrid MPI/pthreads C\texttt{++}11 ray-tracing library to propagate photons on null geodesics within the hierarchy of AMR grids \citep{reverdy2014propagation}. Using adaptive mesh is crucial to fully resolve lensing, potential and velocity profiles near haloes. We note that there has been
a recent resurgence of interest
for a relativistic approach to ray-tracing within cosmological simulations (\citealt{giblin2017general} has solved optical scalar equation with a full general relativistic code at low resolution; \citealt{borzyszkowski2017liger} has solved it at higher resolution but using Born approximation).

A detailed description of our fast and very accurate ray-tracing library based on template-meta-programming is available in \citet{reverdy2014propagation}, we now review some of its specialities. The light-cone provides a regular grid at coarse level with refinement in high density areas. In each cell of the grid we have $\left(a, \phi, \frac{\partial \phi}{\partial x}, \frac{\partial \phi}{\partial y}, \frac{\partial \phi}{\partial z}\right)$ respectively the scale factor of the shell, the potential and derivatives of the potential with respect to spatial Cartesian coordinates of the simulation box. In order to propagate light-rays we interpolate these quantities at the photon space-time location and solve the geodesic equations. 
In this work, we chose the linear interpolation for space and nearest-neighbour interpolation for time.
Higher order interpolation are interesting prospects for future work, but for this paper we will not consider it. 
Photons are launched backward in time (backward ray-tracing) from the light-cone centre at $z = 0$  with a given $k^i$ (spatial part of the wavevector) and initially setting $g_{\mu\nu}dx^{\mu}dx^{\nu} = 0$. We then let \textsc{Magrathea} solve the linearised geodesic equations with the metric given by Eq.~\eqref{eq:metric},
\begin{eqnarray}
   \label{eq:geodesic_equation}
	      \frac{\textrm{d}^2 \eta}{\textrm{d}\lambda^{2}} &=& -\frac{2a'}{a}\frac{\textrm{d}\eta}{\textrm{d}\lambda}\frac{\textrm{d}\eta}{\textrm{d}\lambda} - \frac{2}{c^2}\frac{\textrm{d}\phi}{\textrm{d}\lambda}\frac{\textrm{d}\eta}{\textrm{d}\lambda} + 2\frac{\partial\phi}{\partial\eta}\left(\frac{\textrm{d}\eta}{\textrm{d}\lambda}\right)^2 \\
	      \frac{\textrm{d}^2 x^i}{\textrm{d}\lambda^2} &=& -\frac{2a'}{a}\frac{\textrm{d}\eta}{\textrm{d}\lambda}\frac{\textrm{d}x^i}{\textrm{d}\lambda} + \frac{2}{c^2}\frac{\textrm{d}\phi}{\textrm{d}\lambda}\frac{\textrm{d}x^i}{d\lambda} - 2\frac{\partial\phi}{\partial x^i}\left(\frac{\textrm{d}\eta}{\textrm{d}\lambda}\right)^2
\end{eqnarray}
where $\lambda$ is the affine parameter along the photon path.

As we are interested in source-averaged observables rather than direction-averaged ones \citep{kibble2005average}, we now describe recent modifications of the solver to build a catalogue of sources including all relativistic effects. To find the null geodesic connecting a source and the observer, we launch 
several photons from the observer to the tentative directions of observation of the source. Then using a root finder, the geodesics intersecting both the source and observer world lines are identified. In practice, we assume that sources are present at any time (as opposed to an event which corresponds to a specific space-time location). Moreover, since sources are moving, we use a nearest-neighbour interpolation for the time location of the source. Because sources are moving slowly and light-ray deflections are small, the sources lie very close the null FLRW light-cone. A refinement would be to linearly interpolate the position of particles between two light-cones at different times. Moreover, we only search for one geodesic for each source since we focus on large scales, dominated by the weak lensing regime: generalisation to strong lensing (i.e. multiple geodesics for one source) is straight-forward with enough resolution. We leave these possible refinements for future work. 

\begin{figure} 
	\includegraphics[width=0.9\columnwidth]{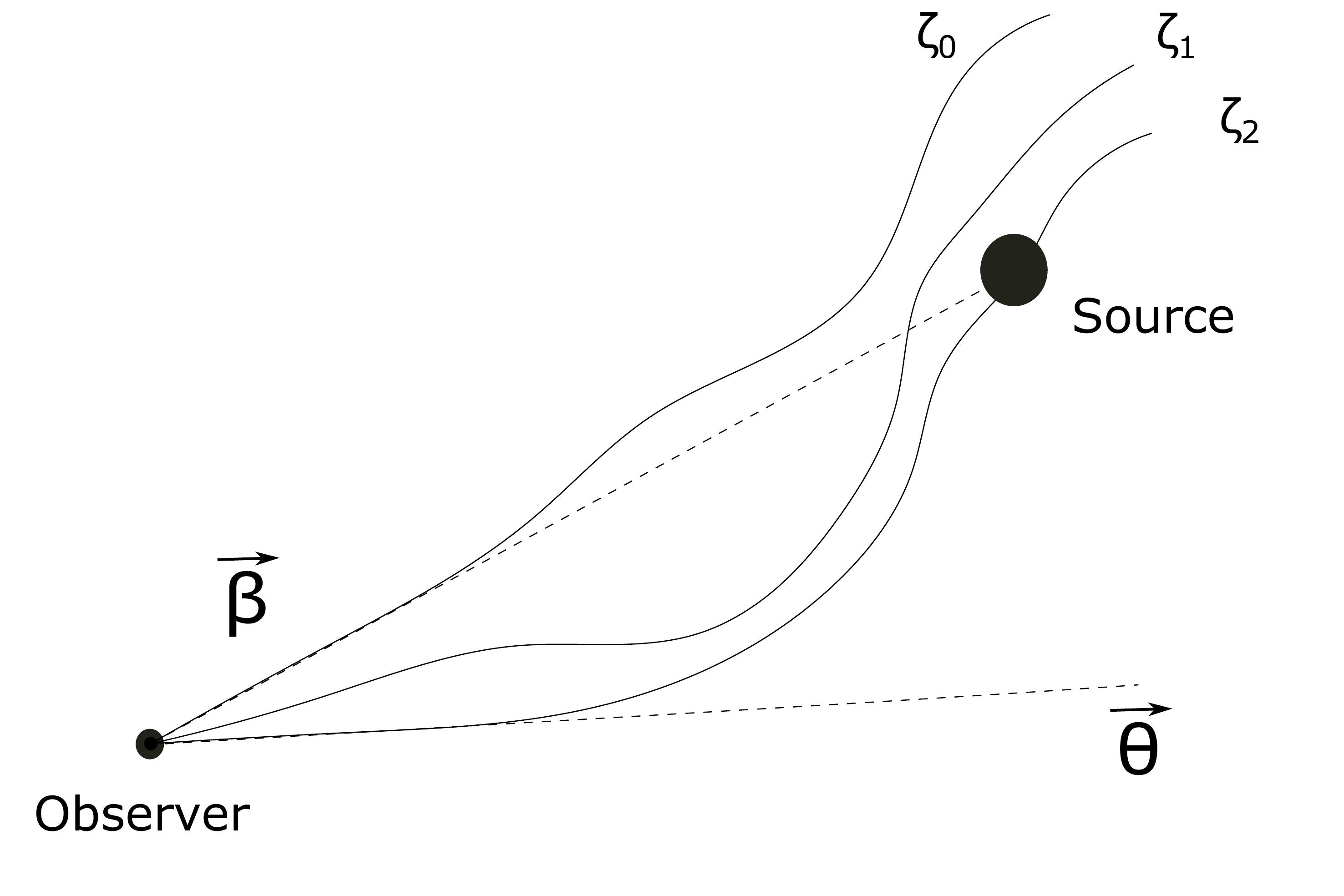}
    \caption{Illustration of the geodesic-finder algorithm. Each tentative photon is designated by $\zeta_n$ with n being the number of iterations. The first photon $\zeta_0$ is launched towards the source with an angle $\beta$. The first ray will generally miss the source, we then iterate using Newton's method in order to get a new initial angle. In this example we iterate twice to find a ray close enough to the source at the same radius, the initial angle of $\zeta_2$ is given by $\theta$ and is interpreted as the \emph{seen} angle.}
    \label{fig:geodesic}
\end{figure}

Let a halo be at location $\left(X, Y, Z\right)$ on the light-cone. 
For an observer at the centre of the simulation, the two components of the \emph{true} angle $\beta$ are  (assuming the same convention as for lensing): $\beta_1 = \arctan\left(Y/X\right)$,  $\beta_2 = \arccos\left(Z/R\right)$ where $R$ is the comoving distance $R = \sqrt{X^2+Y^2+Z^2}$.
We expect the lensing deviation to be small, we thus launch the photon in the direction $\beta$, but the ray is deflected and does not reach the position $\left(X, Y, Z\right)$. As shown in Fig.~\ref{fig:geodesic}, we iterate on the initial launching conditions using a root-finder method (Newton's method in our case) to find the initial angle that minimizes the angle difference between $\beta$ and the position of the photon at same radius. In practice only one or two iterations are needed to get an angle difference lower than $10^{-2}$ arcsec. With this method we know the \emph{true} angle $\beta$ and the \emph{seen} angle $\theta$. We can then directly derive the Jacobian matrix $A_{ij} = \frac{\partial \beta_i}{\partial \theta_j}$, hence the distortion matrix (related to lensing). This way of computing the lensing directly from a beam of light rays (instead of integrating Sachs equation) is similar to the ray-bundle approach \citep{fluke1999raybundle,fluke2011shape}
except that the geodesic equations are directly integrated.

To gain comprehension on the various contributions to the total redshift we decompose it as follows:

\begin{eqnarray}
    \label{eq:firstnewredshift}
	z_{\textrm{0}} &=& \frac{a_0}{a} - 1, \\
	z_{\textrm{1}} &=& \frac{a_0}{a} \left( 1 + \phi_o/c^2 - \phi_s/c^2 \right) - 1, \\
    z_{\textrm{2}} &=& \frac{a_0}{a} \left( 1 + \bm{v}_s\cdot\bm{n}/c \right) - 1, \\
    z_{\textrm{3}} &=& \frac{a_0}{a} \left( 1 + \frac{1}{2}\left(\frac{v_s}{c}\right)^2 \right) - 1,   \\  	    
    z_{\textrm{4}} &=& \frac{a_0}{a} \left( 1 - \frac{2}{c^2}\int^{\eta_o}_{\eta_s} \dot{\phi}\textrm{d}\eta\right) - 1,\\
    z_{\textrm{5}}&=&\frac{(g_{\mu\nu}k^{\mu}u^{\nu})_s}{(g_{\mu\nu}k^{\mu}u^{\nu})_o} - 1, 
    \label{eq:lastredshift}    
\end{eqnarray}
with $g_{\mu\nu}k^{\mu}u^{\nu} = -ack^0 \left( 1 + \phi/c^2 + \bm{v}\cdot\bm{n}/c + \frac{1}{2}\left(\frac{v}{c}\right)^2 \right)$ and the observer velocity set to zero.
 
Each redshift corresponds to a specific contribution. $z_0$ is the redshift directly inferred from the scale factor. However this scale factor is related to the conformal time computed until arriving at the source, using the geodesic equation Eq.~\eqref{eq:geodesic_equation}. It therefore implicitly takes into account time delay.  
$z_1$ only includes the gravitational redshift perturbation, $z_2$ the Doppler perturbation, $z_3$ the transverse Doppler perturbation  and $z_4$ the ISW perturbation. $z_5$ is the exact general relativity redshift computation. It almost corresponds to $z_0$ plus all redshift perturbations above. The ISW effect is hidden in the $k^0$ term, which comes directly from our geodesic integration. 

 Finally, ray-tracing gives us catalogues with $\beta$, $\theta$, $A_{ij}$, various redshifts containing each terms of Eq.~\eqref{eq:perturbedredshift} and the number of dark matter particles for each halo. In these catalogues all the relativistic effects have been computed in a self-consistent way. These catalogues\footnote{\url{http://cosmo.obspm.fr/raygalgroupsims-relativistic-halo-catalogs}}
 will be described in detail in Breton et al. (in prep).

\subsection{Estimation of the correlation function}
\label{sec:estimation_correlationfunction}
The halo-halo two-point cross-correlation function $\xi_{h_1 h_2}(\bm{r})=\langle\delta_{h_1}(\bm{x}) \delta_{h_2}(\bm{x}+\bm{r})\rangle$ is a measure of the excess of probability relative to a Poisson distribution of finding a pair of haloes separated by a vector $\bm{r}$. For a statistically homogeneous and isotropic field the correlation function can be written as $\xi_{h_1 h_2}(r)$ since it only depends on the norm $r$ of the separation. However the presence of an observer breaks the isotropy and one needs to specify two components 
of the separation vector $\bm{r}$, for instance its norm $r$ and projection $\mu$ along the line-of-sight.

To estimate the correlation function we used a modified version of \textsc{Cute} \citep{alonso2012cute} (a parallel tree-code pair-counting algorithm). It implements an LS estimator \citep{landy1993bias,kerscher2000comparison}, which is one of the most commonly used estimator for the correlation function (since its variance is almost Poisson),
\begin{eqnarray}
\xi_{\rm LS} = \frac{D_1D_2 - D_1R_2 - R_1D_2 + R_1R_2}{R_1R_2},
\end{eqnarray}
where $D_1$, $D_2$ refer to different datasets to be cross-correlated while $R_1$ and $R_2$ are the associated random catalogues.
Moreover the pair counts are normalised by the total number of pairs in the catalogues.
Since we are interested in correlation function anisotropies, we bin in ($r,\mu$). Once we compute $\xi(r,\mu)$, we deduce the multipoles as
\begin{eqnarray}
\xi_{\ell}(r) \approx \frac{2\ell + 1}{2}\sum^{1}_{\mu = -1}  \xi(r,\mu)\mathcal{L}_{\ell}(\mu) \Delta \mu,
\end{eqnarray}
where $\mathcal{L}_{\ell}(\mu)$ is the Legendre polynomial of order $\ell$.

We have cross-checked the results of this direct pair-counting method to a grid method. In this method the halo density is estimated on a thin Cartesian grid using a Cloud-in-Cell assignment scheme. The correlation function is then directly computed from its definition as a function of the over-densities of the halo populations. The two methods give very similar results.
With the intention of comparing to linear theory we estimated the linear bias $b_i$ for data-set data$\_$H$_{i}$ (see Appendix \ref{sec:appendix_bias} for more details), 

\begin{eqnarray}
b_{100} & \approx & \sqrt{\frac{\xi^{\ell = 0}_{h_{100}h_{100}}}{\xi^{\ell = 0}_{mm}}},\\
b_i & \approx & b_{100}\frac{\xi^{\ell = 0}_{h_i h_{100}}}{\xi^{\ell = 0}_{h_{100} h_{100}}},
\label{eq:bias}
\end{eqnarray}
where $\xi^{\ell = 0}_{hh}$ and $\xi^{\ell = 0}_{mm}$ are the halo-halo and matter-matter correlation function monopole respectively. The bias is estimated by fitting a constant to Eq.~\eqref{eq:bias} for $r$ between $25$ and $75~h^{-1}$Mpc. Below $25~h^{-1}$Mpc the number of pairs is too low and the correlation function may fluctuate while above $75~h^{-1}$Mpc the Poisson noise becomes non negligible

The last point concerns the estimation of statistical errors. 
Running again the same heavy simulation being much too time consuming,
we estimate the variance using the jackknife method, as it is the internal method that minimizes most of the variance for the linear regime according to \citet{norberg2009statistical}. In this paper we compute the jackknife method with 32 re-samplings.
We then estimate the variance of the correlation function as follows,
\begin{eqnarray}
\label{eq:variance}
\sigma^2_{\ell}(r) = \frac{N-1}{N}\sum^N_{k=1}(\xi^k_{\ell}(r)-\bar{\xi}_{\ell}(r))^2,
\end{eqnarray}
with $N$ the number of re-samplings, $k$ the sample number and $\bar{\xi}_{\ell}(r)$ the mean correlation function given by 
\begin{eqnarray}
\label{eq:jackmean}
\bar{\xi}_{\ell}(r) = \frac{1}{N}\sum^N_{k=1}\xi^k_{\ell}(r).
\end{eqnarray}

It is important to note that the variance estimated with Eq.~\eqref{eq:variance} is good enough in the linear regime where the density field is Gaussian but becomes much less accurate for smaller scales, in the non-linear regime. In this regime error bars should be taken with caution.

In the linear regime, we note that the theoretical predictions for the cross-correlation dipole are proportional to the bias difference (except for evolution effects). Therefore, normalising by this quantity should give the same signal for each pair of populations. We take advantage of this feature by using a weighted average of the normalised dipole for all mass combinations to increase our signal to noise ratio. In the linear regime, the mean signal is computed as

\begin{eqnarray}
\xi_1^{\rm lin}/\Delta b = \frac{\sum\limits_{ij} \frac{\xi_1^{ij}}{b_i - b_j} \frac{1}{\sigma^2_{ij}}}{\sum\limits_{ij} 1/\sigma^2_{ij}},
\end{eqnarray}
where $b_i$ and $b_j$ are the bias of different halo populations. $\xi_1^{ij}$ is the dipole of the cross-correlation between two halo populations of bias $b_i$ and $b_j$, and $\sigma^2_{ij}$ its variance normalised by the bias difference. The variance of the weighted average dipole is

\begin{eqnarray}
\sigma^2 = \frac{1}{\sum\limits_{ij} 1/\sigma^2_{ij}}.
\end{eqnarray}
The error bars are probably underestimated due to the lack of independence.

\section{Data and validation}
\label{sec:Datavalid}
\begin{table}
    \scriptsize
	\centering
	\caption{Summary of the different datasets used: name, number of haloes, range for the number of particles per halo, mean mass, bias at the volume averaged redshift $z=0.341$, and estimated mean halo concentration (taken from \citealt{zhao2009accurate}).}
	\label{tab:datasets}
	\begin{tabular}{lccccr} 
		\hline
		name & nb of haloes & nb of part & mass ($h^{-1}$M$_{\odot}$) & bias & $c^{200m}_{\textrm{zhao}}$\\
		\hline
		data$\_$H$_{100}$ 		& $5.4\times10^6$ 	  & 100-200   & $2.8\times10^{12}$ &1.08&8.2\\
		data$\_$H$_{200}$       & $3.4\times10^6$ 	  & 200-400   & $5.6\times10^{12}$ &1.22&7.7\\
		data$\_$H$_{400}$       & $1.9\times10^6$ 	  & 400-800   & $1.1\times10^{13}$ &1.42&7.1\\    
		data$\_$H$_{800}$       & $1.0\times 10^6$  & 800-1600  & $2.2\times10^{13}$ &1.69&6.6\\   
		data$\_$H$_{1600}$      & $4.0\times10^5$ 	  & 1600-3200 & $4.5\times10^{13}$ &2.07&6.1\\    
		data$\_$H$_{3200}$      & $2.0\times10^5$       & 3200-6400 & $9.0\times10^{13}$ &2.59&5.7\\           
		\hline
	\end{tabular}
\end{table}

We now proceed to the presentation and validation of our datasets. In Section \ref{sec:Datasets} we introduce the halo catalogues and in Section \ref{sec:monopole_quadrupole} we validate our two-point correlation measurements on the monopole and quadrupole. 

\subsection{Datasets}
\label{sec:Datasets}
\begin{figure} 
	\includegraphics[width=\columnwidth]{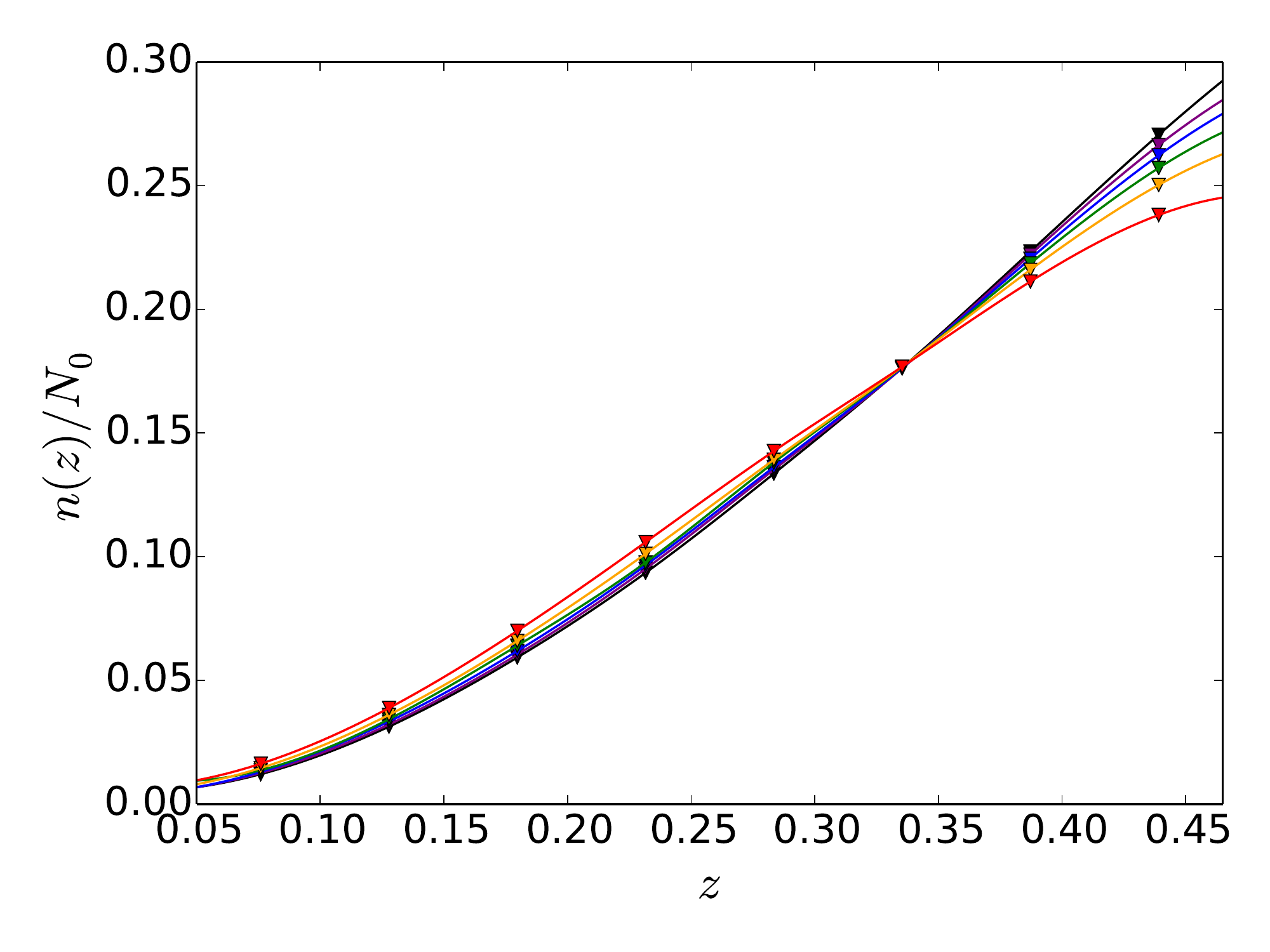}
    \caption{Redshift distribution for each halo dataset in Table~\ref{tab:datasets}. The distributions from least to most massive haloes population are shown in black, purple, blue, green, orange and red. The distributions are normalised so that the integral is unity.}
    \label{fig:dndz}
\end{figure}

We consider haloes between $z_{\textrm{min}} = 0.05$ and $z_{\textrm{max}} = 0.465$ (hereafter, when we refer to the \emph{full} light-cone we mean the light-cone between these two redshifts), leading to a volume of $8.34$ ($h^{-1}$Gpc)$^3$. We choose redshift limits which are not too close to the observer to avoid issues when computing angles, and not too close from the edge of our full-sky light-cone to avoid edge effects.  We also focus on haloes with mass between $1.9\times10^{12}~h^{-1}$M$_{\odot}$ and $1.2\times10^{14}~h^{-1}$M$_{\odot}$. The total number of haloes in this volume is $1.2\times10^{7}$ leading to a mean halo density of $n \approx 5\times 10^{-4}~$Mpc$^{-3}$. We divide the halo catalogue in six logarithmic mass bins as shown Table~\ref{tab:datasets}.

data$\_$H$_{N}$ represents catalogues of haloes sampled by a number of dark matter particles between $N$  and $2N$.
We cross-correlate all the datasets, with 15 linear bins in $r$ going from 0 to 150$~h^{-1}$Mpc for large scales and 8 linear bins from 0 to 32$~h^{-1}$Mpc for smaller scales. We also use 201 bins in $\mu$. 
For computations on the full light-cone we generate a random catalogue for each dataset with more than ten time the number of haloes. The redshift distribution for random catalogues follow the same distribution as the associated data catalogue. To avoid losing information on the clustering along the line-of-sight, the redshift distribution is smoothed using 8 redshift bins as shown in Fig.~\ref{fig:dndz}.

Otherwise, if our computation is done with shell averages we use a random catalogue with $10^8$ object with a uniform distribution.

\subsection{Monopole and quadrupole validation}
\label{sec:monopole_quadrupole}

\subsubsection{Matter auto-correlation monopole on the light-cone}

\begin{figure} 
	\includegraphics[width=\columnwidth]{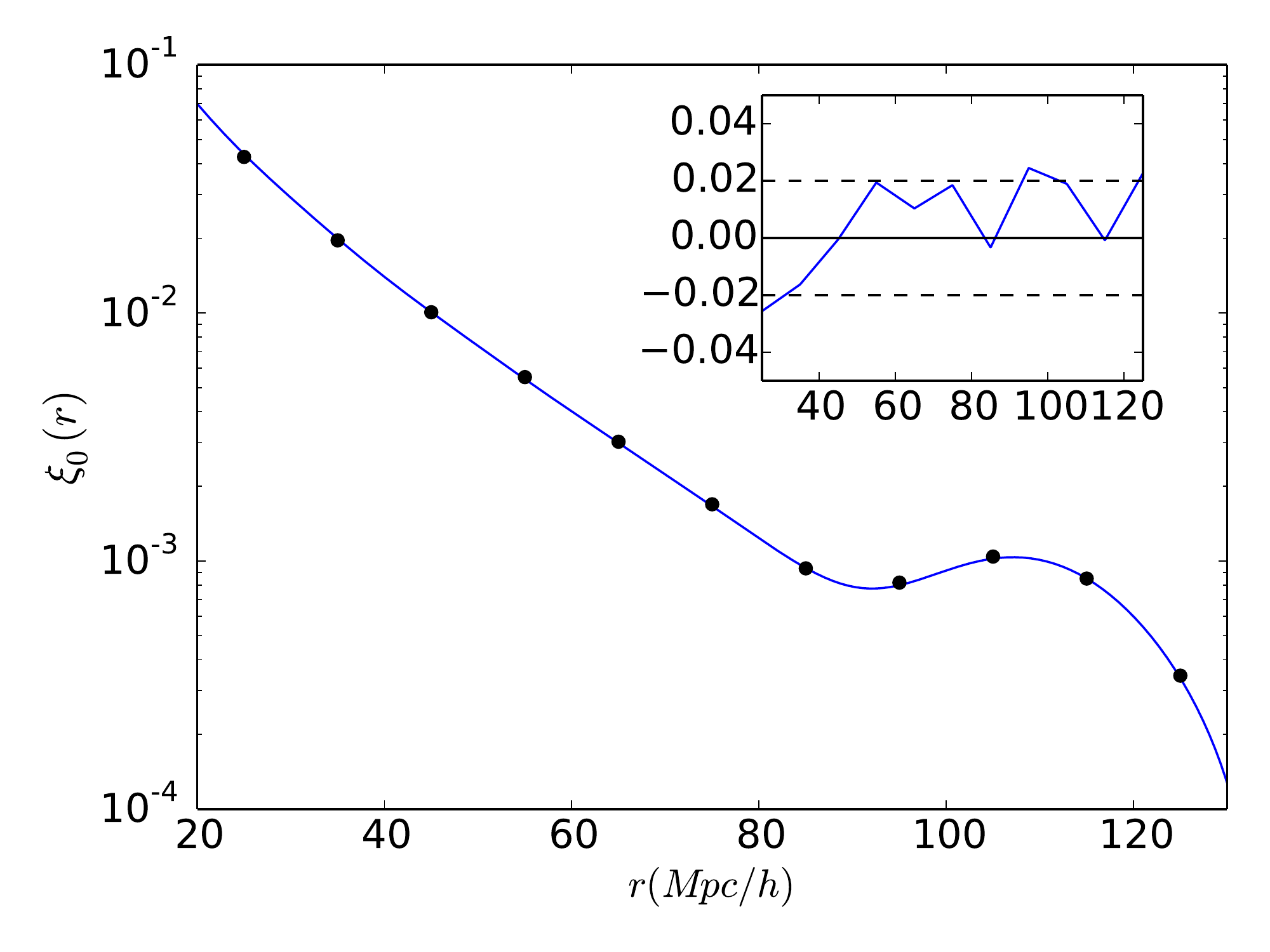}
    \caption{Matter monopole auto-correlation computed on the unperturbed FLRW light-cone compared with \textsc{CosmicEmu} emulator \citep{heitmann2016mira}. Subplot shows the relative difference.}
    \label{fig:monopole}
\end{figure}

We first check our measurement of the well-known matter auto-correlation monopole. Out of the $\approx 30$ billions particles in the light-cone we randomly pick $10^8$ that we ray-trace. We expect this catalogue to be representative of the general distribution of dark matter in the simulation. We compute the monopole on our full light-cone using $10^9$ particles for the random catalogue with a uniform distribution (since the mean matter density does not evolve with redshift), and we compare to the real-space prediction at the volume averaged redshift (the light-cone effect being negligible for this multipole). For the prediction we use the emulator \textsc{CosmicEmu} \citep{heitmann2016mira} which agrees with our power spectrum computation for different snapshots at percent level between roughly $k = 0.02~h$Mpc$^{-1}$ and $k = 2~h$Mpc$^{-1}$. The result is shown in Fig. \ref{fig:monopole}. The monopole of the matter auto-correlation is in good agreement (better than two percent) with the emulator in the range $r=20-120~h^{-1}$Mpc.
 
\subsubsection{Halo auto-correlation monopole and quadrupole in redshift space}

In this section we want to validate our monopole and quadrupole measurements in redshift space taking only into account  the effects of peculiar velocity e.g. the standard RSD effect. 
We compute the correlation on the full light-cone and the errors bars are estimated with the jackknife method.
For the quadrupole in redshift space we have subtracted the real-space quadrupole.

\begin{figure} 
	\includegraphics[width=\columnwidth]{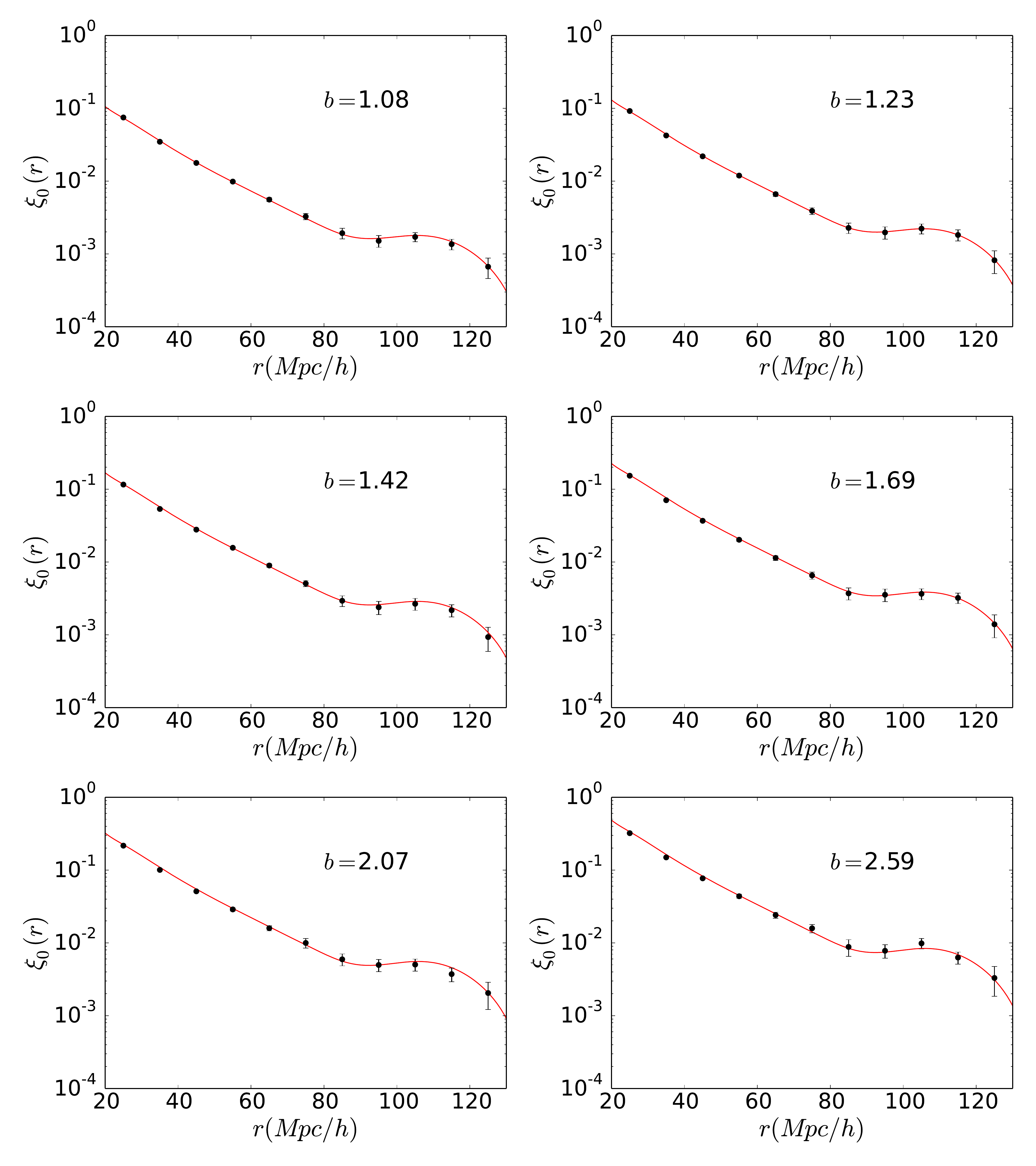}
    \caption{Monopole auto-correlation in redshift space (Doppler term only). Red full lines give the prediction from RegPT+TNS \citep{taruya2010baryon,taruya2012direct,taruya2013precision}. Black circles give the computation on the full light-cone.}
    \label{fig:monopole1209}
\end{figure}

\begin{figure} 
	\includegraphics[width=\columnwidth]{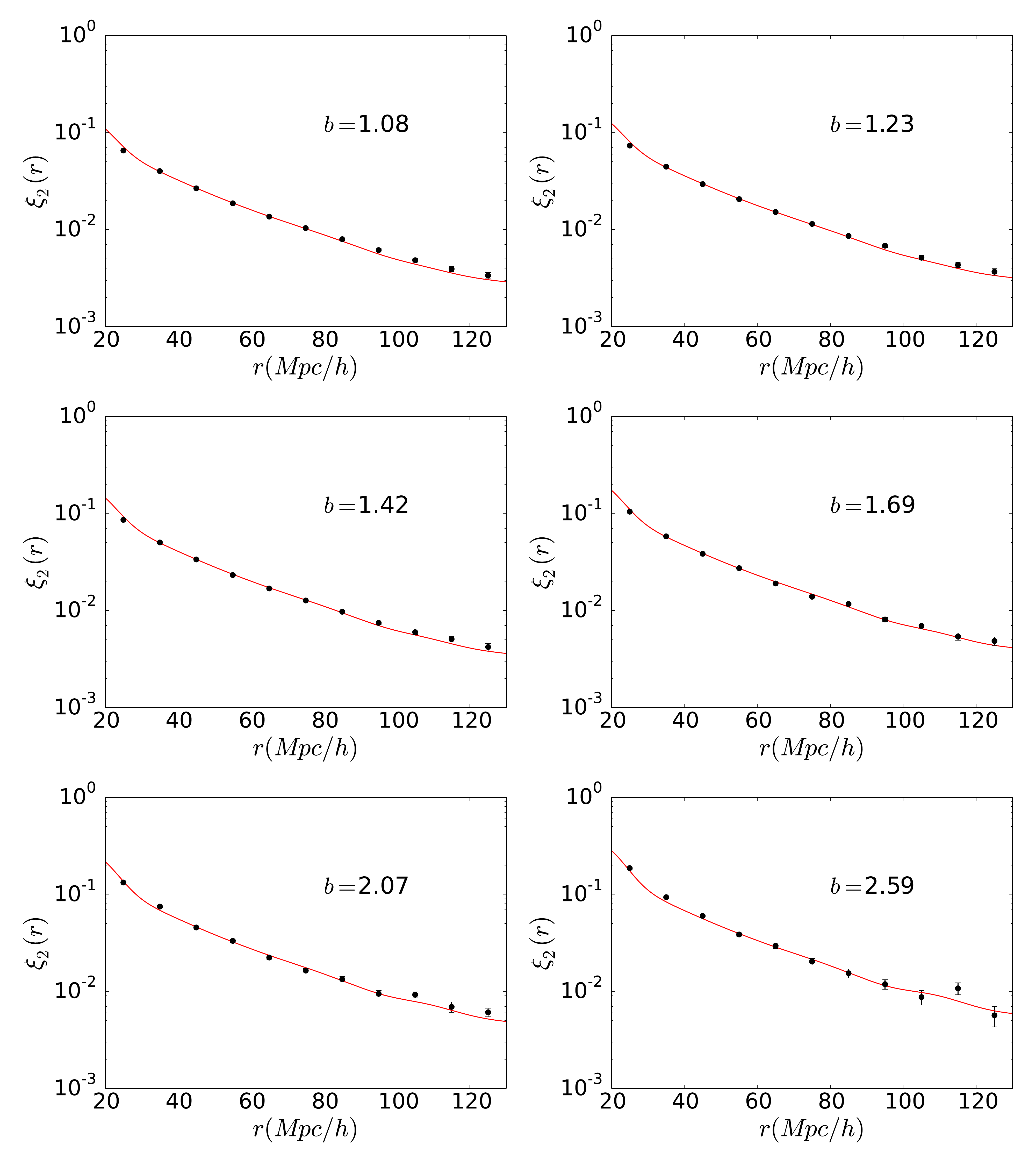}
    \caption{Absolute value of the quadrupole auto-correlation in redshift space (Doppler term only). Red full lines give the prediction from RegPT+TNS \citep{taruya2010baryon,taruya2012direct,taruya2013precision}. Black circles give the computation on the full light-cone.}
    \label{fig:quadrupole1209}
\end{figure}

To predict the monopole (Fig.~\ref{fig:monopole1209}) and quadrupole (Fig.~\ref{fig:quadrupole1209}) in redshift space we use the RegPT+TNS \citep{taruya2010baryon,taruya2012direct, taruya2013precision} model 
with the measured linear bias and the parameter characterising the damping of small-scale clustering $
\sigma_{\rm v}= 5.204$~$h^{-1}$Mpc, which is estimated from the linear theory assuming that haloes trace dark matter flow.
We compute the prediction at the volume averaged redshift.
The prediction is supposed to be accurate in the weakly non-linear regime,  which roughly corresponds to the scales larger than $30~h^{-1}$Mpc for this redshift. Note that the validity of the prediction relies on the assumption of linear bias and on the distant observer approximation.

We have a very good agreement on the redshift-space monopole and a good agreement for the quadrupole,
although we see a small discrepancy at large scales which might be due to the cosmic variance of the measurement results or alternatively to the presence of a finite distance observer.
Our procedure is therefore fully validated.

Finally we have also checked that the relative amplitude of effects beyond the standard Doppler RSD are of order of $\sim 10^{-2}$ (monopole) or $\sim 10^{-3}$ (quadrupole). Disentangling these effects from the main contribution is therefore a challenge.
We now focus on the dipole where the standard Doppler contribution vanishes in the distant-observer limit.

\section{Results}
\label{sec:Results}

We are now interested in the dipole. In order to investigate this multipole we will split it into three parts. The first part is the one generated by statistical fluctuations in real space. This is mostly interesting for the comparison to observations. It can be minimized by increasing the statistics (number and density of haloes). The second part is generated when the halo catalogue is projected from real space to the unperturbed FLRW light-cone. This part is a small contribution to the dipole which has to be taken into account when making accurate predictions. The third part of particular interest in this paper is the dipole generated by the perturbation of the FLRW light-cone (at first order in the weak-field approximation) related to the formation of large-scale structures in the universe.

In what follows, we compare our results with the linear theory predictions presented in Section \ref{sec:lineartheory}, and summarised in Table~\ref{tab:predictiontable}. In the predictions including evolution terms (i.e. Eqs.~\ref{eq:evo1} and \ref{eq:evo2}), the comoving distance derivative of the bias changes the amplitude of dipole correlation, and the resultant prediction does not simply scale as $\xi_1\,\propto\,\Delta b$. We thus plot the averaged prediction as well as the maximum and minimum of the predictions among possible combination of haloes, filling up the interval with gray shade (Fig.~\ref{fig:lightconeeffect}, upper-right panel of Fig.~\ref{fig:lineardipole}, and Fig.~\ref{fig:alleffectslarge}).

\begin{figure*} 
\centering
\includegraphics[width=0.65\columnwidth]{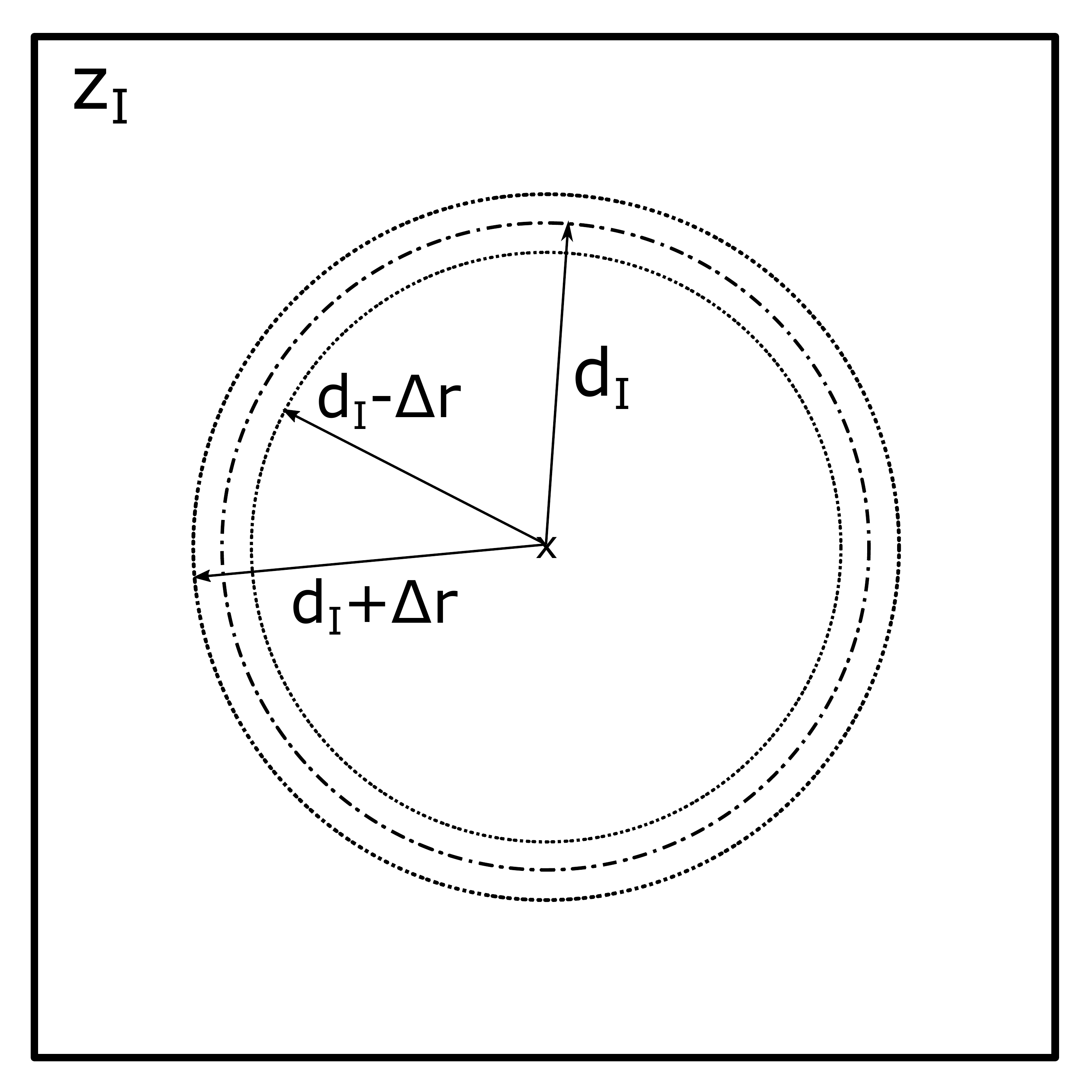}
\includegraphics[width=0.65\columnwidth]{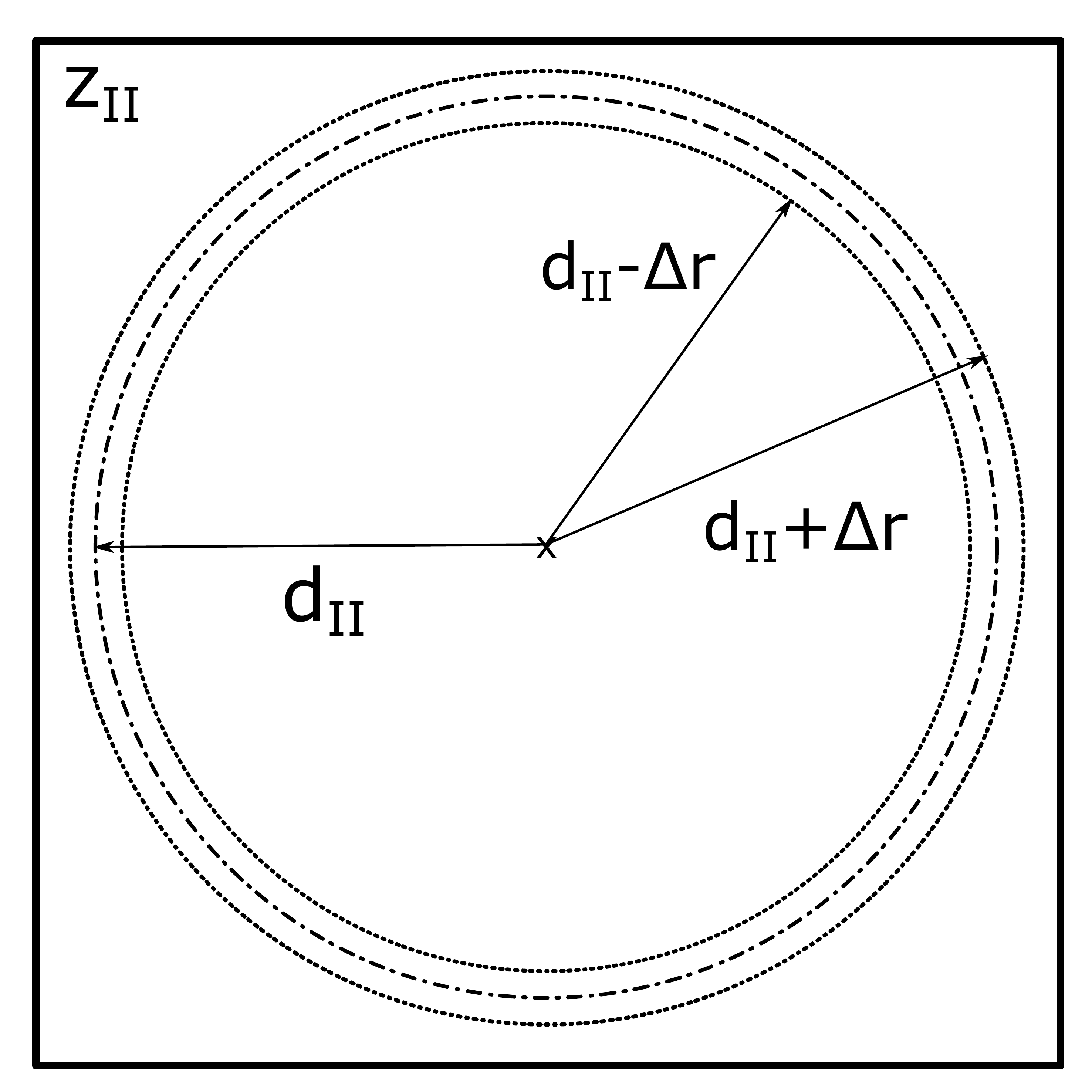}
\includegraphics[width=0.65\columnwidth]{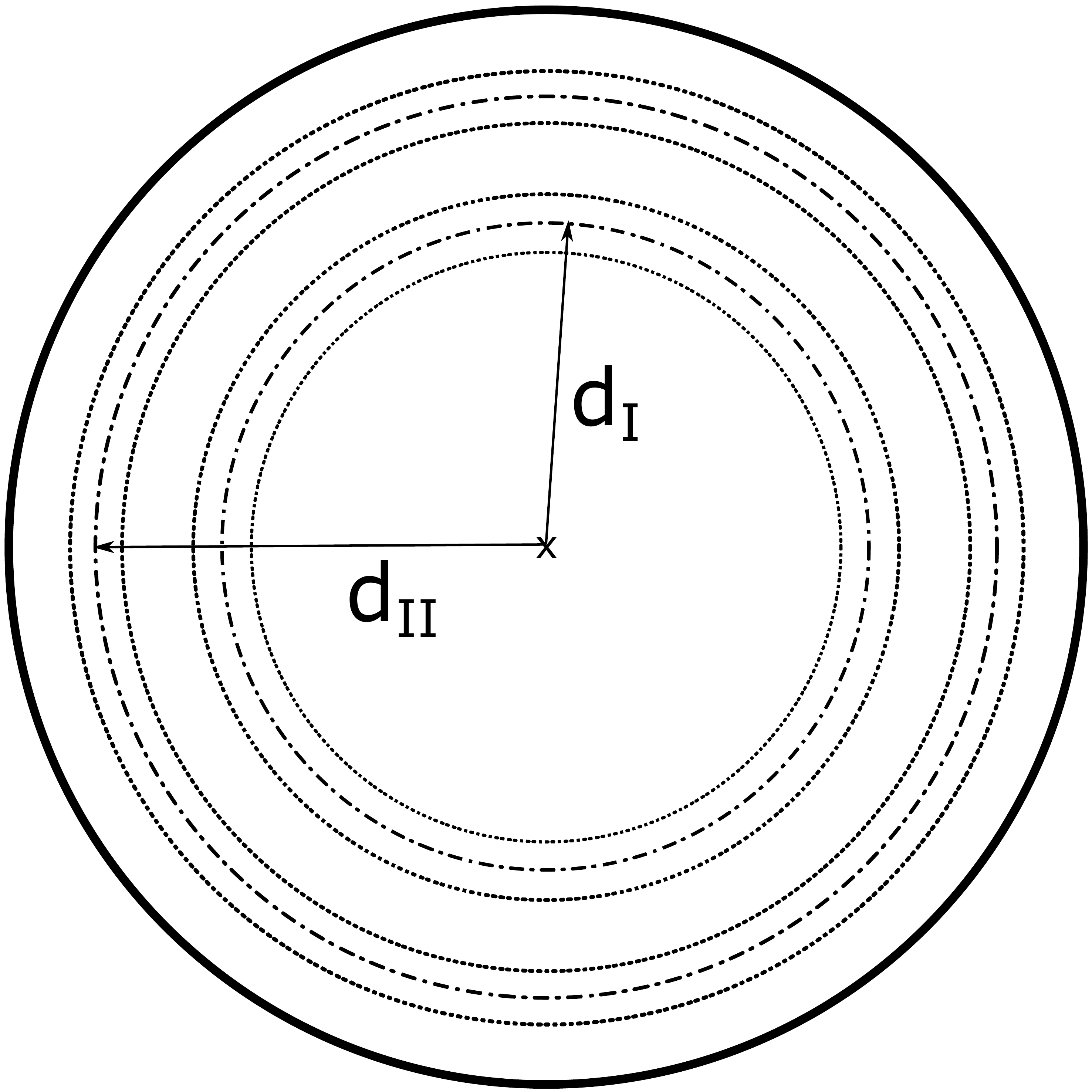}
\caption{Representation of snapshots and light-cone. \emph{Left panel} : snapshot at redshift $z_{\rm I}$. \emph{Middle panel} : snapshot at redshift $z_{\rm II}$. \emph{Right panel} : full-sky light-cone (made of $\sim 300$ small shells of size $\sim 4~h^{-1}$Mpc not represented here). Here $z_{\rm II} > z_{\rm I}$ and $d_{\rm I}$ (respectively $d_{\rm II}$) is the FLRW comoving distance at redshift $z_{\rm I}$ (respectively $z_{\rm II}$). To estimate light-cone effects, we compare large light-cone shells (of size $2\Delta r\sim 300~h^{-1}$Mpc) to the equivalent volume in a snapshot at the corresponding homogeneous redshift.}
\label{fig:snaplight}
\end{figure*}

In Section \ref{sec:statistical_fluctuations} and \ref{sec:linear_regime_dipole} we focus on the dipole  at large scales. At these scales the theoretical predictions are proportional to the bias difference between two halo populations (except for evolution effects). We take full advantage of all the cross-correlations by using them to compute the weighted dipole normalised by the bias difference (see Section~\ref{sec:estimation_correlationfunction}). Each cross-correlation are shown in Appendix \ref{sec:appendix_massdependancelinear}.
In Section \ref{sec:nonlinear_regime_dipole} we investigate the dipole at smaller scales where non-linearities arise. For these scales we show the dipole and its mass dependence.

\subsection{Statistical fluctuations and light-cone effect}
\label{sec:statistical_fluctuations}
In this section, we measure the dipole of the halo distribution using snapshot information (i.e. distribution of haloes at constant time). We consider 7 snapshots in the interval between $z=0.05$ and $z=0.465$. For each snapshot, we build a shell of size of order $\sim 300~h^{-1}$Mpc centred on the comoving distance corresponding to the snapshot redshift. In this way, we are able to compute the mean dipole in real space. We can also compare the dipole computed at constant time to the one computed on the FLRW unperturbed light-cone at the same position (see Fig.~\ref{fig:snaplight}). 

\begin{figure} 
    \includegraphics[width=\columnwidth]{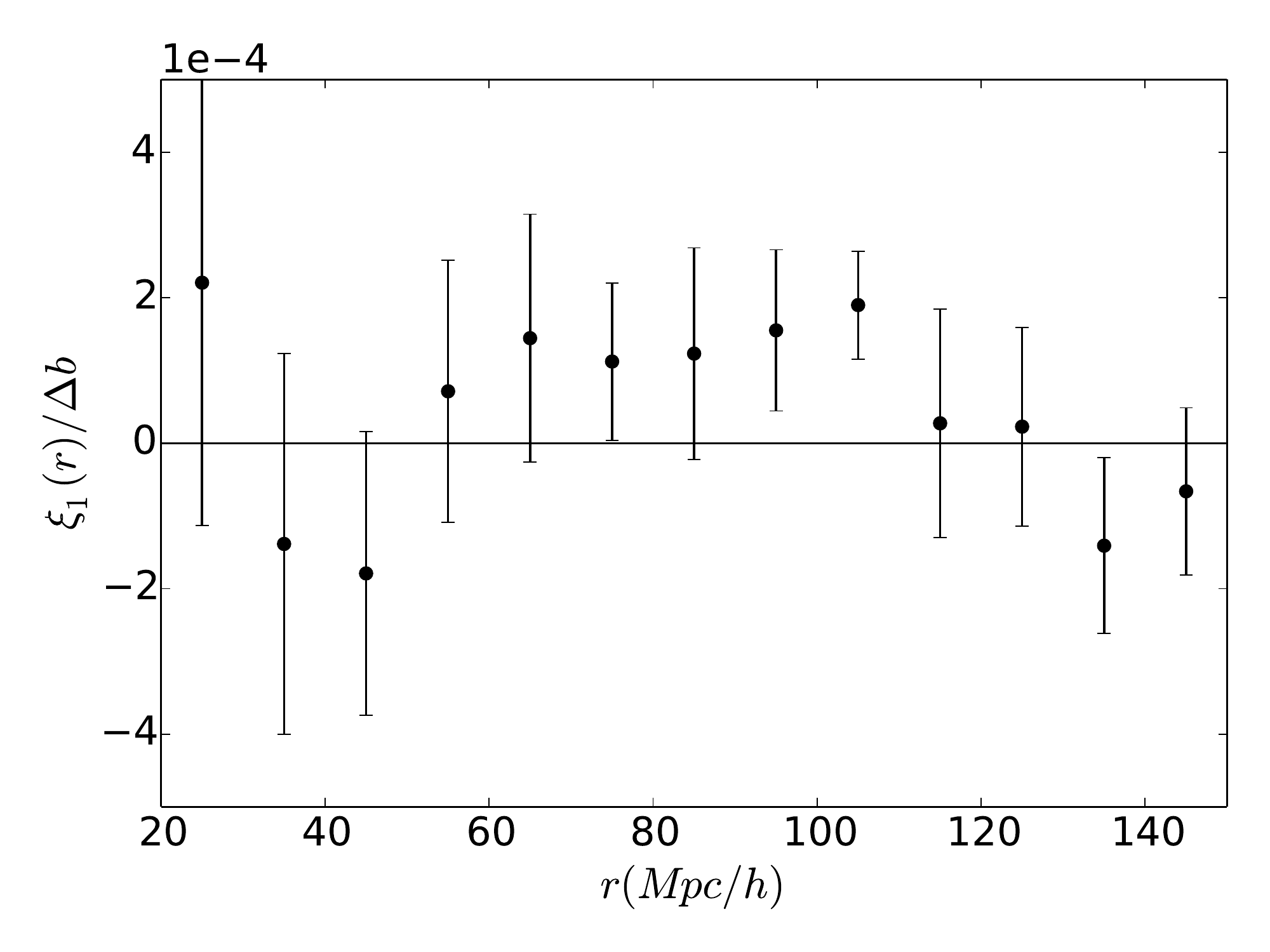}
    \caption{Real-space (snapshot) dipole of the cross-correlation function, normalised by the bias. We use a weighted average of all the cross-correlations as well as a weighted average on 7 redshift shells. The dipole is consistent with zero within the statistical error bars.}
    \label{fig:snapshot}
\end{figure}

\subsubsection{Dipole generated by statistical fluctuations}

After averaging the snapshot dipole computed in each shell (with a weigh given by the volume of each shell), we obtain the mean snapshot dipole.  The standard errors on the mean are computed from the 7 snapshots and should therefore be taken with caution because of the lack of independence. For a wide range of radii ($20<r<150~h^{-1}$Mpc) the dipole shown in Fig.~\ref{fig:snapshot} is roughly compatible with zero within the statistical error bars 
(except for several points at 100 and 130 $h^{-1}$Mpc where the error bars are very likely underestimated). Moreover the error bars are limited to $\sim 2\times10^{-4}$.

In principle, for a very large volume and for a very large density of haloes, the dipole  tends towards zero. For finite volume surveys with finite number of galaxies per unit of volume such a fluctuation might blur the dipole. However, this noise can be minimized by increasing the size of the surveys and the density of pairs of haloes (smaller haloes, more halo populations, see \citealt{bonvin2016optimising}). As we will see for our light-cone, the noise is below the signal but it can sometimes reach the same order of magnitude as the signal. Increasing the halo statistics by a factor of $\sim 10$ should be enough to boost the signal-to-noise ratio. In our simulation, we have simply subtracted this noisy contribution to extract the physical part of the dipole signal.

\begin{figure} 
	\includegraphics[width=\columnwidth]{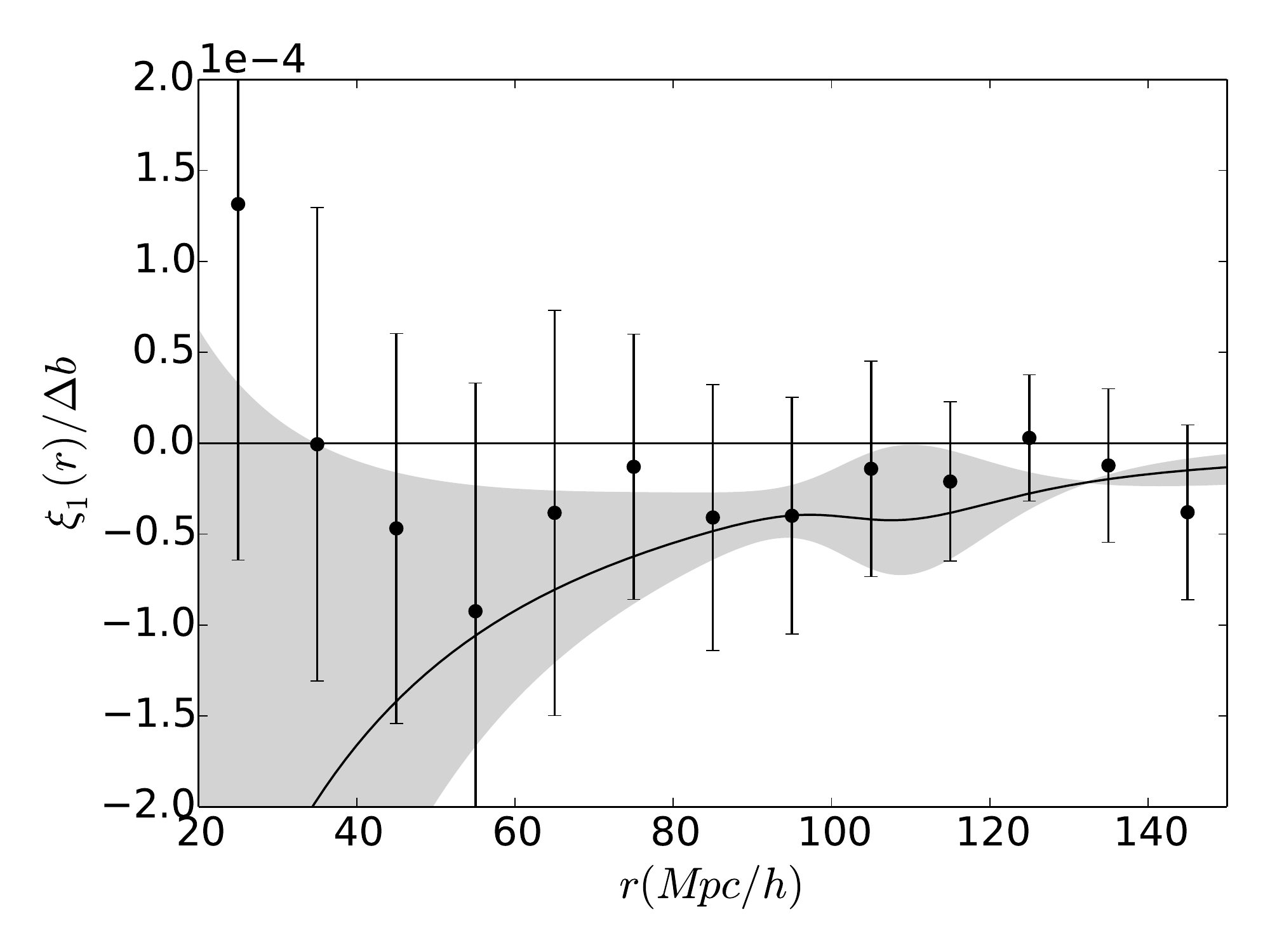}
    \caption{Light-cone-effect contribution to the dipole of the cross-correlation function, normalised by the bias. We use a weighted average of all the cross-correlations as well as a weighted average on 7 redshift shells. The linear prediction given by the first line of Table~\ref{tab:predictiontable} is shown by the grey filled curve (as the prediction is not completely proportional to the bias difference) while the averaged prediction is shown with black solid line.}
    \label{fig:lightconeeffect}
\end{figure}

\subsubsection{Light-cone and evolution effects}

\begin{figure} 
	\includegraphics[width=\columnwidth]{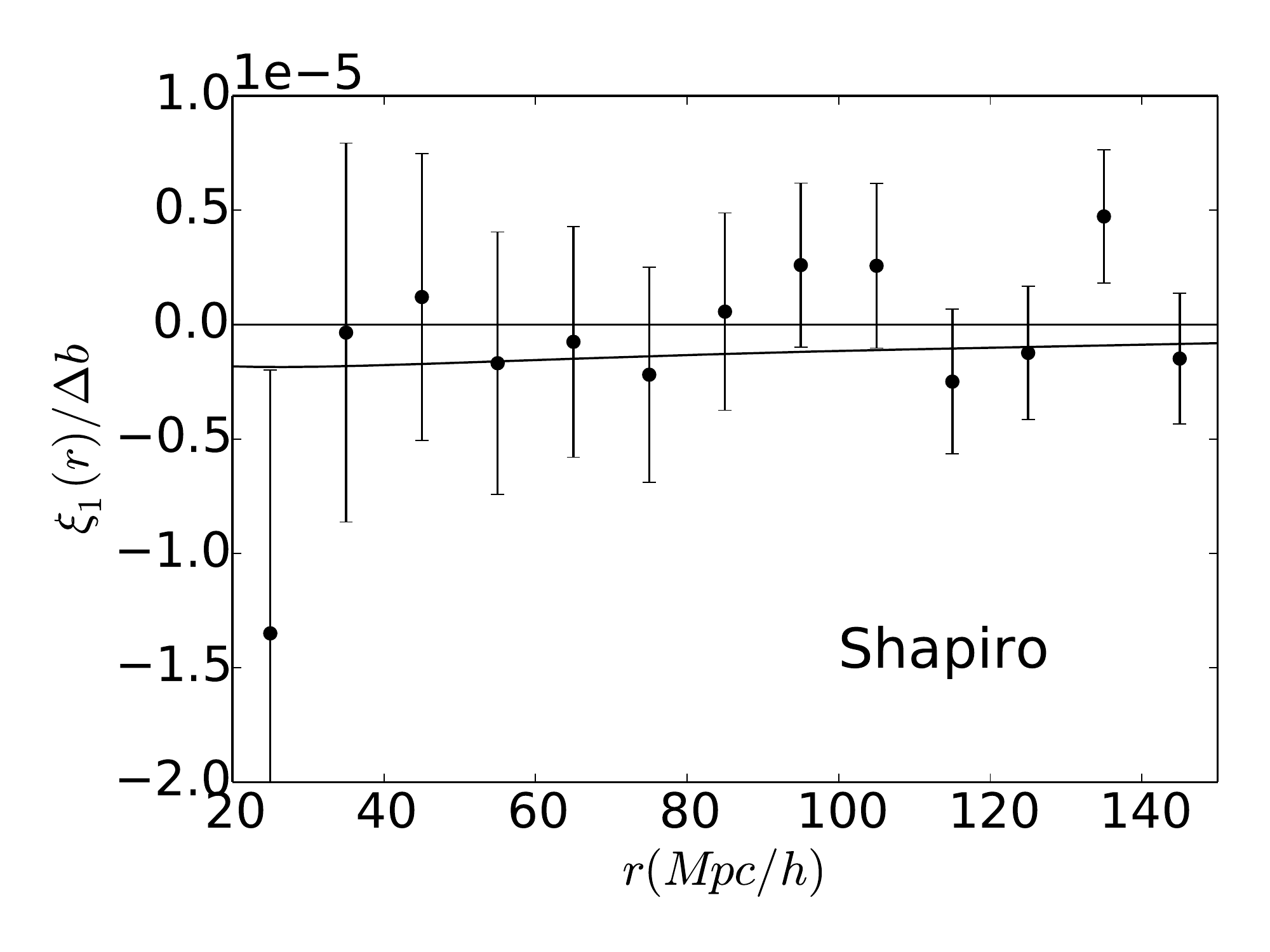}
    \caption{The dipole of the cross-correlation function normalised by the bias induced by the Shapiro effect. The prediction for this effect is shown in Section~\ref{sec:lineartheory}. This contribution is very small and will be neglected afterwards.}
    \label{fig:shapiroeffect}
\end{figure}

In this section, we compute the dipole within the shells in the light-cone and the corresponding shells in the snapshots. By subtracting the two we can extract the so-called light-cone effect (\citealt{kaiser2013measuring,bonvin2014asymmetric}). The main contribution to this effect is related to the peculiar motions of haloes: haloes are not at the same position in the snapshot (constant time) and in the FLRW light-cone. Another contribution comes from the evolution effects: haloes are not exactly the same in the snapshot and light-cone as they experience merging.

While the light-cone effect has already been taken into-account in simulations \citep{cai2017gravitational,zhu2017nbody},
our approach is more sophisticated since we have directly built the light-cone on the fly during the simulation (at each coarse time-step). In previous work, the light-cone effect was added as a post-processing procedure on top of the snapshots. This approach usually neglects the variation of the potential during the evolution. Moreover, evolution effect are not easily captured while they are a direct outcome of our approach.

As shown in Fig.~\ref{fig:lightconeeffect}, the light-cone effect (plus evolution effect) is in agreement with the linear expectation (see first line of Table~\ref{tab:predictiontable}) while error bars remain important and the points at small scales seem to depart from the averaged prediction. 
The normalised dipole is of order $10^{-4}$ at most.
As we will see later it is a subdominant contribution to the full dipole.
The linear prediction is broad due to evolution effects which are not proportional to the bias difference. For the cross-correlation of the two most massive halo populations the evolution effect is of the same order as the light-cone effect. This shows the importance of an accurate modelling as well as precise bias measurement to disentangle these two effects.

Last, there is a difference between the particle positions on the FLRW light-cone given by the simulation and the particle position on the perturbed FLRW light-cone seen by photons due to time delay (Shapiro effect). However this contribution is too small to be detectable (inferior to $10^{-5}$) as seen in Fig.~\ref{fig:shapiroeffect} and will therefore be neglected.

\begin{figure*} 
	\includegraphics[width=\columnwidth]{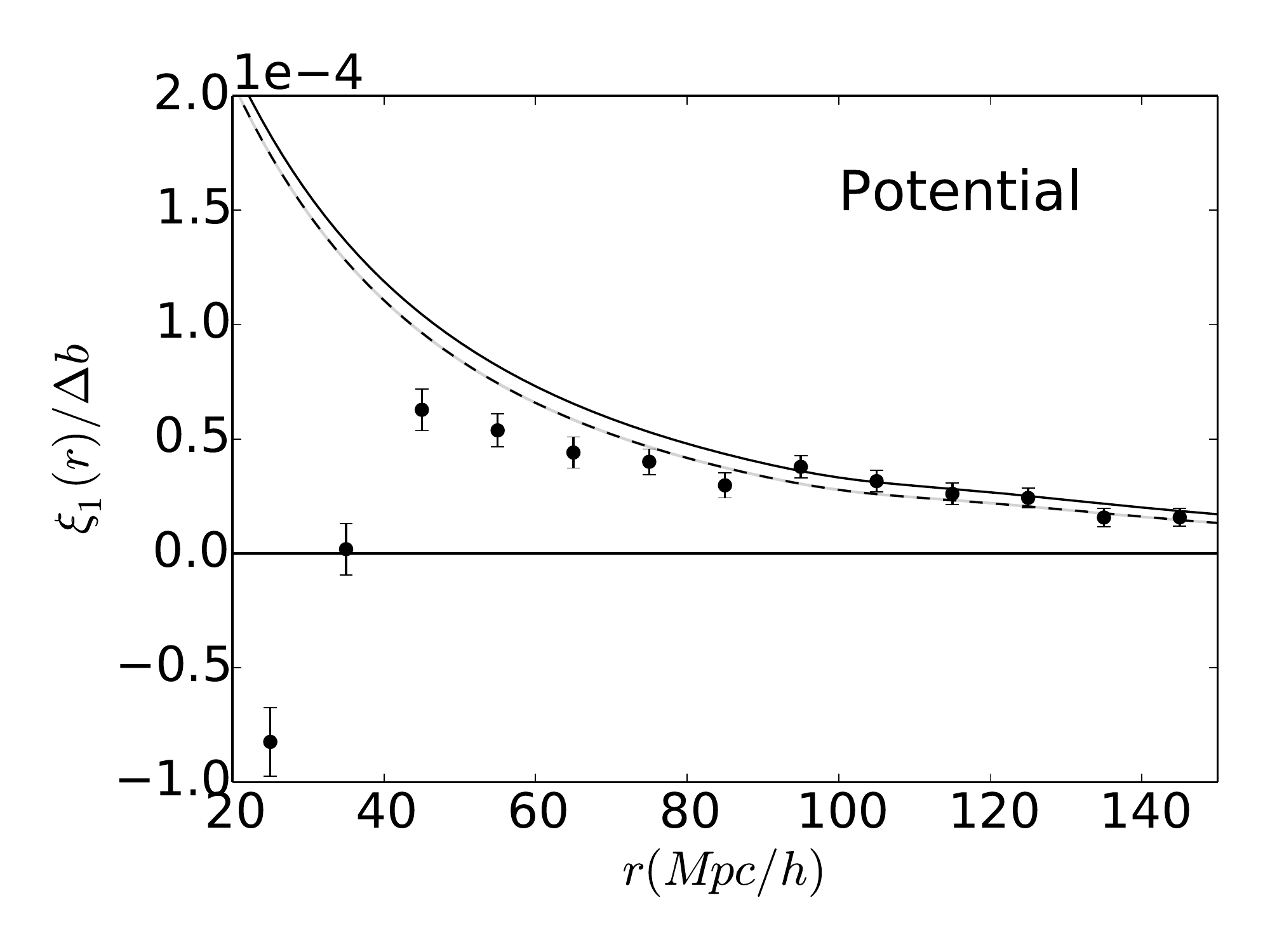}
	\includegraphics[width=\columnwidth]{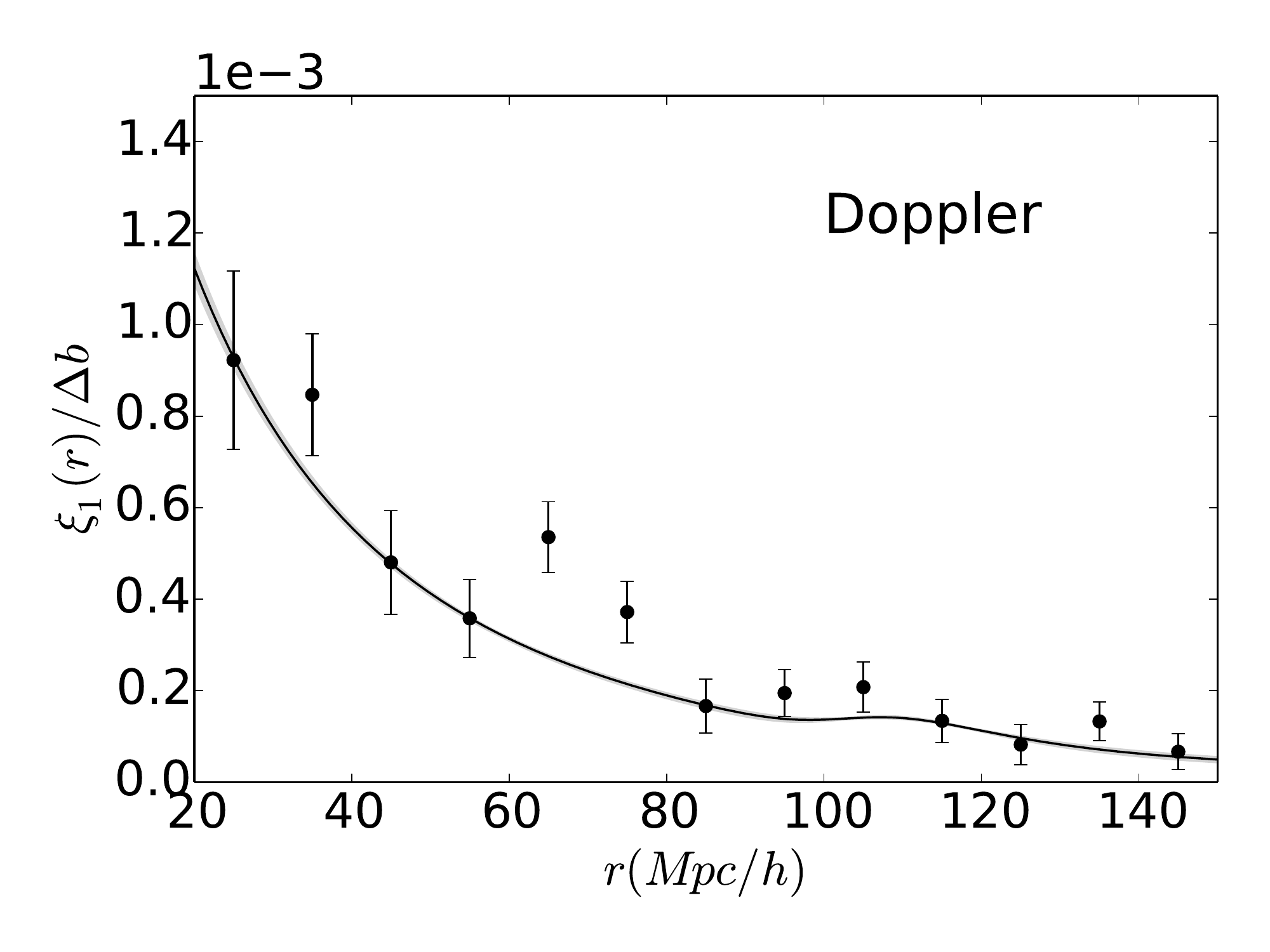}
	\includegraphics[width=\columnwidth]{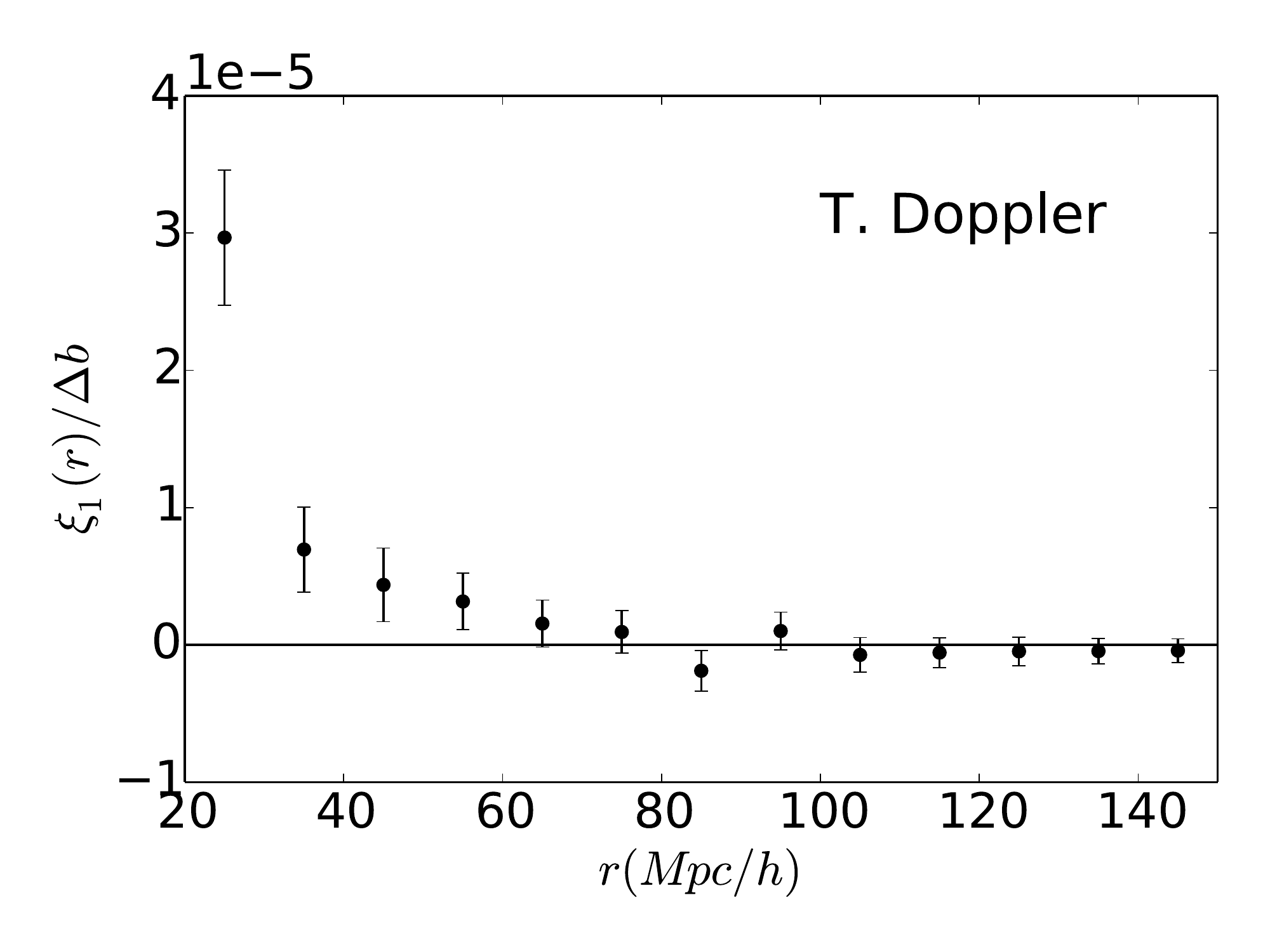}
	\includegraphics[width=\columnwidth]{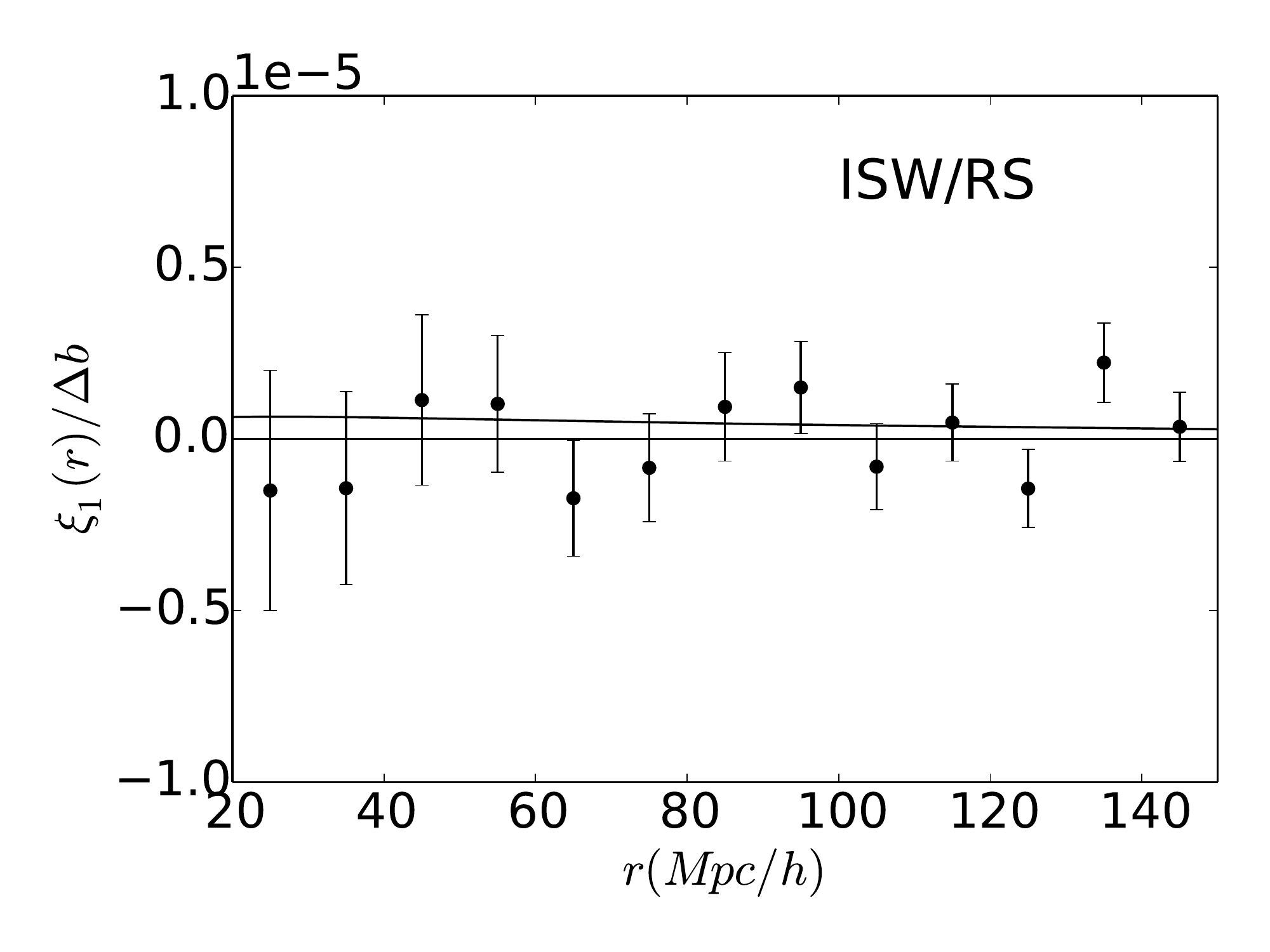}
	\includegraphics[width=\columnwidth]{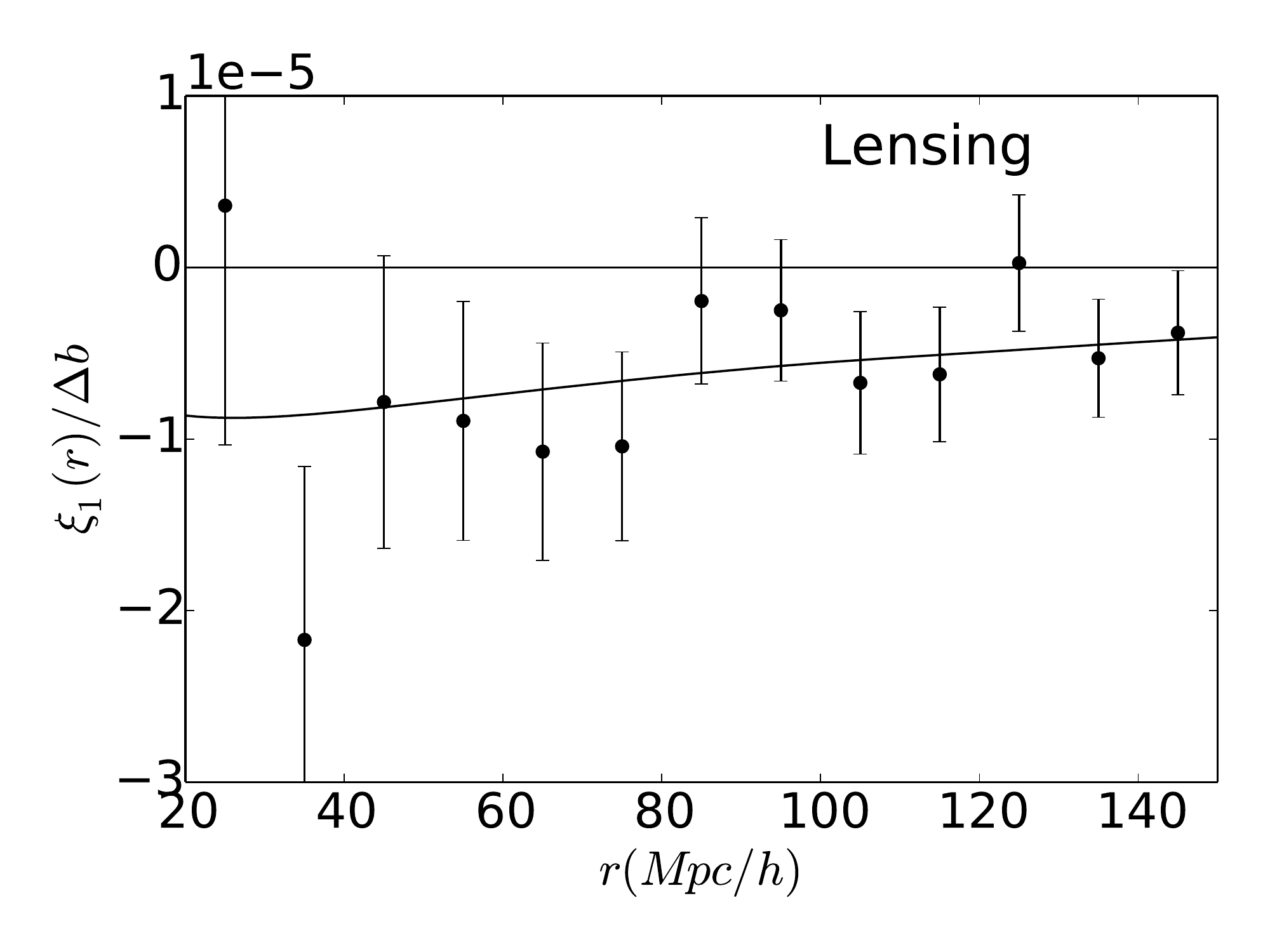}
	\includegraphics[width=\columnwidth]{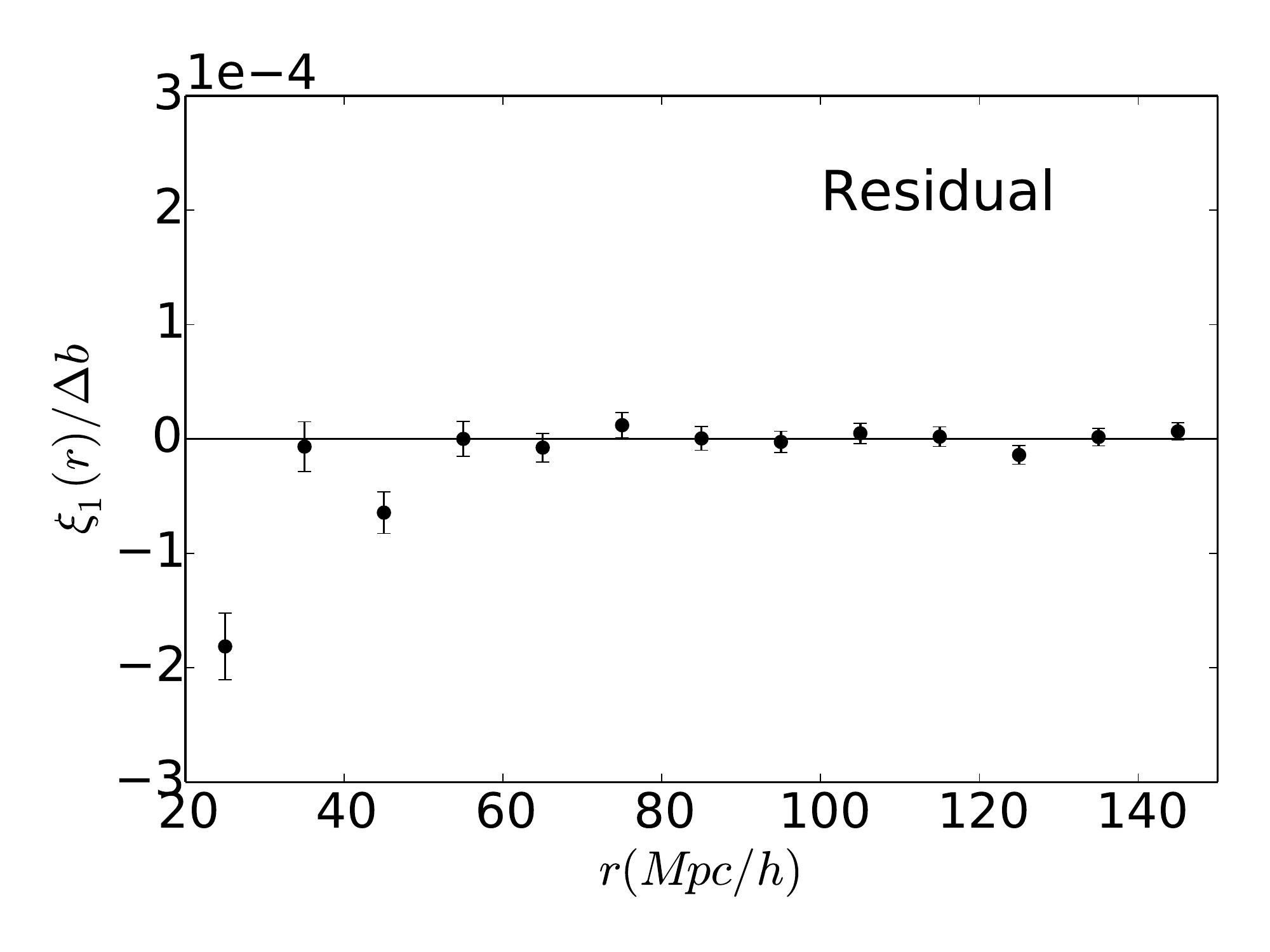}
    \caption{Dipole of the cross-correlation function normalised by the bias, at large scales, for different perturbations of the observed halo number count.  
   This leads to: \emph{upper left panel} only the contribution from gravitational potential was taken into account as a source of RSD, in black dashed line we have the prediction when accounting for leading terms in $\left(\mathcal{H}/k\right)^2$. \emph{Upper right panel} Doppler only, \emph{middle left panel} transverse Doppler only, \emph{middle right panel} ISW/RS only, \emph{bottom left panel} weak lensing  only, and finally \emph{bottom right panel} the \emph{residual} where we subtract all the previous effects to the full dipole taking into account all the effects at once. In black we have the averaged prediction using linear theory at first order in $\mathcal{H}/k$.}
    \label{fig:lineardipole}
\end{figure*}

\subsection{The linear regime and its breakdown}

\label{sec:linear_regime_dipole}
From now on, we subtract the effects from statistical fluctuations as well as the light-cone effect described above. In this section we investigate the dipole at large scales from $20~h^{-1}$Mpc to $150~h^{-1}$Mpc. This corresponds to the linear regime (Section~\ref{sec:lineartheory}) and the beginning of the quasi-linear regime where, in principle, linear theory does not hold any more. 
We focus on the weighted average of the normalised dipole.
Each cross-correlation is shown in Appendix~\ref{sec:appendix_massdependancelinear}.

\subsubsection{Contributions to the dipole at large scales}

At first order in the metric perturbation, redshift-space distortions are the sum of five contributions: four redshift perturbations (see Eqs.~\ref{eq:firstnewredshift} to~\ref{eq:lastredshift}) plus the lensing effect which affects the apparent position of haloes (we neglect the small Shapiro term which is already subtracted). To investigate each of these effects we have produced five catalogues where only one effect is present at a time.  This allows us to compare our measurements to the linear predictions summed up in Table \ref{tab:predictiontable} as shown in Fig.~\ref{fig:lineardipole}. 

The upper left panel shows the dipole when we only consider the gravitational potential as a source of RSD. We find a good agreement at large scales with linear theory (even better with higher-order terms in $\mathcal{H}/k$) down to a radius $r \approx 60-80~h^{-1}$Mpc where $\xi_1/\Delta b \approx 5\times10^{-5}$. The dipole drops sharply at smaller scales and even becomes negative at scales smaller than $30~h^{-1}$Mpc. Linear theory fails to predict this drop. As we shall see, it is the second most important contribution to the total dipole at large scales. The upper right panel shows the dipole for the Doppler term only. 
The signal is much larger rising from $\xi_1/\Delta b \approx 2\times10^{-4}$ at $100~h^{-1}$Mpc up to $\xi_1/\Delta b \approx 10^{-3}$ at $20~h^{-1}$Mpc. Data points and linear prediction are in agreement when looking at the scatter (except for two points at 65 and 75$~h^{-1}$Mpc which are at about $3-\sigma$ above the linear expectation). 
This is the dominant term and it is mostly related to the divergence of the line-of-sights \emph{at this redshift} (this term is inversely proportional to the comoving distance and is therefore fainter at higher redshift). In the middle left panel we only consider the transverse Doppler term which should be null in the linear regime. Our measurement is consistent with zero between $r \approx 60~h^{-1}$Mpc and $r \approx 150~h^{-1}$Mpc at a precision better than $2\times10^{-6}$. Below $60~h^{-1}$Mpc the data show a positive dipole. The transverse Doppler yields an overall redshift of the lighter galaxies w.r.t to the more massive ones which is explained (at smaller scales) by \citet{zhao2013testing}. In the middle right panel we see the dipole of the ISW/RS effect only. This effect is integrated and suppressed by a factor $\mathcal{H}/k$ compared to the other terms and is therefore expected to be small. Here we only check the consistency with zero that is given on the full scales of interest with a precision of order  $\lesssim 5\times10^{-6}$. Bottom left shows the effect of weak lensing only. 
This effect is much fainter than the potential and Doppler terms and the data clearly favours a negative dipole of order of a few $10^{-6}$ which follows well the linear prediction.

Lastly, the bottom right panel shows the \emph{residual}, i.e. the full dipole including all effects (full redshift perturbations and lensing) minus the individual contributions mentioned above. It includes all the cross-terms (potential-Doppler, potential-lensing, etc...) as well as the non-linear mapping that was ignored by the linear calculation.  
We see that the \emph{residual} is consistent with zero to a good precision beyond $50~h^{-1}$Mpc. However below this threshold the \emph{residual} gives a negative contribution to the total dipole. We expect this term to be dominated by the potential-Doppler cross-correlation term as they are the two dominant terms. We can see that the departure of the \emph{residual} from zero occurs at approximately the same scale as the departure of the dipole for the potential only term from the linear prediction. It indicates a failure of the linear regime at this scale. In Appendix \ref{sec:appendix_massdependancelinear} we show the behaviour of the \emph{residual} for a wide range of cross-correlations.
Depending on the values of bias and bias difference, the amplitude of the signal as well as the scale at which the \emph{residual} departs from zero can change.

\begin{figure} 
	\includegraphics[width=\columnwidth]{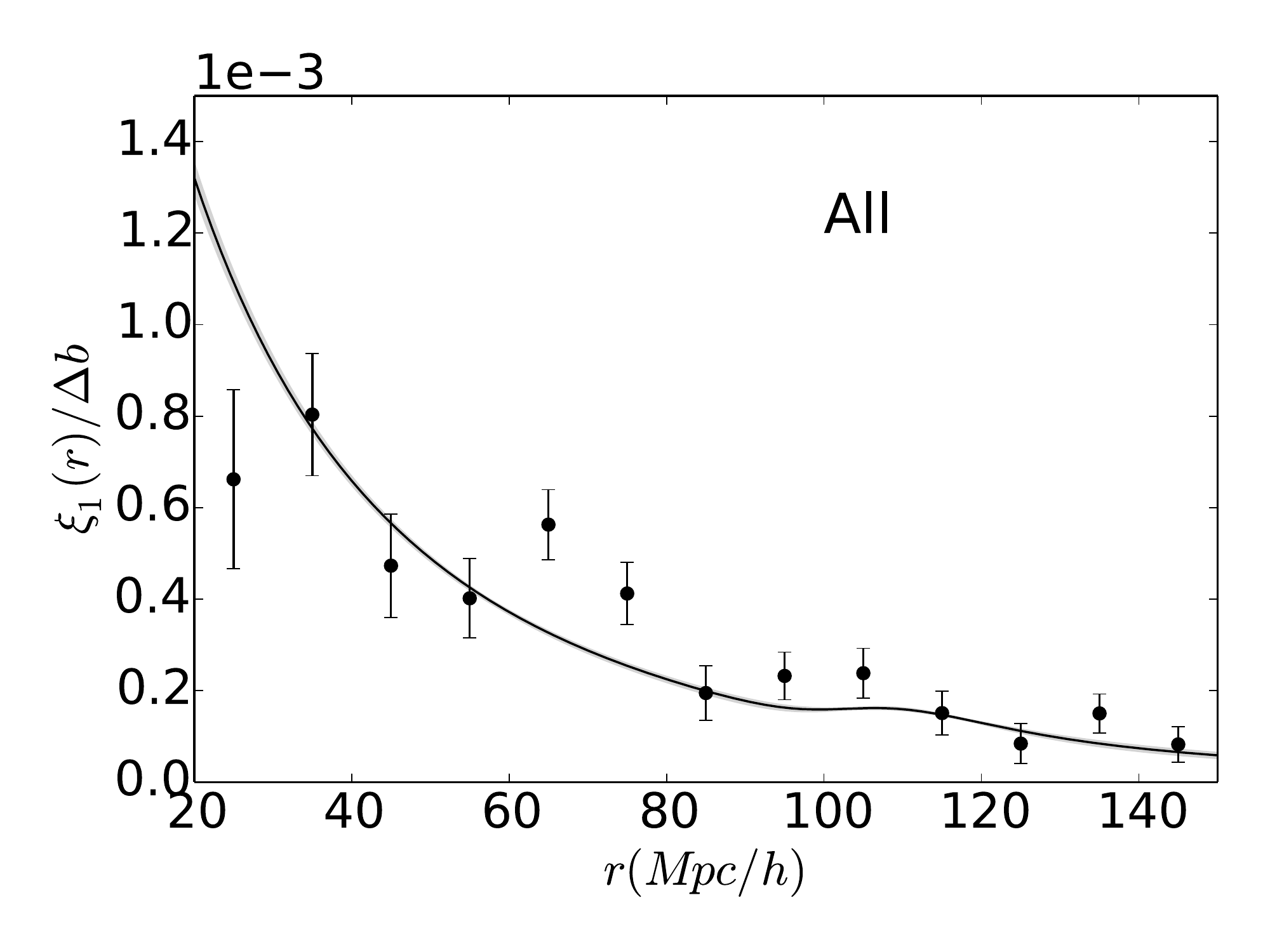}

    \caption{Full dipole of the cross-correlation function normalised by the bias. The dipole is dominated by the Doppler contribution.}
    \label{fig:alleffectslarge}
\end{figure}

\subsubsection{Total dipole}
Now that we have seen all the individual contributions to the dipole, we show the final result that is the sum of all the previous contributions (see Fig.~\ref{fig:alleffectslarge}). As we expected the dipole is dominated by the Doppler term as the effects of the others terms are small. 
However the dipole departs from linear theory near $30 h^{-1}$Mpc due to the non-linear contribution from the potential and residual.
In \citet{bonvin2014asymmetric} the authors claim that a measurement of the total dipole in the linear regime will allow us to probe velocity field and to test general relativity (through the Euler equation). As we have seen to reach this goal, one has to pay attention to real-space statistical fluctuations of the dipole (therefore huge sample and survey volume are mandatory), evolution effect (as we need to properly model bias evolution while for the moment we are limited to simple phenomenological models) and the bias itself (as the linear prediction is proportional to the bias difference between the population and simple scale-independent bias models are considered).

\subsection{From quasi-linear scales to non-linear scales}
\label{sec:nonlinear_regime_dipole}
After investigating the linear regime (beyond $50~h^{-1}$Mpc) and its breakdown (at 40-60$~h^{-1}$Mpc scales for the potential contribution and the \emph{residual}), we now focus on the transition from quasi-linear to non-linear scales between $30~h^{-1}$Mpc and $5~h^{-1}$Mpc. We use the conservative lower bound of $5~h^{-1}$Mpc because it stands well above the coarse grid size ($0.6~h^{-1}$Mpc), beyond halo's virial radii ($\sim 0.9~h^{-1}$Mpc for large group-size haloes) and it is larger than the light-cone shell size ($\sim 4~h^{-1}$Mpc). In this regime, we expect baryonic effects to be negligible although this has to be further investigated with dedicated simulations.

The theoretical predictions are not necessarily proportional to the bias difference, thus showing the weighted average of the normalised dipole would not make much sense. Here we focus on the cross-correlation of data$\_$H$_{1600}$ (halo mass $M_{h1} \approx 4.5\times10^{13} h^{-1}$M$_{\odot}$) with
data$\_$H$_{100}$ (halo mass $M_{h2} \approx 2.8\times10^{12} h^{-1}$M$_{\odot}$), which gives a bias difference $\Delta b \approx 1$.

\begin{figure*} 
	\includegraphics[width=\columnwidth]{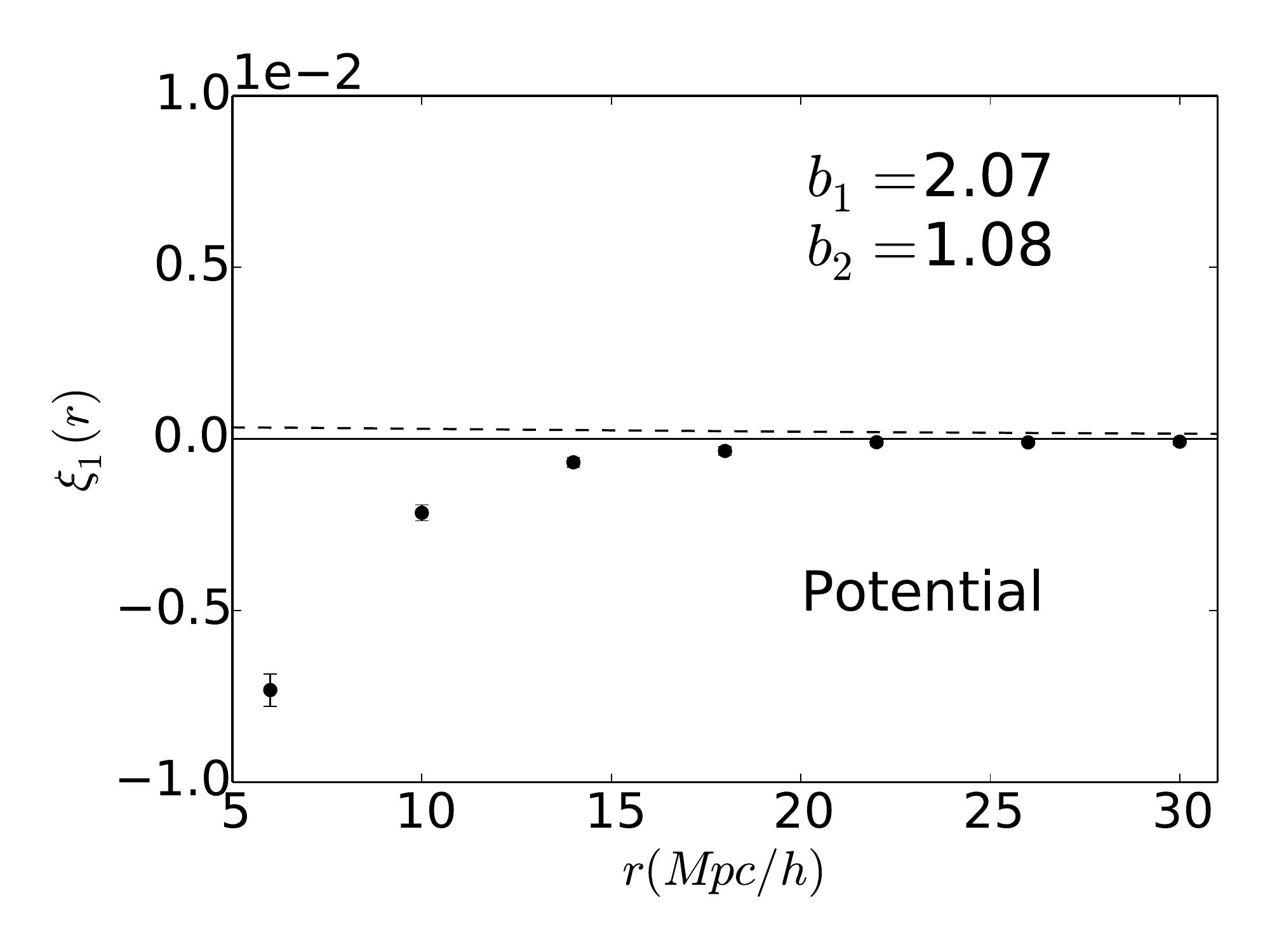}
	\includegraphics[width=\columnwidth]{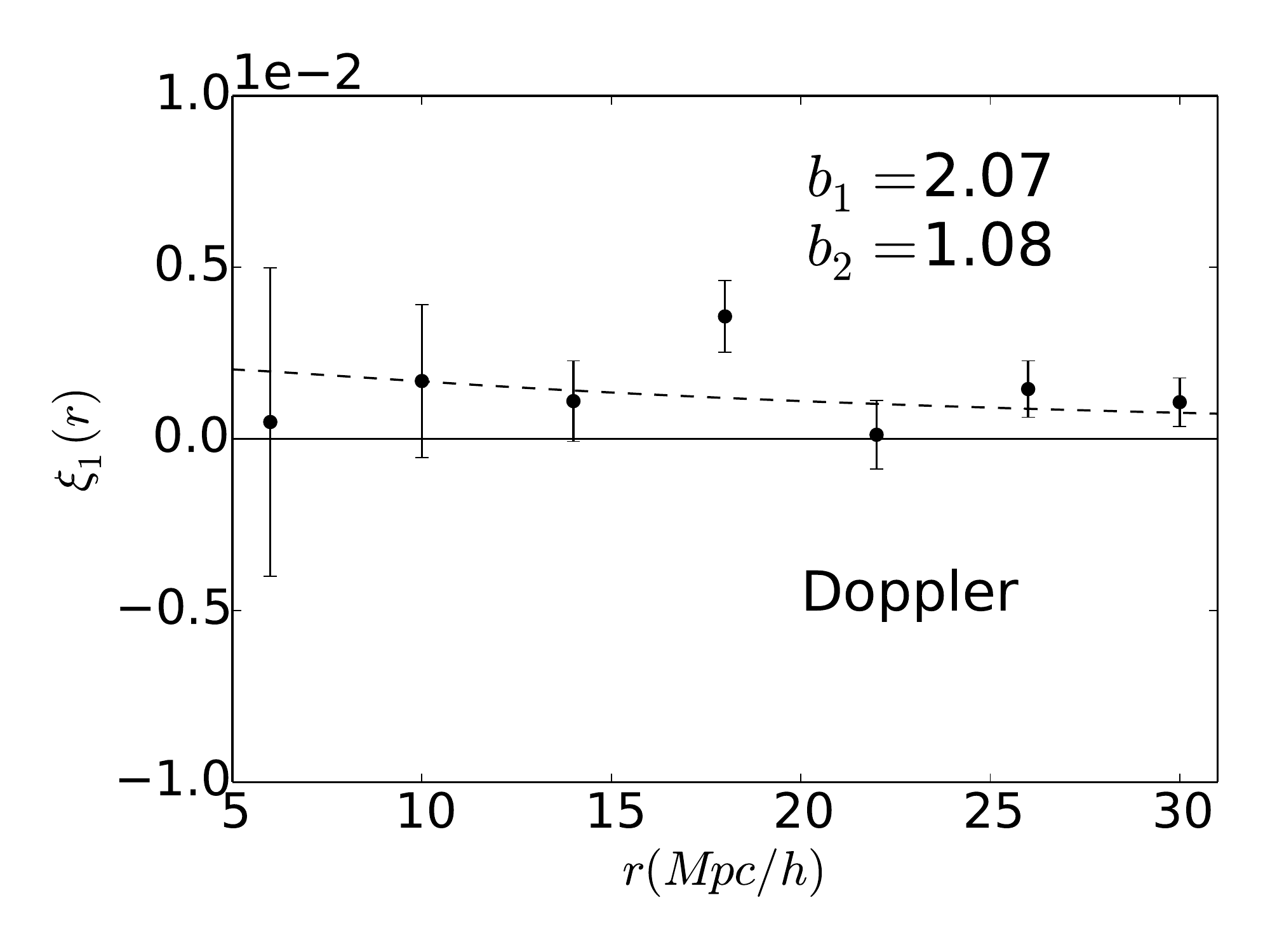}
	\includegraphics[width=\columnwidth]{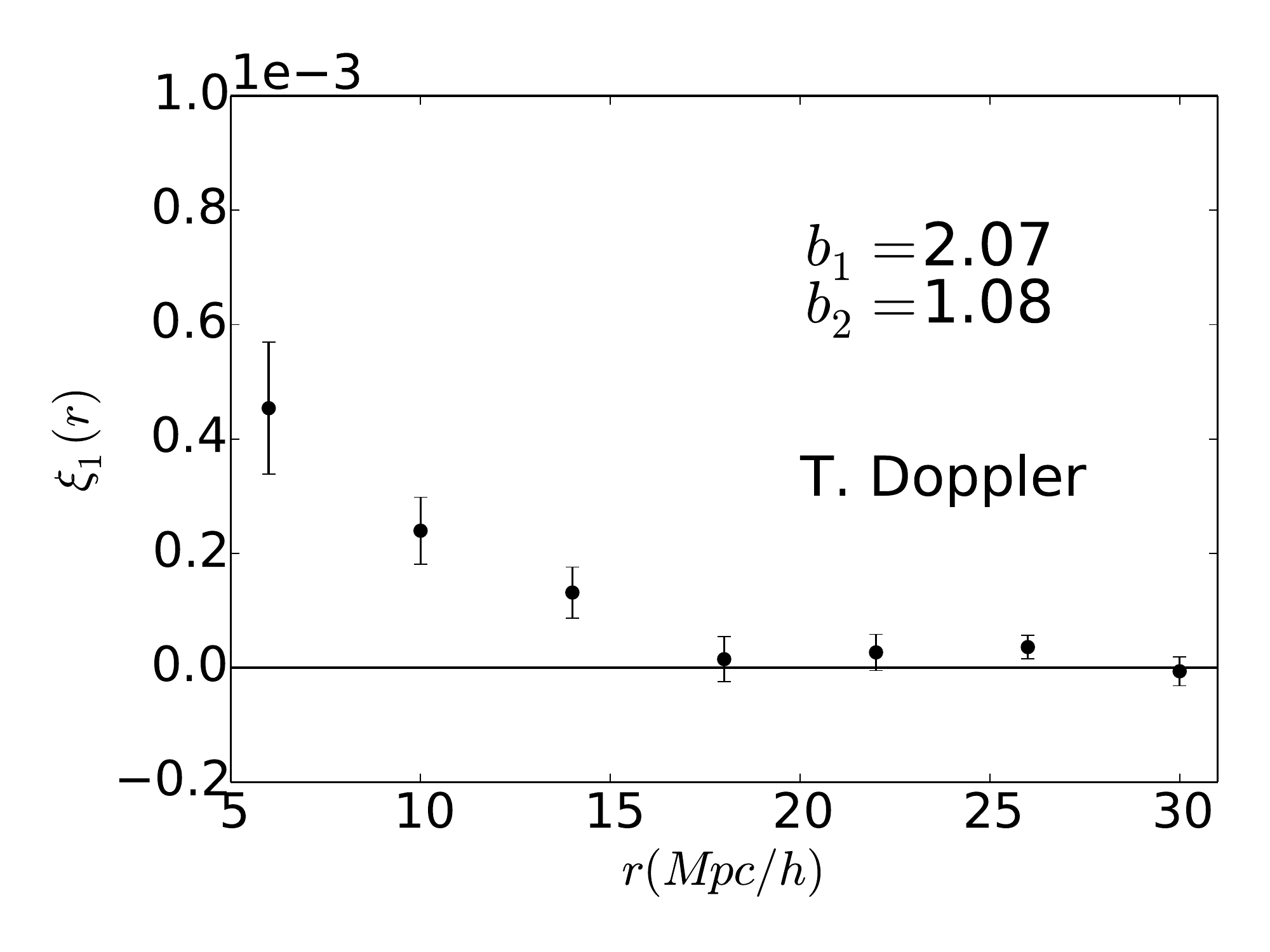}
	\includegraphics[width=\columnwidth]{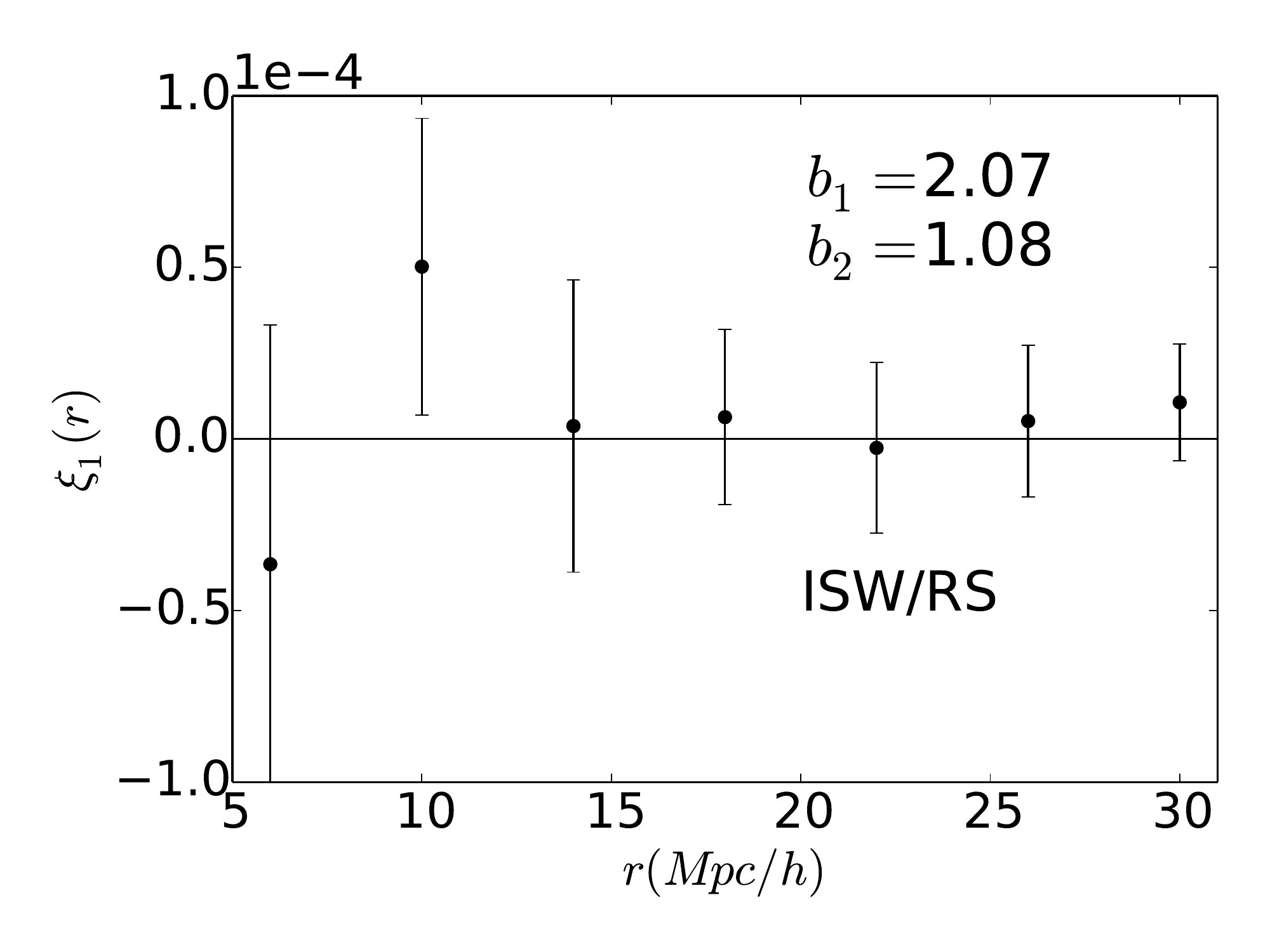}
	\includegraphics[width=\columnwidth]{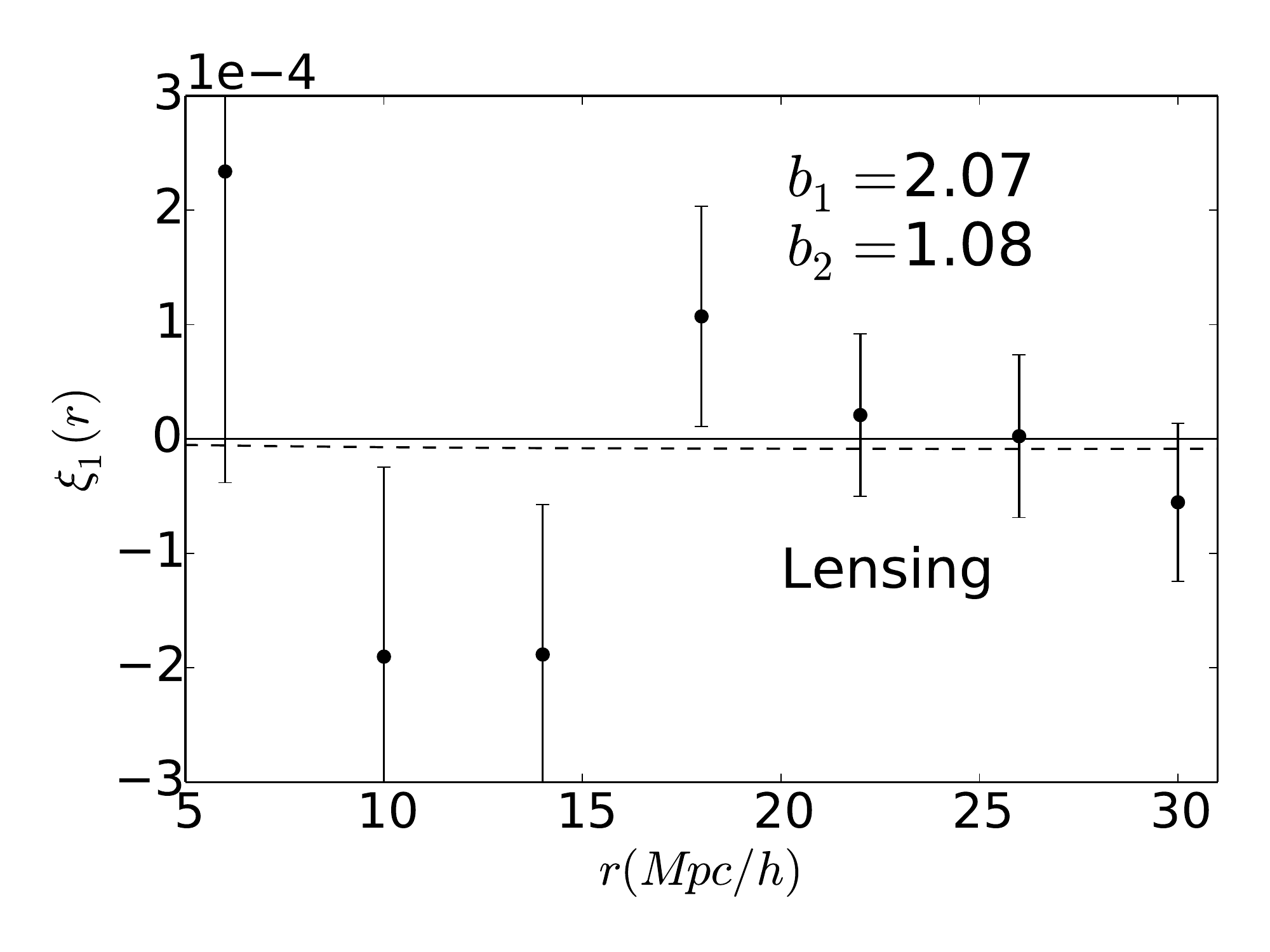}
	\includegraphics[width=\columnwidth]{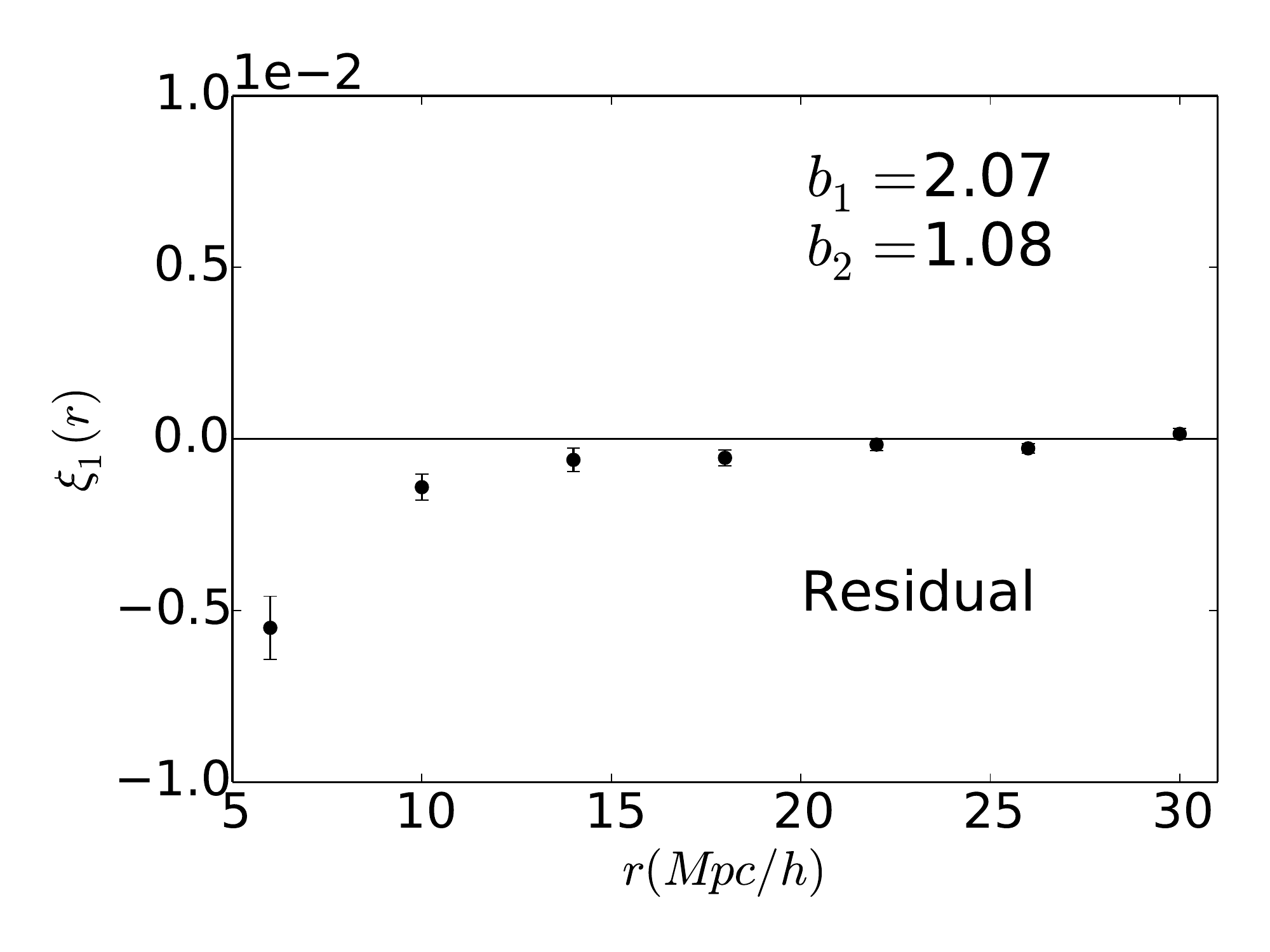}
    \caption{Dipole of the cross-correlation function between data$\_$H$_{1600}$ and data$\_$H$_{100}$, at small scales, for different perturbations of the halo number count.  
    This leads to: \emph{upper left panel} only the contribution from gravitational potential was taken into account as a source of RSD, \emph{upper right panel} Doppler only, \emph{middle left panel} transverse Doppler only, \emph{middle right panel} ISW/RS only, \emph{bottom left panel} weak lensing  only, and finally \emph{bottom right panel} the residual for which we subtract all the previous effects to the full dipole taking into account all the effects at once.}
    \label{fig:nonlineardipole}
\end{figure*}

\subsubsection{Contributions to the dipole at small scales}

The various contributions to the dipole are shown Fig.~\ref{fig:nonlineardipole}. The dominant contributions at non-linear scales (especially at r $<$ $10~h^{-1}$Mpc) are very different from the ones at linear scales. Here the dominant contributions are the potential term (upper left) and the \emph{residual} (bottom right) while in the linear regime the dominant contribution is the Doppler term (upper right). Moreover both contributions are very negative resulting in a negative dipole.

The potential contribution (upper left) drops slowly from $\xi_1 \simeq 0$ near $30~h^{-1}$Mpc to $\xi_1 \simeq -1\times10^{-4}$ at $20~h^{-1}$Mpc. The fall is then much steeper at smaller separations, from $\xi_1 \simeq -5\times10^{-4}$ at $14~h^{-1}$Mpc down to $\xi_1 \simeq -6\times10^{-3}$ at $6~h^{-1}$Mpc. We also note that the measurement is very robust since the statistical error bars are very small (although error bars should be taken with caution at these scales). 
The linear prediction completely fails. The dipole of the halo-halo cross-correlation is a sensitive probe of the gravitational potential up to about ten virial radii for this halo mass.

The velocity contribution (upper right) remains high $\xi_1 \simeq 5-20\times10^{-4}$ between $30$ and $6~h^{-1}$Mpc. At smaller scales the error bars increase from $\sigma_{\xi} \simeq 5\times10^{-4}$ to $\sigma_{\xi} \simeq 5\times10^{-3}$ at smaller scales. Interestingly the Doppler-only dipole remains close to the linear expectation.

The transverse-Doppler contribution to the total redshift in the vicinity of galaxy clusters has originally been highlighted by \citet{zhao2013testing}. However it was restricted to the region $r<2~R_{\rm vir}$ inside or close to the virial radius $R_{\rm vir} \sim 1-2~h^{-1}$Mpc of the clusters.
Interestingly, the transverse-Doppler contribution to the dipole (middle-left) is non-zero even at very large radii ($r>2~R_{\rm vir}$). It remains positive of order $\xi_1 \simeq 2-6\times10^{-5}$ at radii $14<r<30~h^{-1}$Mpc. At smaller scales there is strong increase from $\xi_1=2\times10^{-4}$ at $14~h^{-1}$Mpc to $\xi_1=5\times10^{-4}$ at $6~h^{-1}$Mpc. The ratio to the potential contribution to the dipole is of order $-10$ at this scale.

The ISW contribution (middle right) and lensing contribution (bottom left) are consistent with zero at small scales. The size of the error bars provide an upper limit for the signal of $\xi_1<5 \times 10^{-5}$ for ISW and $\xi_1<10^{-4}$ for lensing. It is still in agreement with the linear prediction which is of the same order of magnitude, however the fluctuations are too important to measure the signal.

Surprisingly, the residual (bottom right) is of the same order
as the potential contribution (from $\sim -10^{-4}$ at $30~h^{-1}$Mpc to $\sim -6\times10^{-3}$ at $6~h^{-1}$Mpc). This is an important result of this paper. It means that at these scales and especially below $15~h^{-1}$Mpc, one cannot add up all the contributions one by one. On the contrary, there are some important contributions involving both potential terms and  velocity terms together.

\begin{figure} 
	\includegraphics[width=\columnwidth]{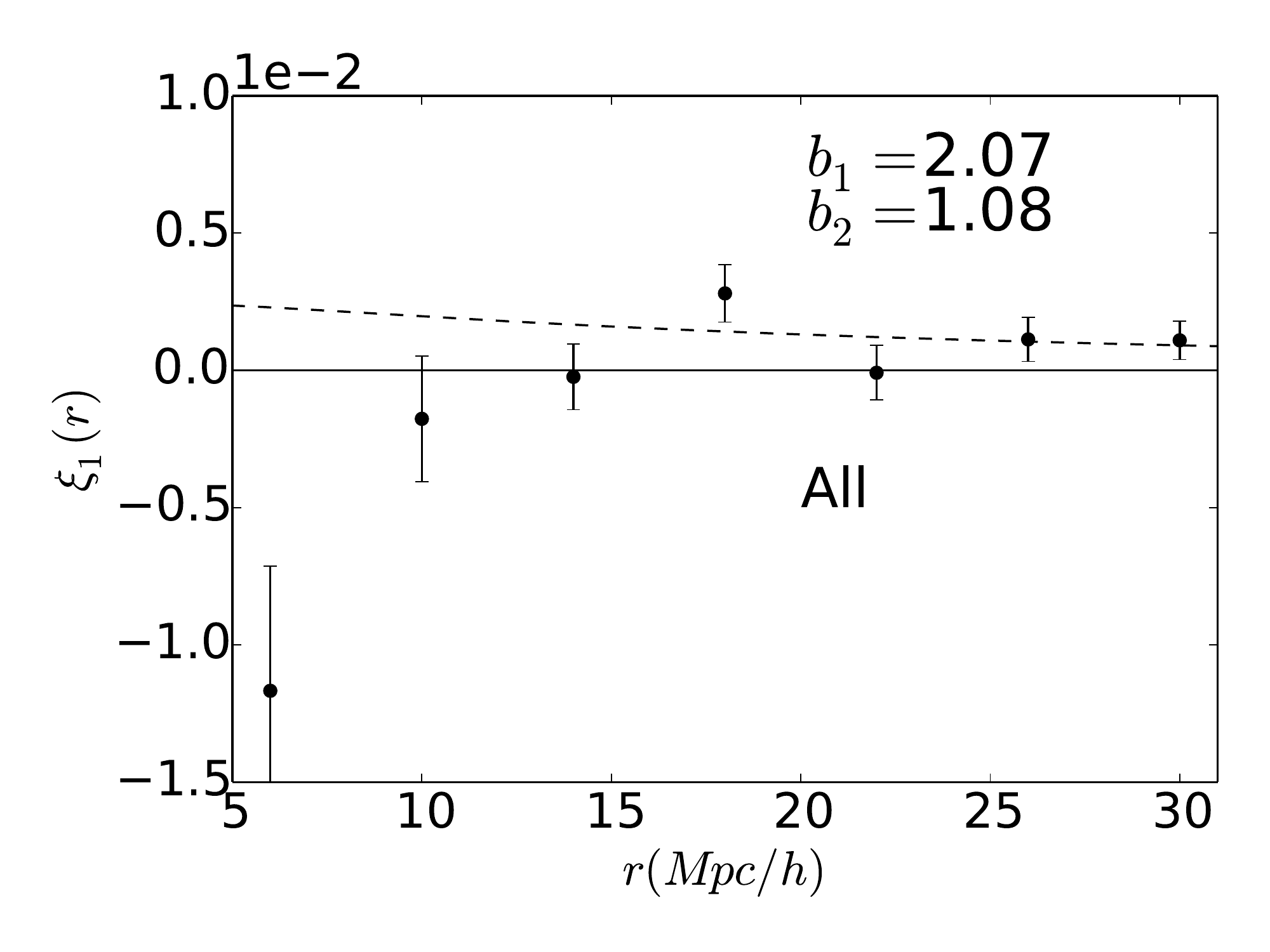}
    \caption{Full dipole of the cross-correlation function between data$\_$H$_{1600}$ and data$\_$H$_{100}$. The deviation from linear theory is  governed by the potential contribution and the  ``residual" (mostly related to the coupling between potential and velocity terms). The dipole is a sensitive probe of the potential well beyond the virial radius of haloes.}
    \label{fig:alleffectssmall}
\end{figure}

\subsubsection{Total dipole}
The total dipole at non-linear scales is presented Fig.~\ref{fig:alleffectssmall}.   It remains slightly positive of order $\xi_1 \sim 1\times10^{-3}$ above $15~h^{-1}$Mpc. As shown in the previous section, this is related to the velocity contribution which remains positive in this region. At smaller scales, the potential contribution dominates over the velocity contribution. The total dipole is then falling down quickly to $\xi_1 \sim - 1\times10^{-2}$ at $6~h^{-1}$Mpc. Moreover within our simulated survey of $8.34~(h^{-1}$Gpc)$^3$, error bars (mostly related to the fluctuations of the velocity field) are smaller than the signal at this scale.  The dipole of the group-galaxy cross-correlation function is therefore a good probe of the potential far outside of the group virial radii. Interestingly, deviations from linear theory are mostly governed by the potential and by the \emph{residual}. The interpretation of the dipole is therefore non-trivial because of correlations between potential and velocity terms. However the dipole carries important information about the potential.

\begin{figure*}
    \includegraphics[width=2\columnwidth]{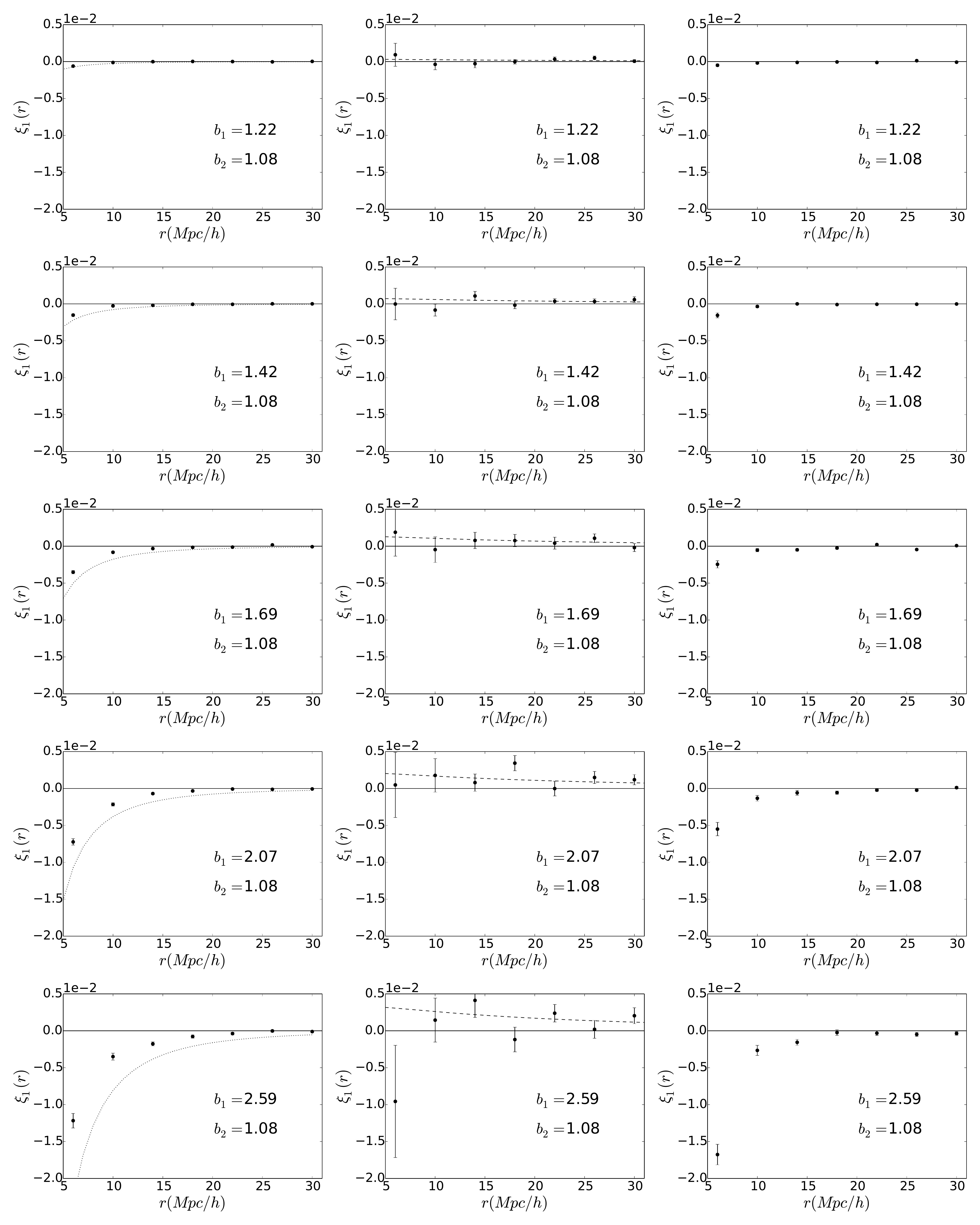}
    \caption{Dipole of the cross-correlation function between different datasets and data$\_$H$_{100}$ (no auto-correlation). \emph{Left panels}: gravitational potential only, dotted lines gives the spherical prediction computed using Eq.~\eqref{eq:ksipotstreamsingle}. \emph{Middle panels}: Doppler only. For massive enough halo the negative potential contribution dominates over the positive Doppler contribution.The linear prediction is given by dashed lines. \emph{Right panels}: residual term.}
    \label{fig:massdependance}
\end{figure*}

\subsubsection{Mass dependence of the contributions}
So far, we have focused on the cross-correlation between haloes of mass $\sim 4.5\times10^{13}~h^{-1}$M$_{\odot}$ and haloes of mass $\sim 2.8\times10^{12}~h^{-1}$M$_{\odot}$. In Fig.~\ref{fig:massdependance}, we investigate the halo mass dependence of the main dipole contributions (velocity, potential). The mass dependence on the \emph{residual} is shown in Appendix \ref{sec:appendix_massdependancelinear}. We explore various configurations by cross-correlating all the different halo populations with the lightest halo population. At large linear scales the variation of the dipole is mostly governed by the bias difference between the two halo populations, however at small non-linear scales the evolution of the dipole is less trivial. The velocity contribution to the dipole does not evolve strongly with halo mass. It stays bounded in the range $0<\xi_1< 1\times10^{-3}$. On the other hand, the potential contribution becomes more negative at larger mass from $\xi_1 \simeq -5~\times~10^{-4}$ to $\xi_1 \simeq -1~\times~10^{-2}$ at $6~h^{-1}$Mpc.  It means that for massive enough haloes the potential contribution dominates over the velocity contribution for a wide range of scales (as seen previously). However for haloes lighter than $\sim 10^{13}~h^{-1}$M$_{\odot}$ the velocity-contribution dominates. The \emph{residual} also departs from 0 at larger radii for heavier haloes. Interestingly it is mostly following the potential contribution. \\

The prediction of the potential effect from Eq.~\eqref{eq:ksipotstreamsingle} (assuming spherical symmetry) reproduces the trend at a qualitative level. 
However the potential contribution is overestimated.
Taking into account the dispersion around the potential deduced from spherical symmetry as in Eq.~\eqref{eq:ksipotstream} should improve the agreement with the measured dipole \citep{cai2017gravitational}. Note that we have checked (see Appendix~\ref{sec:appendix_b01}) that our conclusions still hold for a very different halo definition (i.e. linking length $b = 0.1$). The main difference is a slightly better agreement with the spherical predictions for the potential contribution to the dipole.

\section{Conclusions}

In this work we explored the galaxy clustering asymmetry by looking at the dipole of the cross-correlation function between halo populations of different masses (from Milky-Way size to galaxy-cluster size). We took into account all the relevant effects which contribute to the dipole, from lensing to multiple redshift perturbation terms. At large scales we obtain a good agreement between linear theory and our results.
At these scales the dipole can be used as a probe of velocity field (and as a probe of gravity through the Euler equation). However one has to consider a large enough survey to overcome important real-space statistical fluctuations. It is also important to take into account the light-cone effect and to accurately model the bias and its evolution.

At smaller scales we have seen deviation from linear theory. Moreover the gravitational redshift effect dominates the dipole below $10~h^{-1}$Mpc. It is therefore possible to probe the potential outside groups and clusters using the dipole. By subtracting the linear expectation for the Doppler contribution it is in principle possible to probe the potential to even larger radii.  
This is a path to explore in order to circumvent the disadvantages of standard probes of the potential, usually relying on strong assumptions (such as hydrostatic equilibrium) or being only sensitive to the projected potential (lensing).
A simple spherical prediction allows to predict the global trend of the dipole but not the exact value. Moreover as we have seen the \emph{residual} (i.e all the cross terms and non-linearities of the mapping) is of the same order as the gravitational potential contribution and should be taken into account properly. At small scales the pairwise velocity PDF is also highly non-Gaussian, leading to high peculiar velocities and Finger-of-God effect. Coupled to gravitational potential and possibly wide-angle effect we expect this to be a non-negligible contribution to the dipole. To fully understand and probe cosmology or modified theories of gravity at these scales using the cross-correlation dipole we therefore need a perturbation theory or streaming model which takes into account more redshift perturbation terms and relaxes the distant observer approximation. This will be the focus of a future paper.

There are multiple possible extensions to this work. At large Gpc scales current analysis are limited by the volume of the simulation as well as gauge effect. 
At smaller scales the baryons as well as the finite resolution effect might play a role.
Extension of this work in these two directions can open interesting perspectives. When analysing future surveys, it is also important to consider observational effects. One possibility would be to populate haloes with galaxies and to incorporate effects such as magnification bias, absorption by dust, redshift errors, alignment of galaxies, etc. 
Another straight-forward extension is to explore the influence of cosmology, dark energy, dark matter and modified gravity on the dipole of the halo cross-correlation to shed light on the nature of the dark sector with future large scale surveys.

\section*{Acknowledgements}

MAB and YR acknowledge financial support from the DIM ACAV of the Region Ile-de-France, the Action f\'ed\'eratrice \emph{Cosmologie et structuration de l'univers} as well as the JSPS Grant L16519. AT acknowledges financial support from MEXT/JSPS KAKENHI Grant Numbers JP15H05889 and JP16H03977. SS acknowledge financial support from Grant-in-Aid for JSPS Research Fellow Number 17J10553.

This work was granted access to HPC resources of TGCC/CINES through allocations made by GENCI (Grand Equipement National de Calcul Intensif) under the allocations 2016-042287, 2017-A0010402287 and 2018-A0030402287. We thank Fabrice Roy for important technical support and providing pFoF, and Vincent Reverdy for providing \textsc{Magrathea}. We also thank C. Murray for comments on the catalogues, R. Teyssier for \textsc{Ramses}, D. Alonso for \textsc{Cute} and B. Li for sharing his TSC routine.
MAB thanks Pier-Stefano Corasaniti, Stefano Anselmi and Paul de Fromont for fruitful discussions. YR thanks Ichihiko Hashimoto and Julian Adamek.

To complete this work, discussions during the workshop, YITP-T-17-03, held at Yukawa Institute for Theoretical Physics (YITP) at Kyoto University were useful. Numerical calculations for the present work have been carried out  partly at the Yukawa Institute Computer Facility. 



\bibliographystyle{mnras}
\bibliography{biblio} 



\newpage
\appendix

\section{Bias measurement and evolution}
\label{sec:appendix_bias}

\begin{figure} 
	\includegraphics[width=\columnwidth]{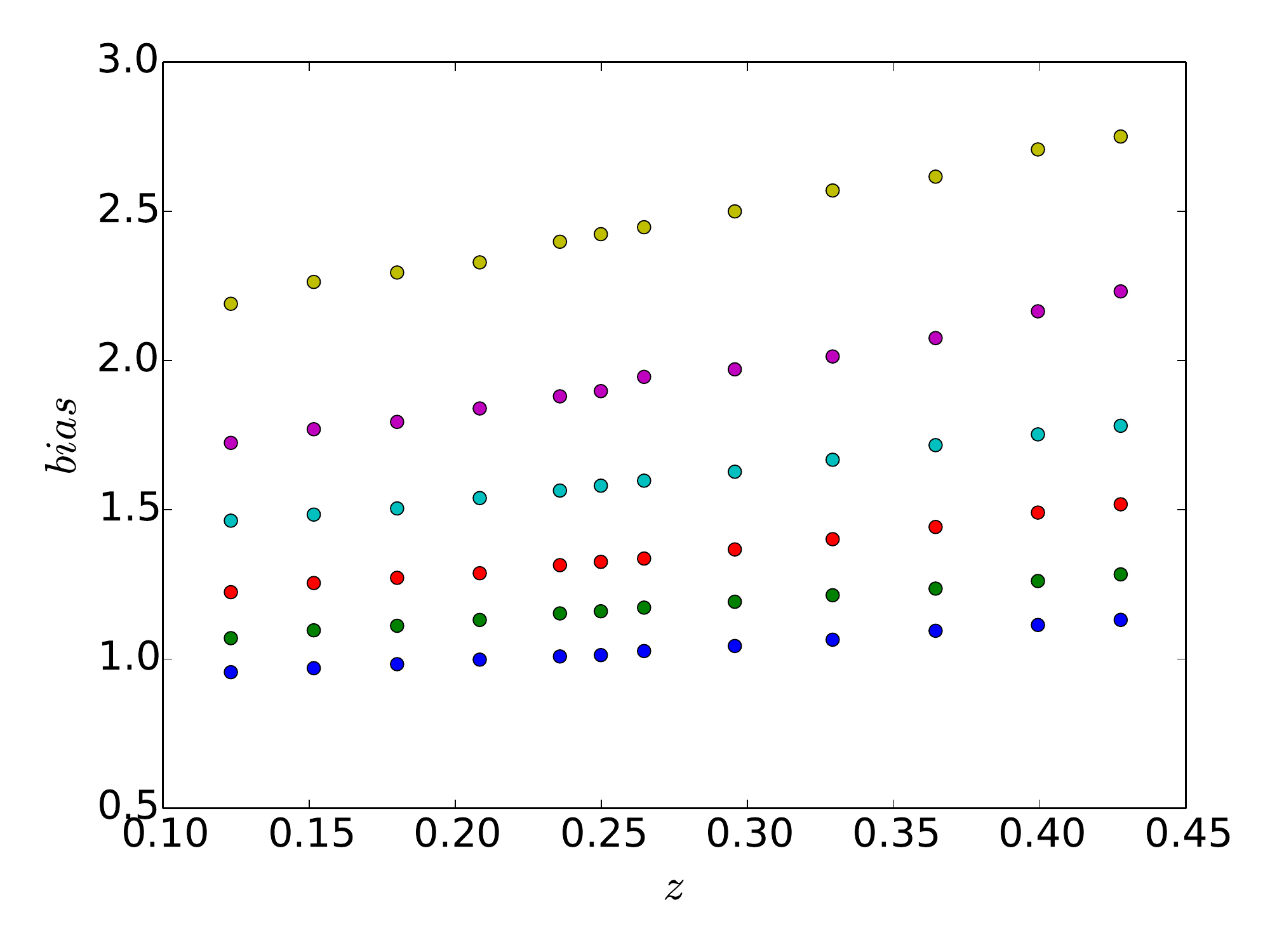}
     \includegraphics[width=\columnwidth]{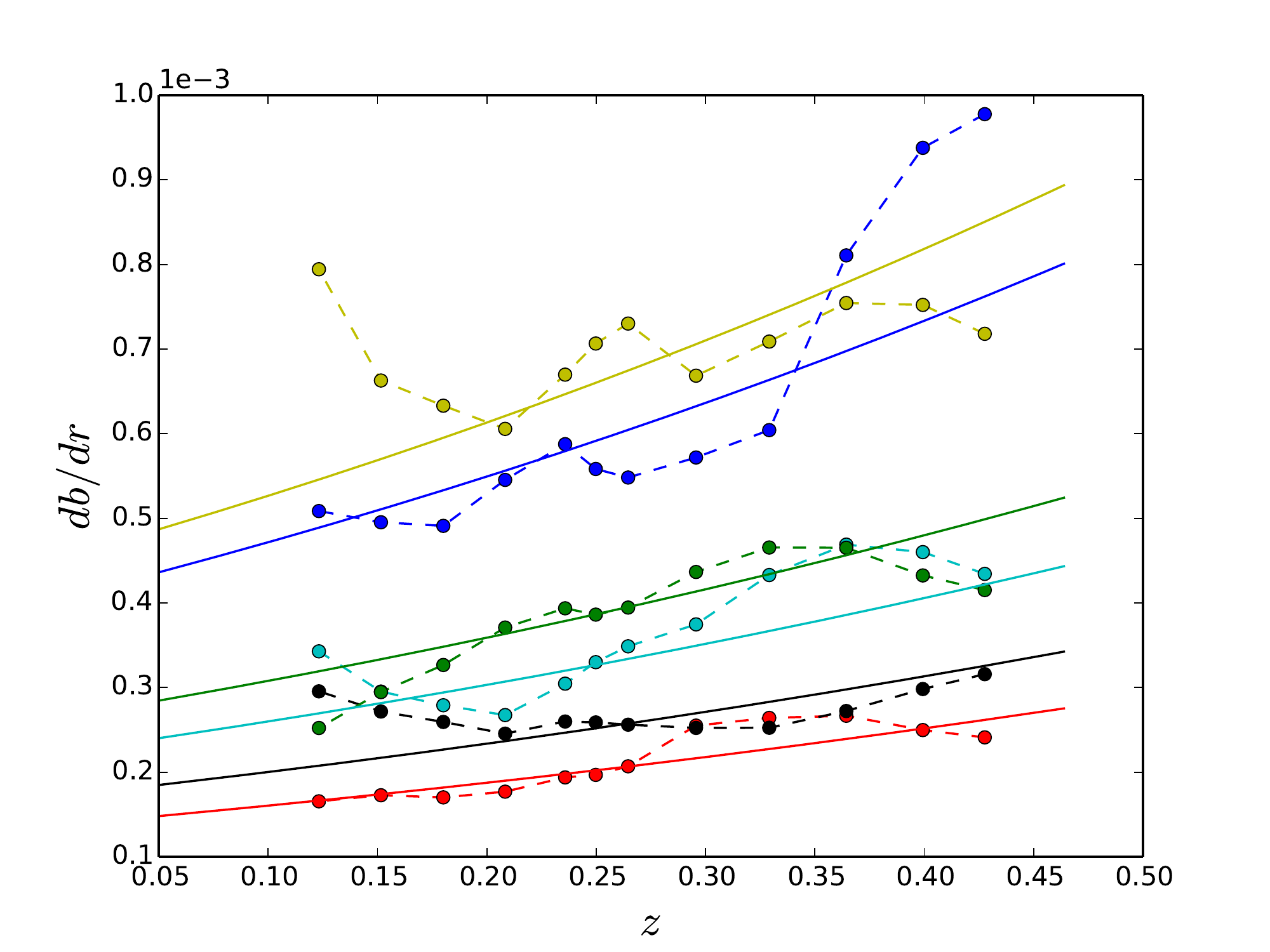}
    \caption{\emph{Top panel}: bias of the halo populations summarised in Table~\ref{tab:datasets} computed on snapshots. \emph{Bottom panel}: Derivative of bias w.r.t comoving distance, comparison of our computation (circles, dashed lines) with the prediction from the \citet{sheth1999large} model Eq.~\eqref{eq:derivbiassheth} (full lines). The associated halo populations are, from top to bottom: data$\_$H$_{3200}$, data$\_$H$_{1600}$, data$\_$H$_{800}$, data$\_$H$_{400}$, data$\_$H$_{200}$, data$\_$H$_{100}$}
    \label{fig:snapshotbias_derivbias}
\end{figure}

In this section we present the methodology used to compute the bias for each halo population summarised in Table~\ref{tab:datasets}. In our case we will consider a scale-independent bias which can be directly fitted with a constant using Eq.~\eqref{eq:bias}. Even if we perform our analysis on the light-cone, computing the bias from the full light-cone monopole would give very poor results (see Fig.~\ref{fig:monopole1209}) for haloes with increasing mass. We therefore compute the bias using snapshots with the same comoving volume as our full light-cone, and interpolating between snapshots to find the bias of a given population at a given redshift. From a numerical point of view, it would be more accurate to use the halo-matter cross-correlation instead of square root of the halo-halo auto-correlation since we would have more particles than haloes and therefore less noise. However this would mean saving all the particles for each snapshot which is unrealistic for our simulation. Instead, we compute $\xi_{mm}$ using \textsc{CosmicEmu} \citep{heitmann2016mira} which is in good agreement which the matter auto-correlation monopole we have on the light-cone (see Fig.~\ref{fig:monopole}). We also checked that the emulator power spectrum is in agreement with the power spectrum we compute in each snapshot at one percent level between roughly $k = 0.02$ and $2$ $h$Mpc$^{-1}$. Finally, we cross-correlate each halo population with the least massive one in order to increase the statistics instead of computing the halo auto-correlation. For the data$\_$H$_{100}$ bias we indeed use 
\begin{eqnarray}
b_{100} \approx \sqrt{\frac{\xi^{\ell = 0}_{hh}}{\xi_{mm}}}.
\end{eqnarray}
For other populations we then take
\begin{eqnarray}
b_1 = b_{100} \frac{\xi^{\ell = 0}_{h_1h_2}}{\xi^{\ell = 0}_{mm}},
\end{eqnarray}
where $h_1$ denotes the halo population from which we want to know the bias, $h_2$ denotes the data$\_$H$_{100}$ population with $b_{100}$ its bias.
This was done on 12 snapshots which covers well our full-sky light-cone. The associated redshift are : $z$ =  0.123, 0.152, 0.180, 0.208, 0.236, 0.250, 0.265, 0.296, 0.329, 0.364, 0.399 and 0.428. The results are shown in Fig.~\ref{fig:snapshotbias_derivbias}.

For bias derivative shown in Fig.~\ref{fig:snapshotbias_derivbias}, we use the prediction from \citet{sheth1999large}, which gives us as a result Eq.~\eqref{eq:derivbiassheth}.
In our case, for each mass bin (increasing in mass) we set $\sigma(M,0) \equiv \sigma_0$ = 1.60, 1.44, 1.27, 1.17, 0.95 and 0.9.

For the most massive bins the \citet{sheth1999large} computation of the bias can be very different from our numerical results, therefore for the dipole predictions we keep the bias from our computation on snapshots but we take the bias derivative from Eq.~\eqref{eq:derivbiassheth} as it is smoother.

\section{Different halo definition}
\label{sec:appendix_b01}
In this appendix we perform the same analysis as in core of the paper but with a different halo definition. 
In the paper we detected haloes using the friend-of-friend algorithm with $b = 0.2$.
To check the sensitivity of our results about the halo definition we here consider a very different linking length, namely $b=0.1$. While $b=0.2$ corresponds to an enclosed over-density very roughly of order of 200 times the mean density of the universe, $b=0.1$ corresponds to an enclosed over-density much larger (approximately 8 times more). Most of the usual halo definitions lie somewhere in between these two definitions.
The datasets are shown in Table.~\ref{tab:datasets_fofb01000m}, where the bias is computed on the full light-cone contrarily to previous datasets where the bias was estimated by interpolating between snapshots. For previous datasets these two methods agree at the $1\%$ level.

\begin{figure} 
	\includegraphics[width=\columnwidth]{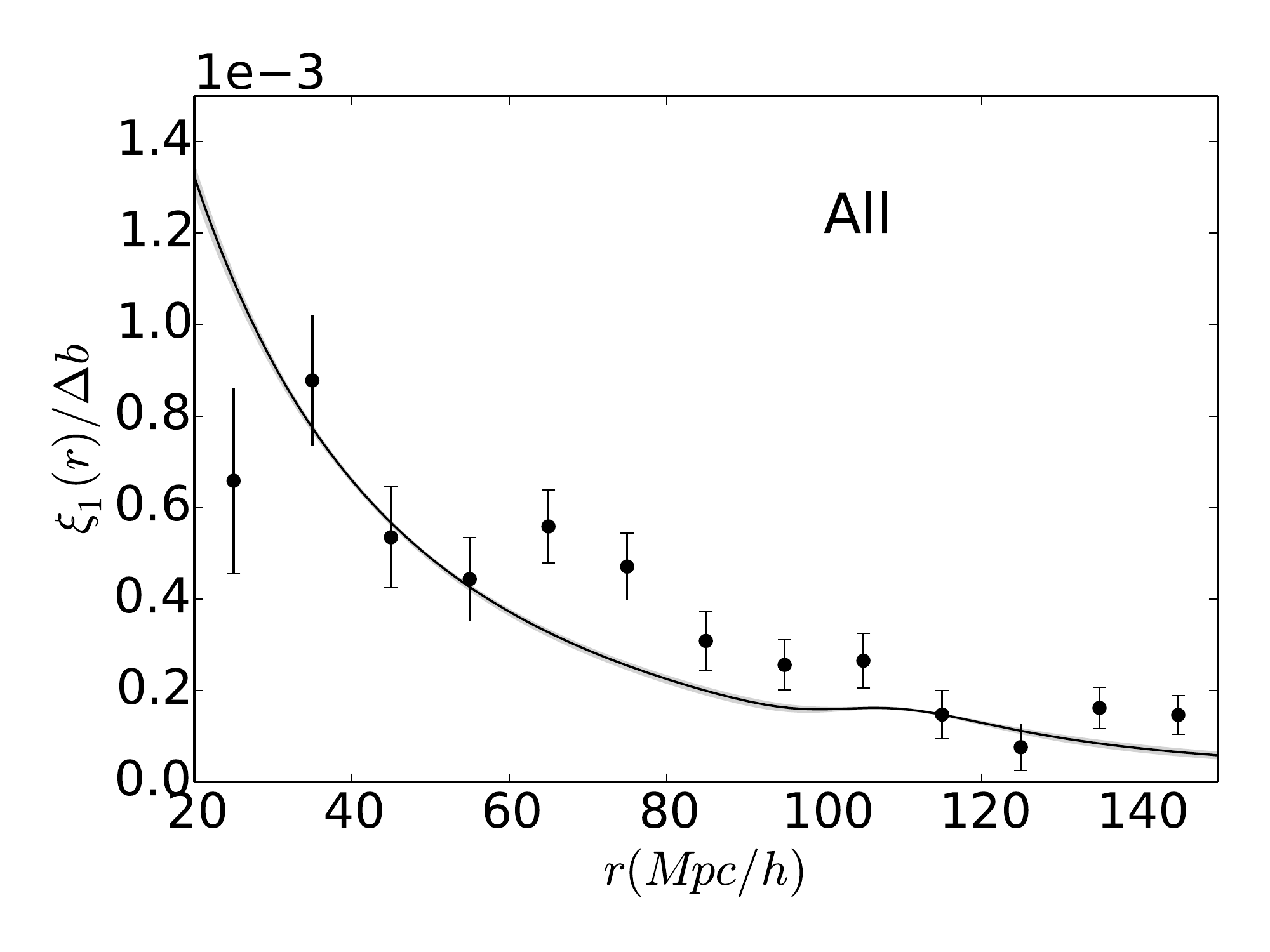}
    \caption{Full dipole of the cross-correlation function normalised by the bias for another halo definition ($b = 0.1$). The dipole is dominated by the Doppler contribution.}
    \label{fig:alleffectslarge_fofb01000m}
\end{figure}

\begin{figure*}
    \includegraphics[width=2\columnwidth]{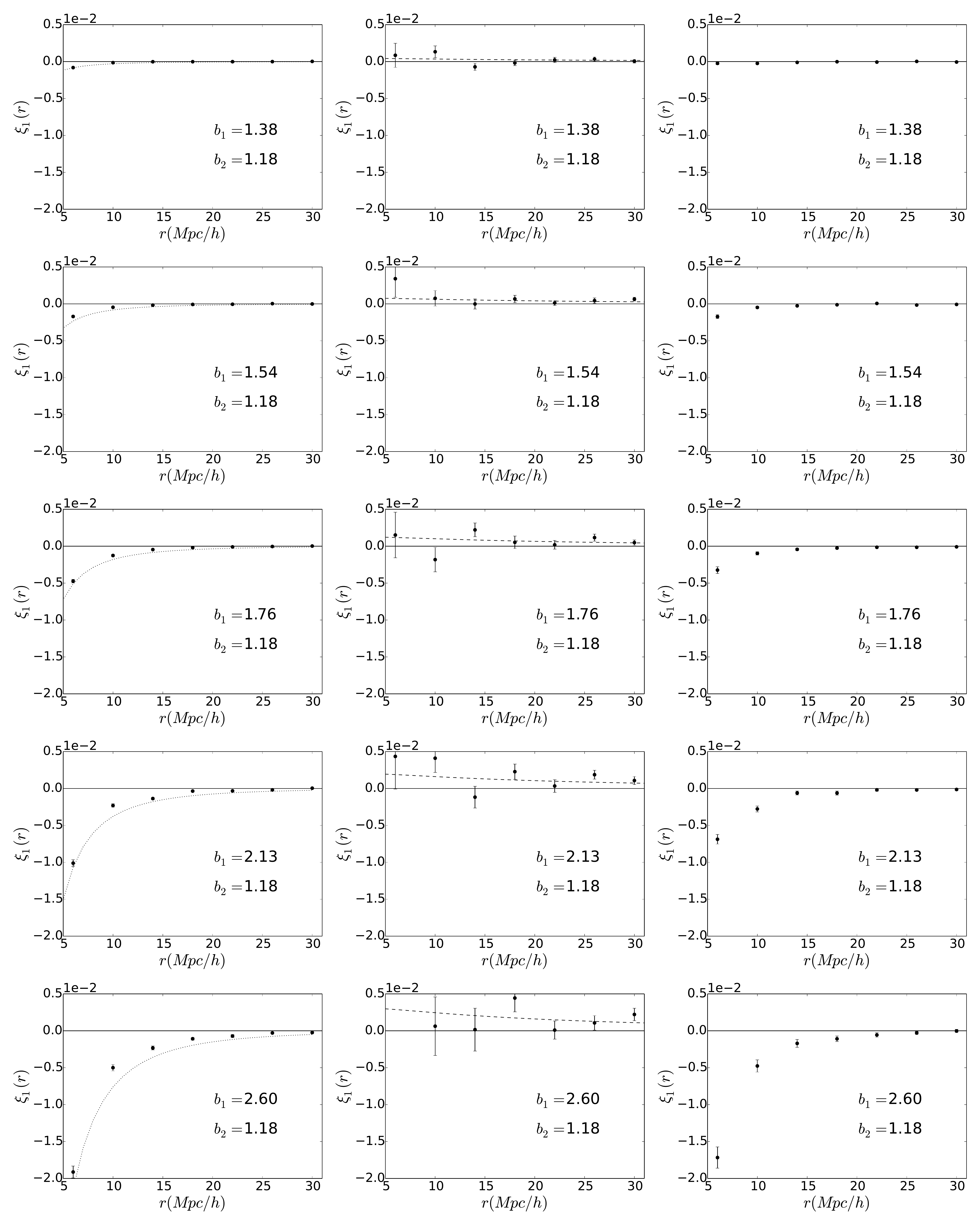}
    \caption{Dipole of the cross-correlation function between different datasets and data2$\_$H$_{050}$ (i.e. for another halo definition $b = 0.1$). \emph{Left panels}: gravitational potential only, dotted lines gives the spherical prediction computed using Eq.~\eqref{eq:ksipotstreamsingle}. \emph{Middle panels}: Doppler only. For massive enough halo the negative potential contribution dominates over the positive Doppler contribution.The linear prediction is given by dashed lines. \emph{Right panels}: residual term. In the bottom plot of the middle column, the point at $6~h^{-1}$Mpc is at $\xi_1 = -0.03$.}
    \label{fig:massdependance_fofb01000m}
\end{figure*}

For linear scales, on a qualitative level the results are similar to the ones using another halo definition. Quantitatively, it seems that we slightly under-estimate the Doppler effect. The remarks on the full dipole shown in Fig.~\ref{fig:alleffectslarge_fofb01000m} are also similar. More interestingly, for the quasi-linear and non-linear regime in Fig.~\ref{fig:massdependance_fofb01000m} we can notice three things. First the results are qualitatively similar the haloes with $b = 0.2$. Second there is a better agreement with $b = 0.1$ with the spherical prediction for the non-linear dipole for the potential only term. This is due to the fact that haloes are now more clustered and sit in deeper potential wells, enhancing the amplitude of the dipole. Still we see that it does not match completely the theoretical prediction. Last, we see that similarly to $b = 0.2$, the point at $6~h^{-1}$Mpc for Doppler only is very negative for the last correlation (most massive halo population with the lightest). This may be due to a coupling between Finger-of-God and wide-angle effects.
\begin{table}
	\centering
	\caption{Summary of the different datasets used: name, number of haloes, range for the number of particles per halo, mean mass, bias at the volume averaged redshift $z=0.341$.}
	\label{tab:datasets_fofb01000m}
	\begin{tabular}{lcccr} 
		\hline
		name & nb of haloes & nb of part & mass ($h^{-1}$M$_{\odot}$) & bias \\
		\hline
		 data2$\_$H$_{050}$ 		& $6.9\times10^6$ 	  & 50-100   & $1.4\times10^{12}$ &1.18\\        
		 data2$\_$H$_{100}$ 		& $3.7\times10^6$ 	  & 100-200   & $2.8\times10^{12}$ &1.38\\
		 data2$\_$H$_{200}$ 		& $2.0\times10^6$ 	  & 200-400   & $5.6\times10^{12}$ &1.54\\
		 data2$\_$H$_{400}$ 		& $1.0\times10^6$ 	  & 400-800   & $1.1\times10^{13}$ &1.76\\    
		 data2$\_$H$_{800}$ 		& $4.3\times10^5$  & 800-1600  & $2.2\times10^{13}$ &2.13\\   
		 data2$\_$H$_{1600}$ 		& $1.7\times10^5$ 	  & 1600-3200 & $4.5\times10^{13}$ &2.60\\    
         \hline
	\end{tabular}
\end{table}

\begin{figure*} 
	\includegraphics[width=\columnwidth]{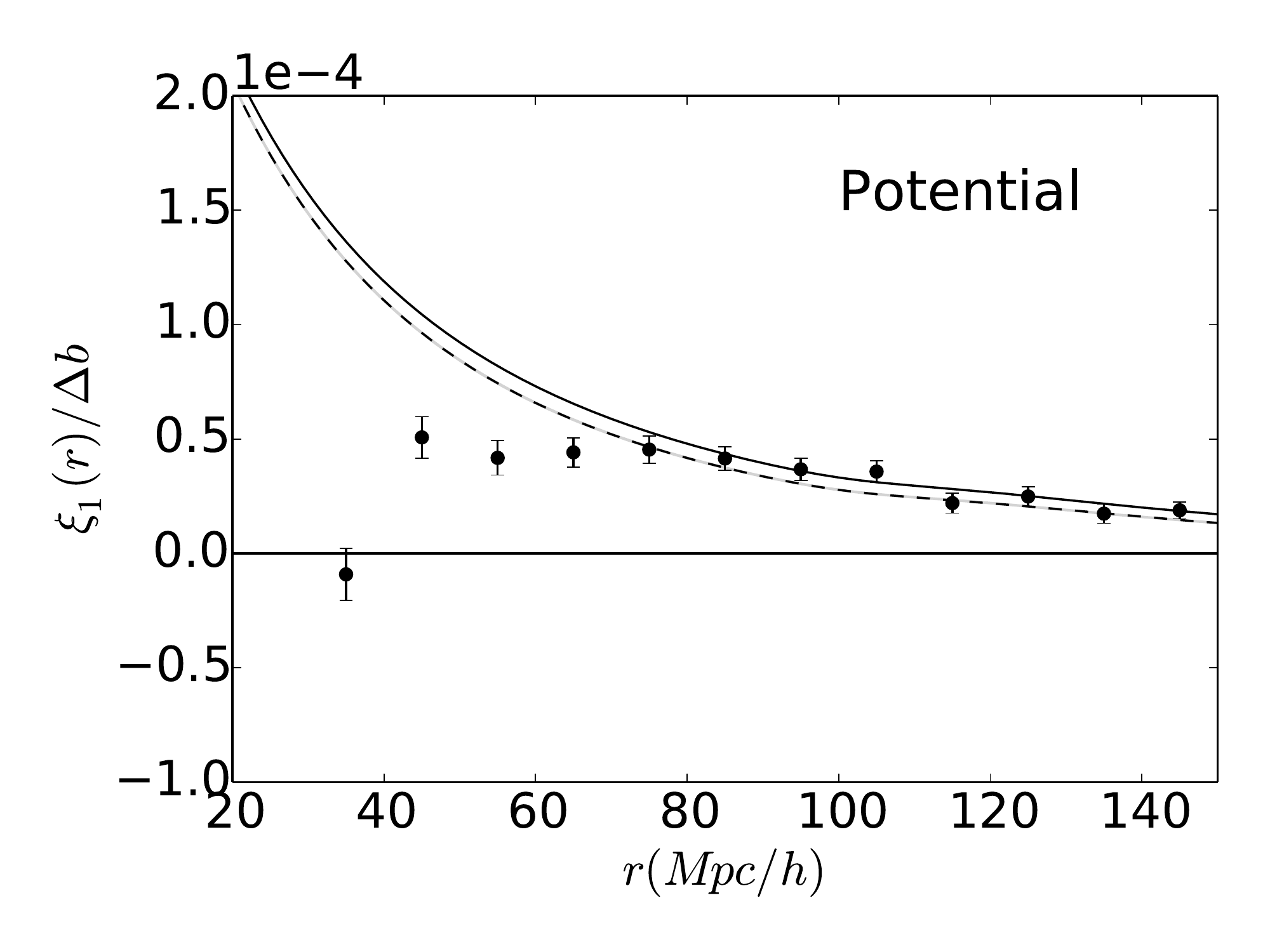}
	\includegraphics[width=\columnwidth]{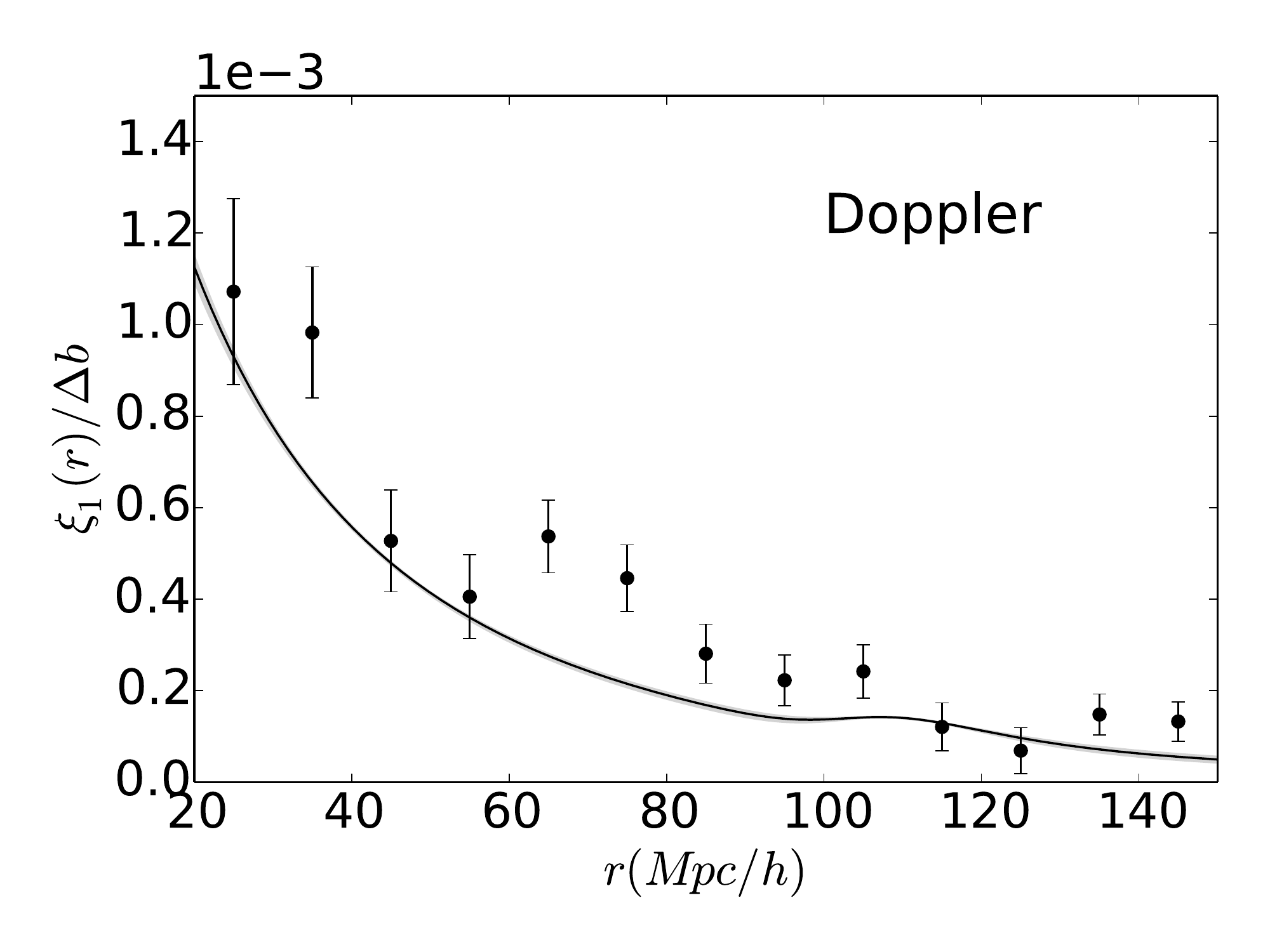}
	\includegraphics[width=\columnwidth]{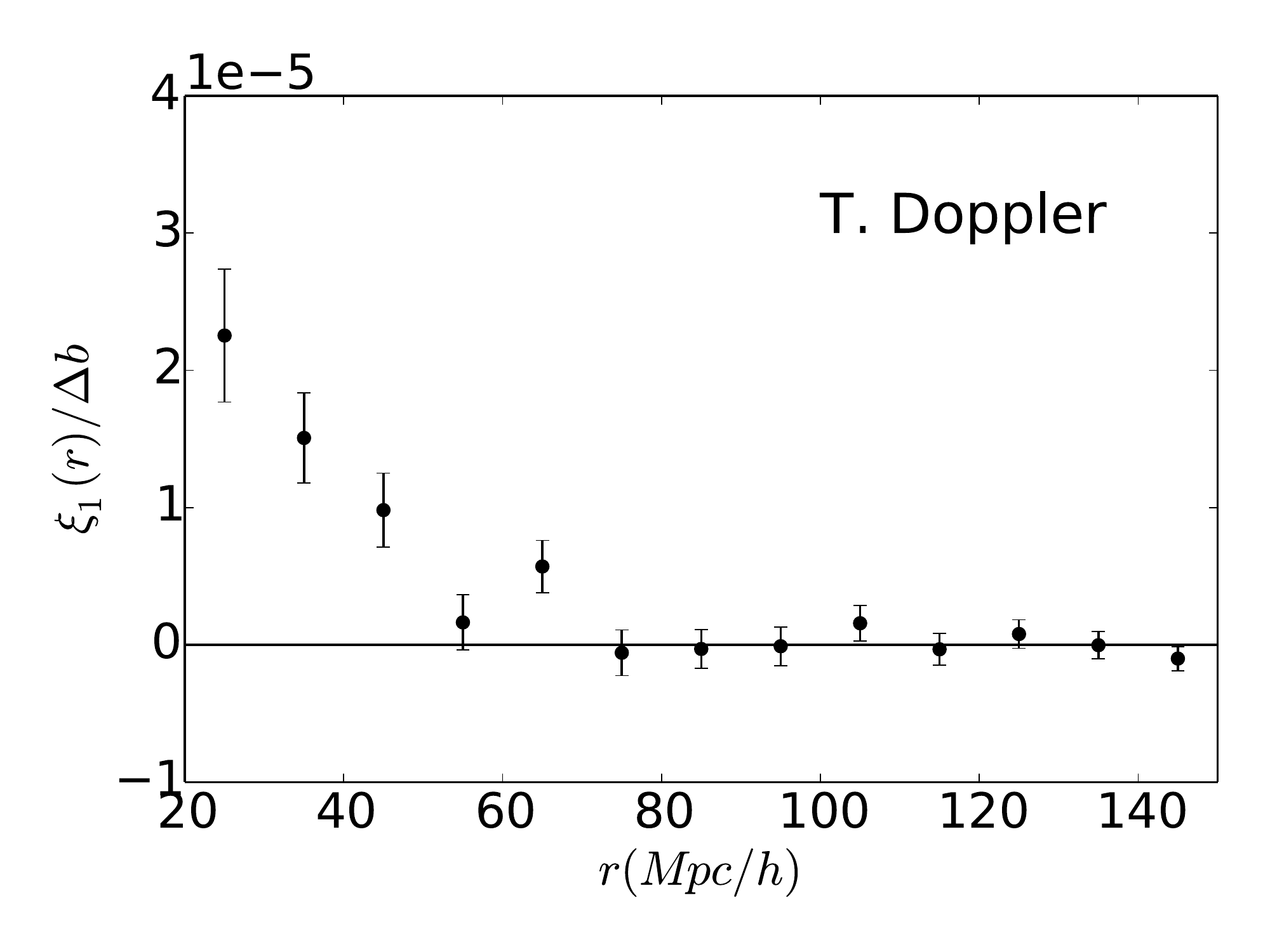}
	\includegraphics[width=\columnwidth]{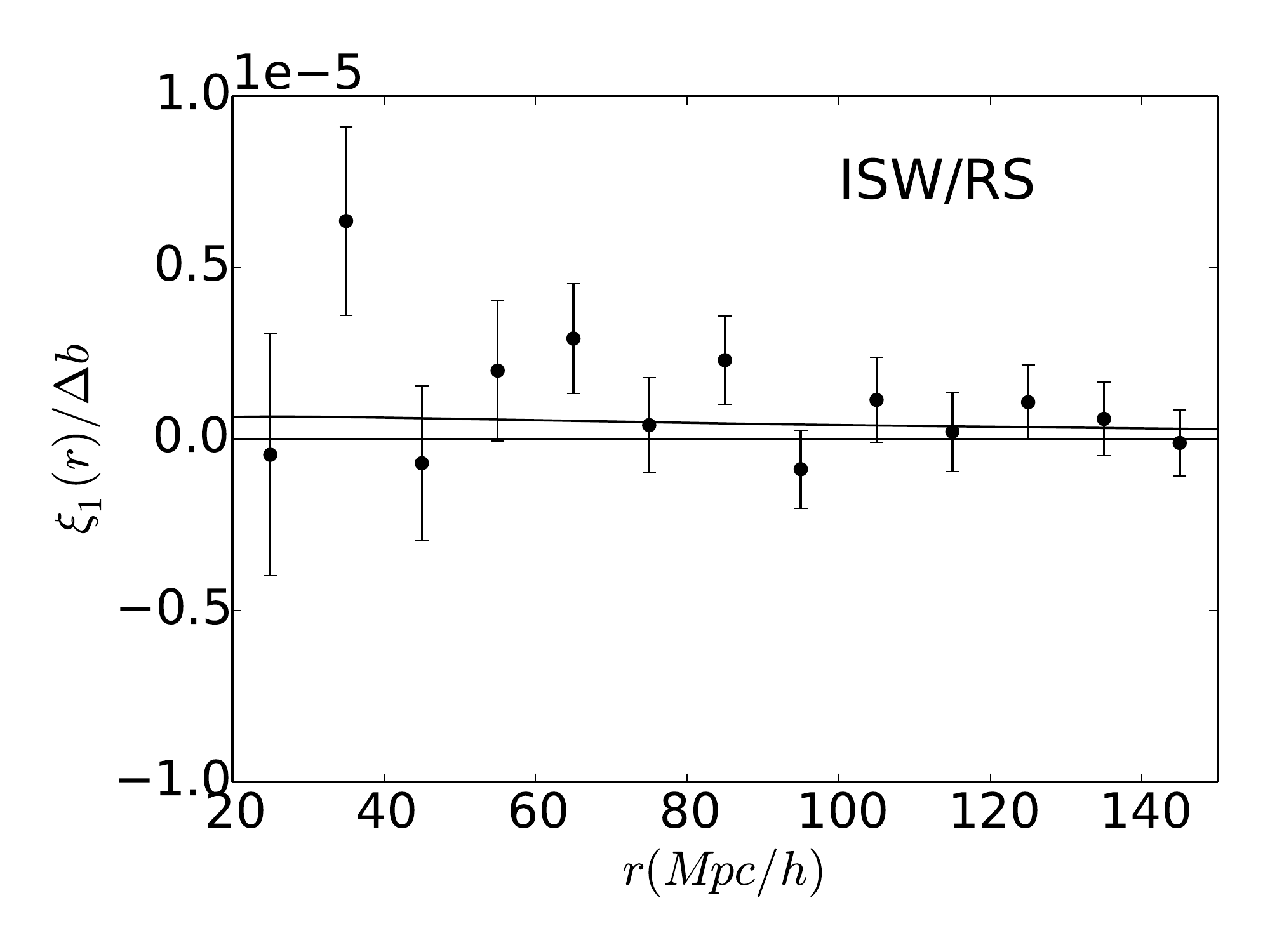}
	\includegraphics[width=\columnwidth]{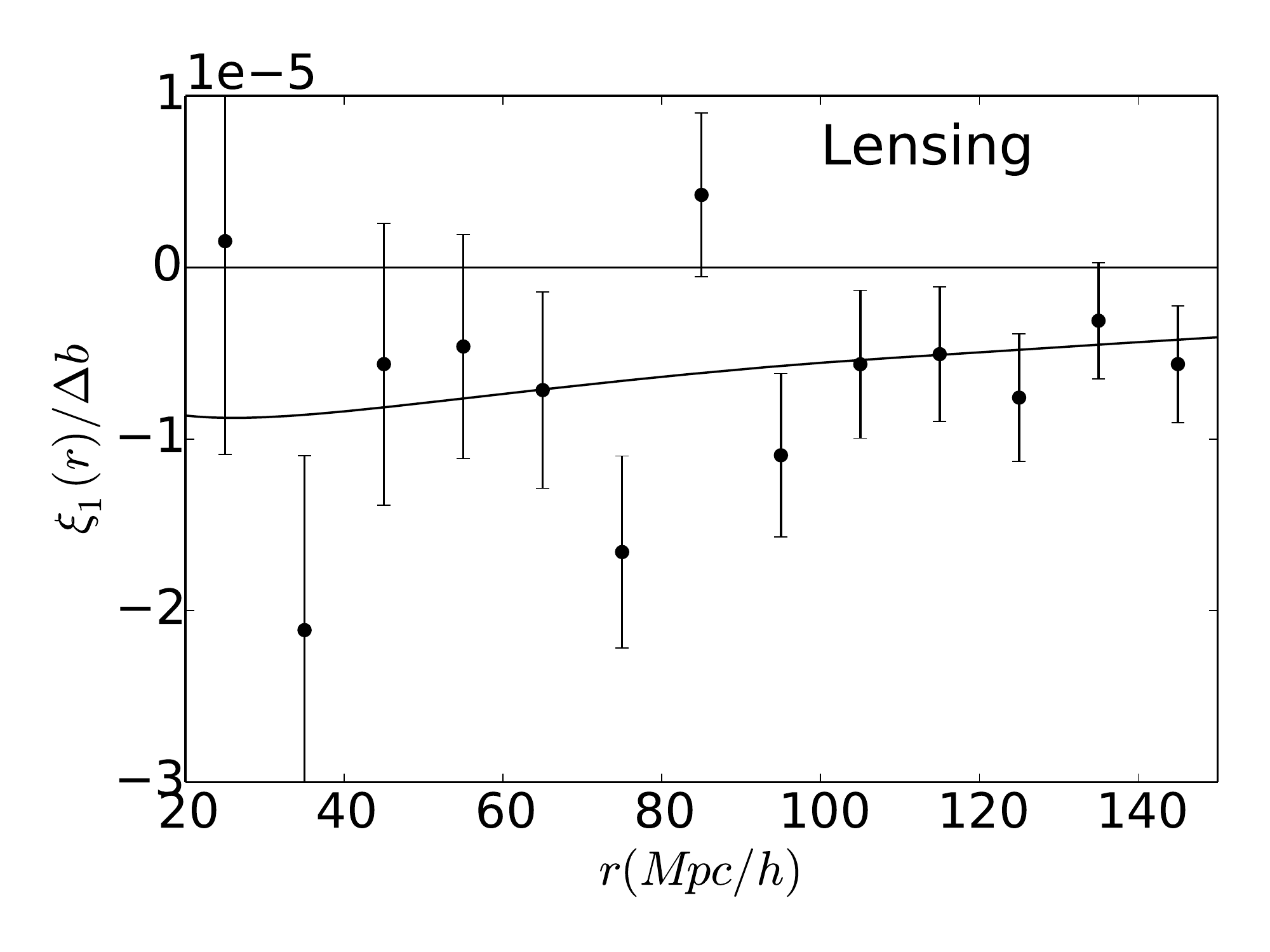}
	\includegraphics[width=\columnwidth]{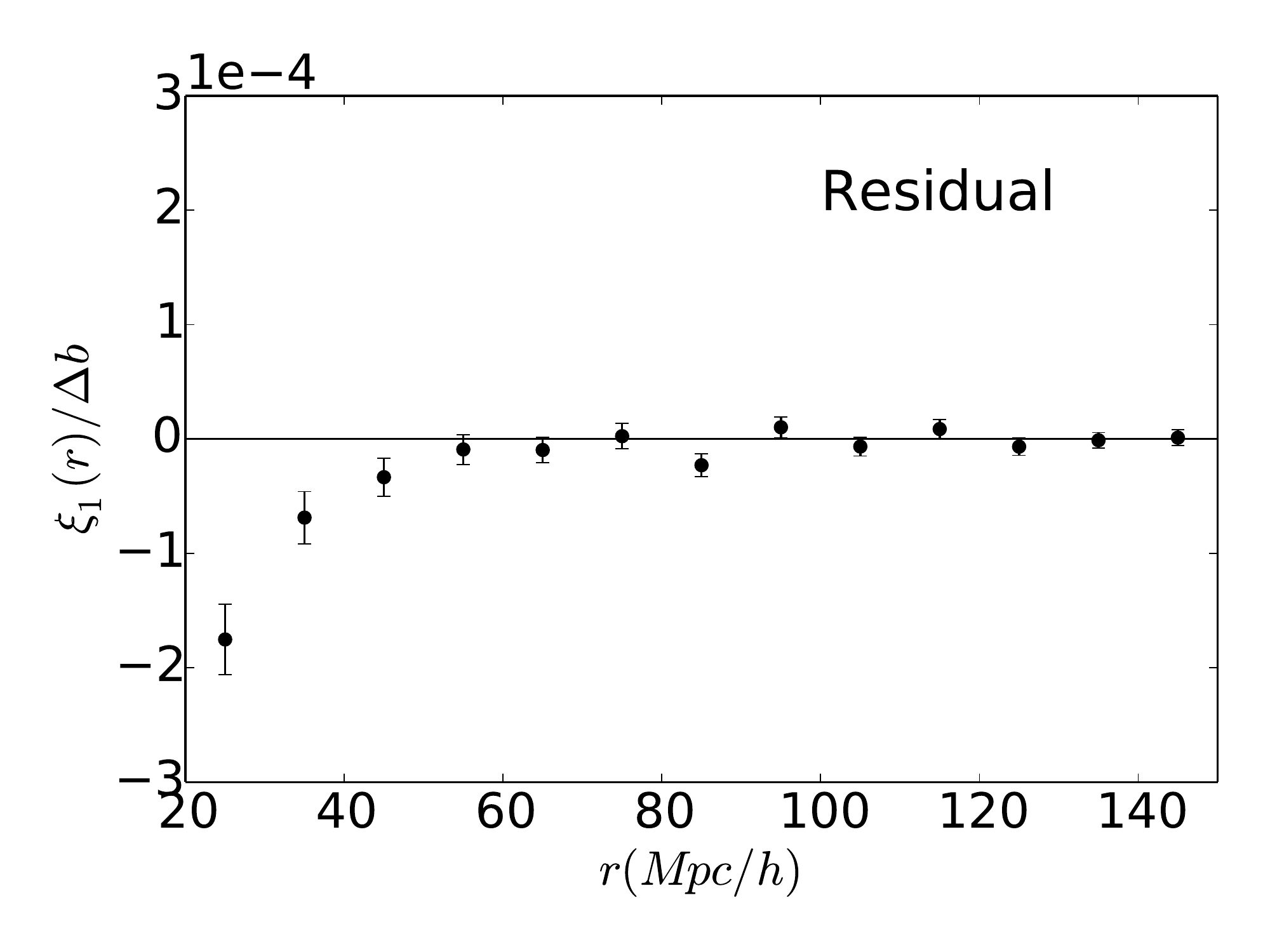}

    \caption{Dipole of the cross-correlation function normalised by the bias, at large scales, for different perturbations of the observed halo number count for another halo definition ($b = 0.1$). 
   This leads to: \emph{upper left panel} only the contribution from gravitational potential was taken into account as a source of RSD, in black dashed line we have the prediction when accounting for leading terms in $\left(\mathcal{H}/k\right)^2$. \emph{Upper right panel} Doppler only, \emph{middle left panel} transverse Doppler only, \emph{middle right panel} ISW/RS only, \emph{bottom left panel} weak lensing  only, and finally \emph{bottom right panel} the \emph{residual} where we subtract all the previous effects to the full dipole taking into account all the effects at once. In black solid lines we show the averaged prediction using linear theory at first order in $\mathcal{H}/k$.}
\end{figure*}

\section{mass dependence of the dipole}
\label{sec:appendix_massdependancelinear}
In Section \ref{sec:statistical_fluctuations} and \ref{sec:linear_regime_dipole}, we presented the computation of the dipole normalised by the bias difference. To do so we used all the cross-correlations available with our datasets shown in Table~\ref{tab:datasets}. We then performed a sum on the dipoles, weighted by the inverse of their variance (see Section~\ref{sec:estimation_correlationfunction}). In this Section we show the different cross-correlations for each perturbation effect, and for every combination of populations at large scales. 
We show the results for the computation of the cross-correlation on the full light-cone using jackknife re-sampling.

\begin{figure*} 
\includegraphics[width=2\columnwidth]{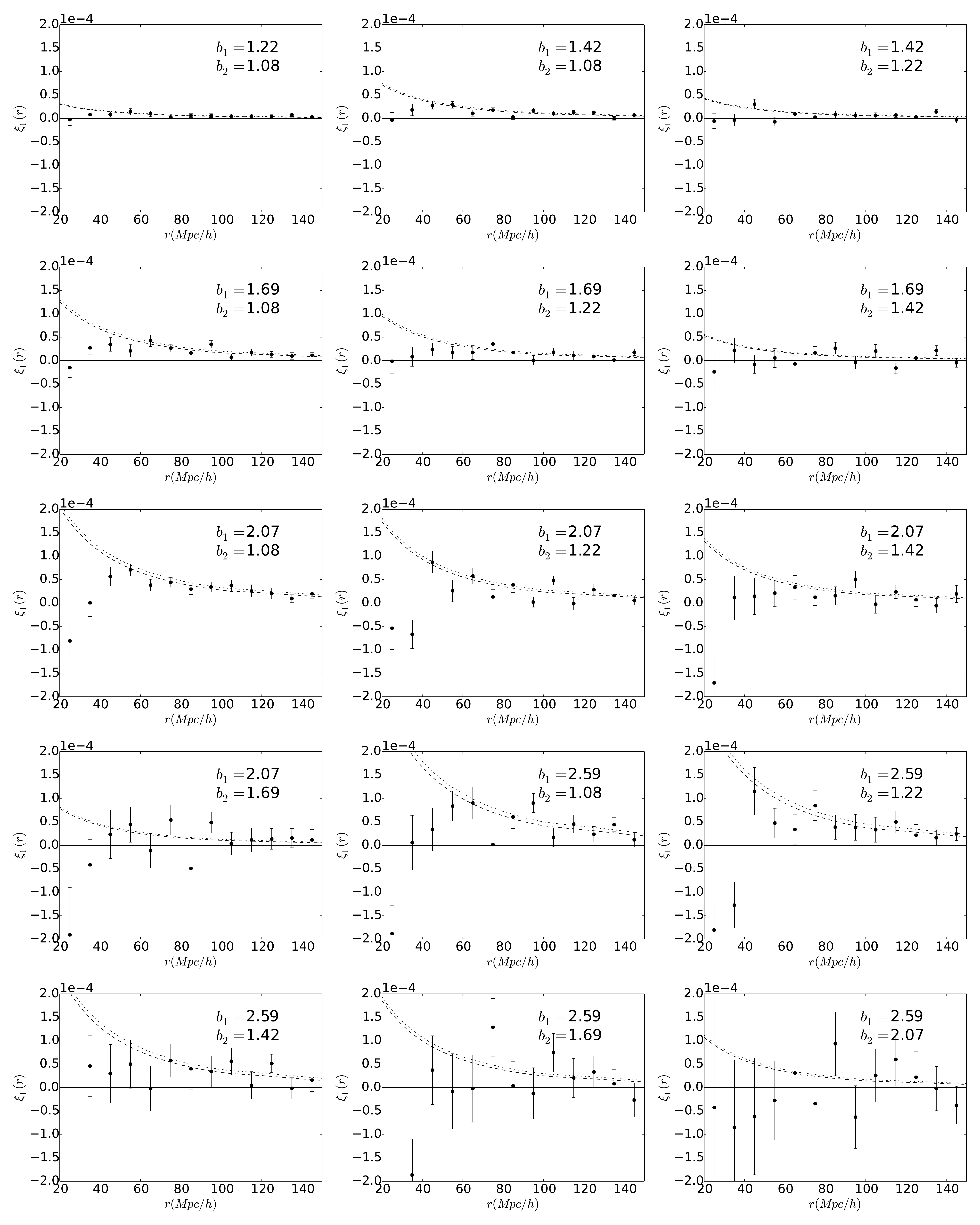} 
    \caption{Potential only term of the dipole of the cross-correlation function on the full light-cone at large scales. The linear predictions at first order in $\mathcal{H}/k$ are shown in dash-dotted lines while the prediction with the dominant $\left(\mathcal{H}/k\right)^2$ terms is shown in dashed lines.}
\end{figure*}

\begin{figure*} 
\includegraphics[width=2\columnwidth]{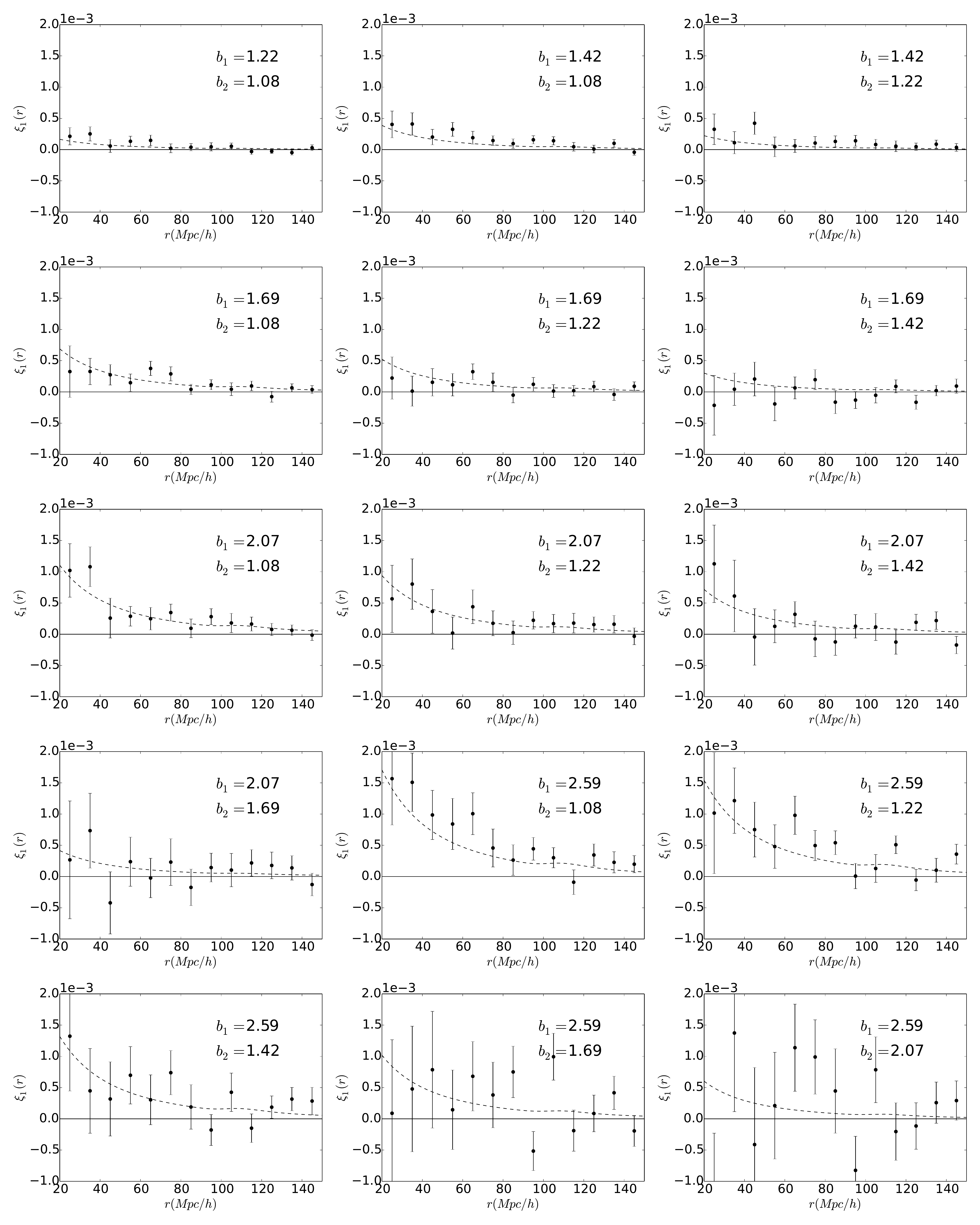} 
    \caption{Doppler only term of the dipole of the cross-correlation function on the full light-cone at large scales. The linear predictions are shown in dashed lines.}
\end{figure*}

\begin{figure*} 
\includegraphics[width=2\columnwidth]{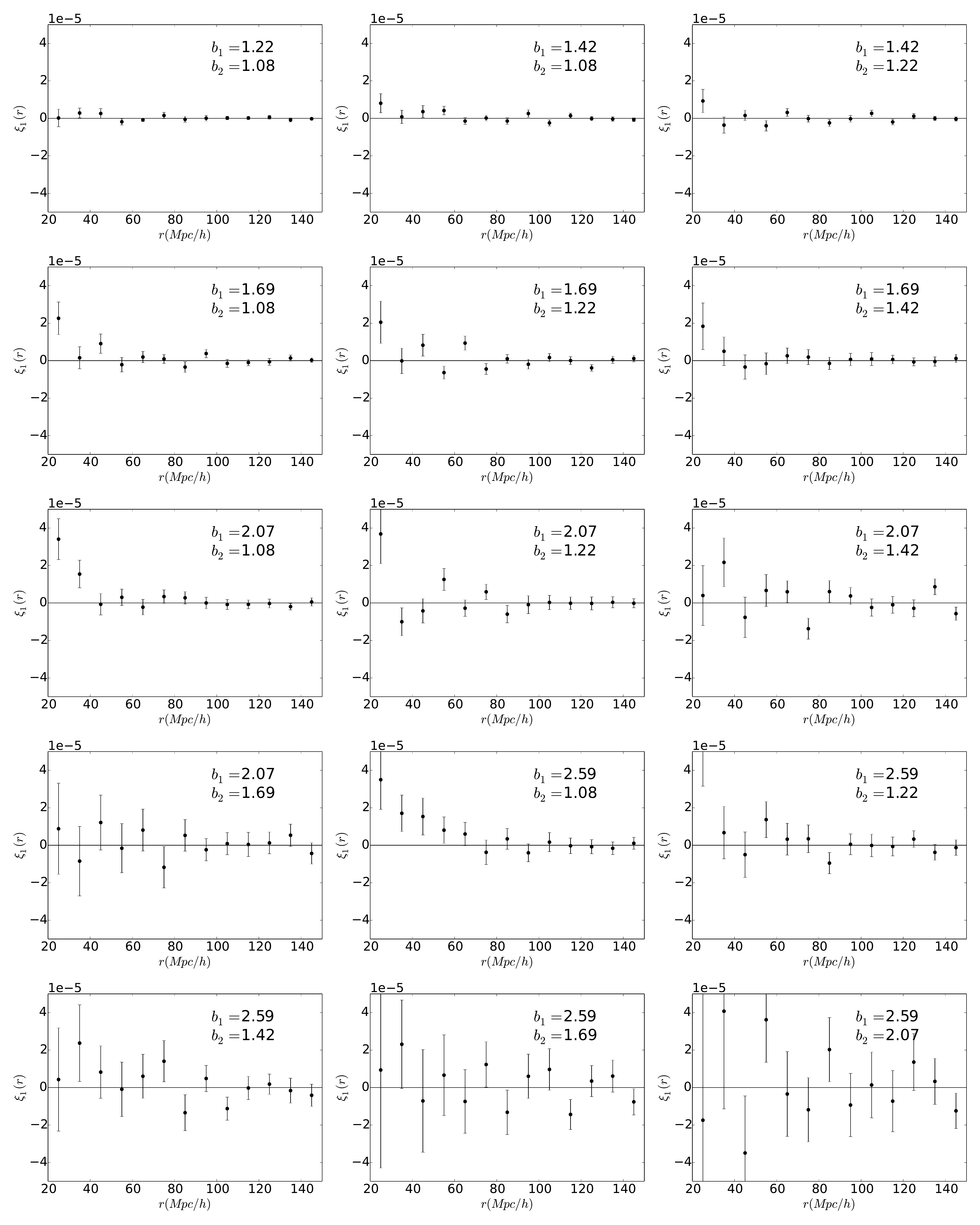} 
    \caption{Transverse Doppler only term of the dipole of the cross-correlation function on the full light-cone at large scales.}
\end{figure*}

\begin{figure*} 
\includegraphics[width=2\columnwidth]{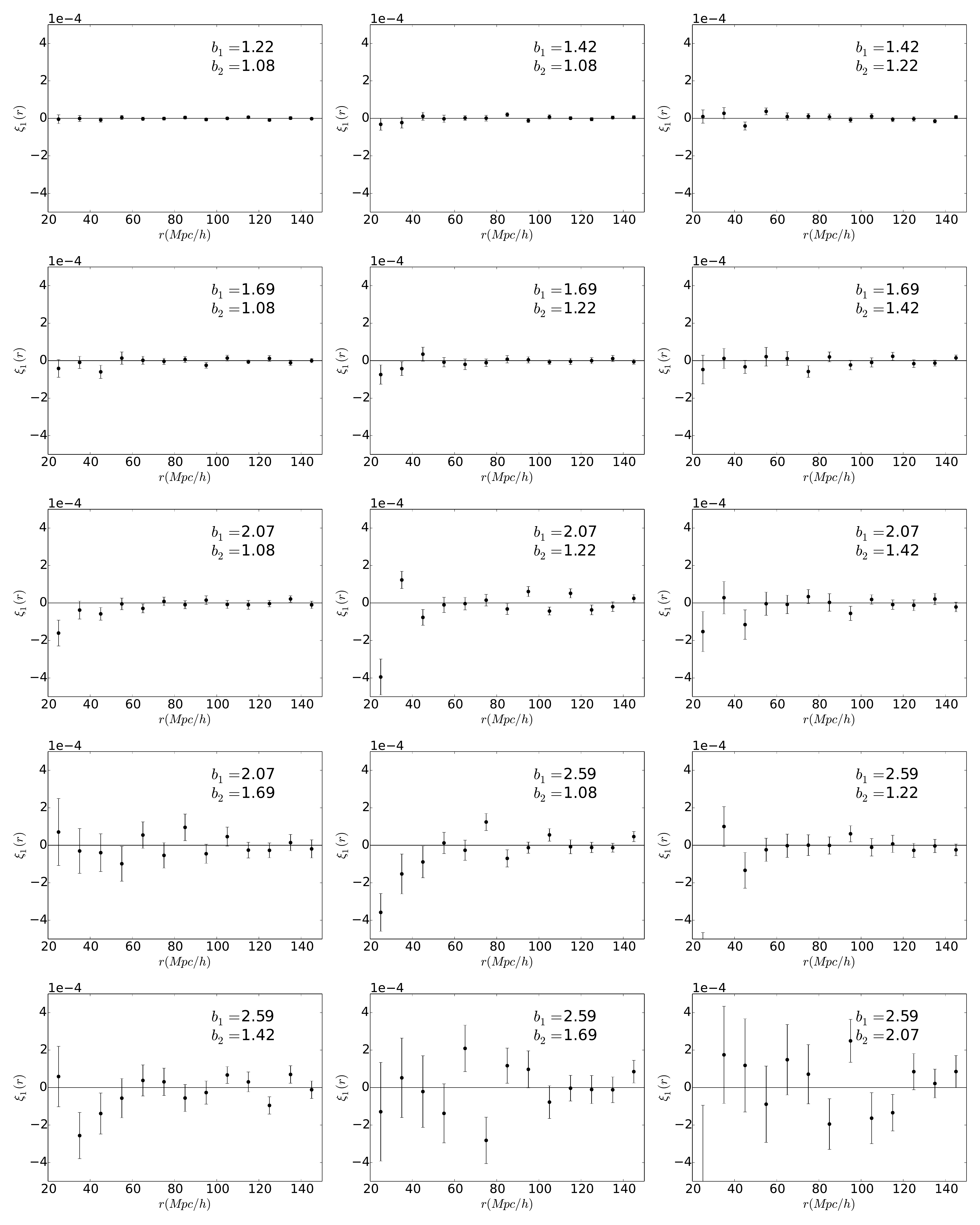} 
    \caption{\emph{Residual} term of the dipole of the cross-correlation function on the full light-cone at large scales. }
\end{figure*}

\begin{figure*} 
\includegraphics[width=2\columnwidth]{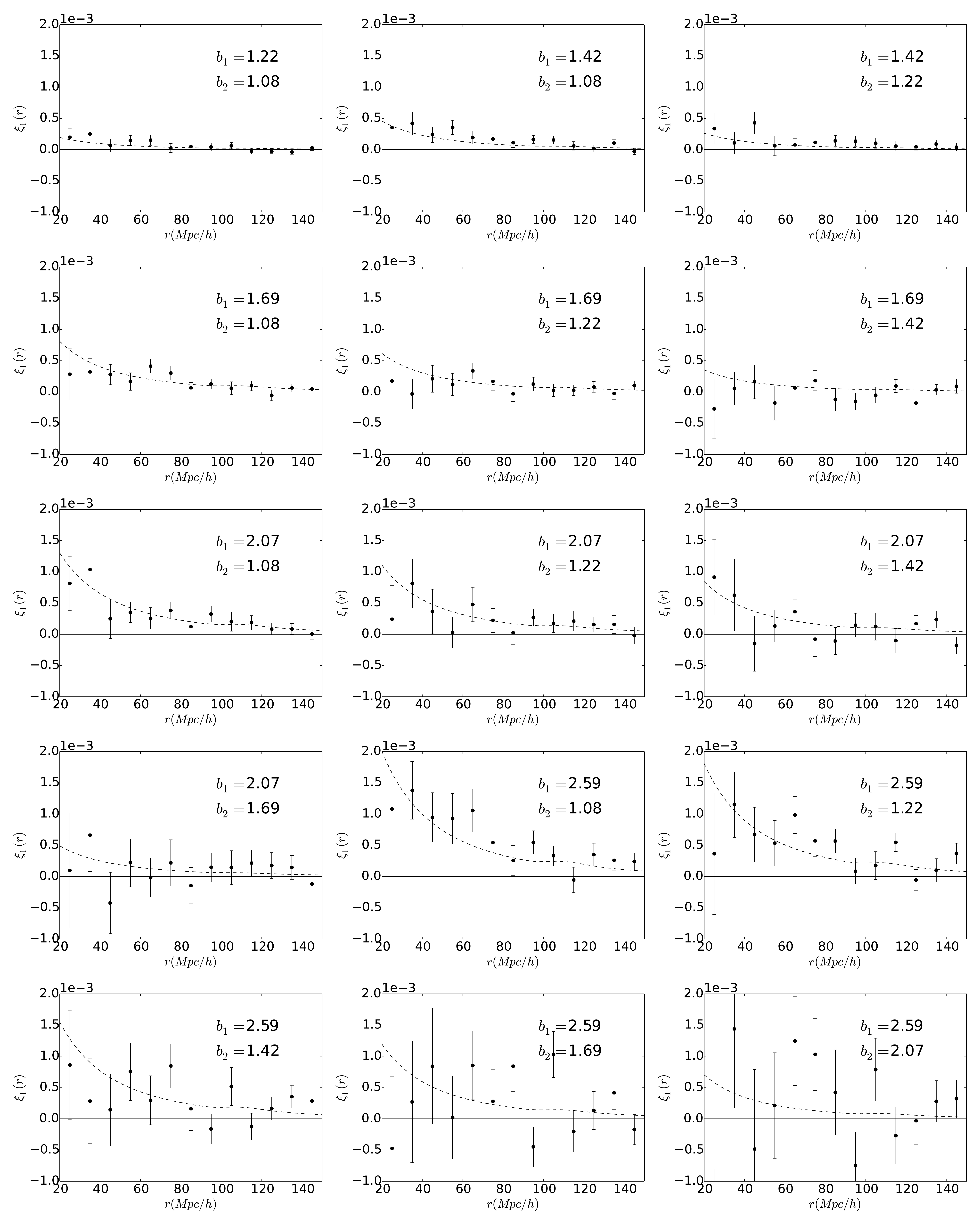} 
    \caption{Full dipole of the cross-correlation function on the full light-cone at large scales. The linear predictions are shown in dashed lines.}
\end{figure*}


\bsp	
\label{lastpage}
\end{document}